\documentclass[11pt,a4paper]{article}
\pdfoutput=1
\usepackage{jheppub}
\usepackage{amsthm,amsbsy,amsfonts,mathrsfs,enumerate,float,wrapfig,amsmath}
\usepackage[utf8]{inputenc}
\usepackage{subfigure}

\usepackage{float}
\usepackage{tikz}
\newcommand{\nn}{\nonumber}
\def\a{{\alpha}}

\def\b{{\beta}}

\def\r{{\gamma}}

\def\0{{\emptyset}}
\def\u{{\mu}}
\def\v{{\nu}}
\def\h{{\eta}}
\def\la{{\lambda}}
\def\bx{{\textbf{x}}}
\def\by{{\textbf{y}}}
\def\p{{\rho}}
\def\tZ{{\tilde{Z}}}

\def\sqqt {{ \sqrt{\frac{q}{t}}}}
\def\sqtq {{ \sqrt{\frac{t}{q}}}}

\def\inf{{\infty}}

\def\l{{\ell}}

\def \l({\left(}
\def \r){\right)}
\def \pert{{\textbf{pert}}}
\def \pertI{{\textbf{pert-I}}}
\def\pertII{{\textbf{pert-II}}}

\def\tA{{\tilde{A}}}
\def\ty{{\tilde{y}}}
\def\tu{{\tilde{u}}}

\def\mAS{{\mathbf{AS}} }
\def\mF{{\mathbf{F}} }
\def\sinh{{\text{sinh}}}
\def\cosh{{\text{cosh}}}

\def\>{{  \succcurlyeq} }
\newcommand{\bsmall}{\begin{small}   }
	\newcommand{\esmall}{ \end{small}   }

\def\oneinst{{\textbf{one-instanton}}}
\def\twoinst{{\textbf{two-instanton}}}

\title{
	Refined topological vertex for a 5D $Sp(N)$ gauge theory with antisymmetric matter
}

\author[a,b]{Shi Cheng,}
\author[a]{Sung-Soo Kim}

\affiliation[a]{School of Physics, University of Electronic Science and Technology of China, \\
No. 2006 Xiyuan Ave, West Hi-Tech Zone, Chengdu, Sichuan 611731, China}
\affiliation[b]{Faculty of Physics, University of Warsaw, ul. Pasteura 5, 02-093 Warsaw, Poland}

\emailAdd{scheng@fuw.edu.pl}
\emailAdd{sungsoo.kim@uestc.edu.cn}

\abstract{
We consider Type IIB 5-brane web diagrams for a 5D $Sp(N)$ gauge theory with an antisymmetric hypermultiplet and $N_f$ fundamental hypermultiplets. The corresponding 5-branes can be obtained by Higgsing a 5-brane web for quiver gauge theory. We use the refined topological vertex formalism to compute Nekrasov partition functions of 5D $Sp(2)$ theories with one antisymmetric hypermultiplet and flavors. Our results agree with the known results obtained from the ADHM method. We also discuss a particular tuning of K\"ahler parameters associated with this Higgsing.
}

\begin{document}
\maketitle

\section{Introduction} \label{sec:intro}
A large class of five-dimensional (5D) $\mathcal{N}=1$ supersymmetric gauge theories \cite{Seiberg:1996bd} 
can be constructed in 
Type IIB 5-brane webs \cite{Aharony:1997ju, Aharony:1997bh} or in M-theory on Calabi-Yau threefolds  \cite{Morrison:1996xf, Douglas:1996xp, Ganor:1996pc, Intriligator:1997pq}. The duality between a 5-brane web in type IIB string theory and a toric Calabi-Yau threefold in M-theory \cite{Leung:1997tw} enables one to utilize the topological string partition function on a toric Calabi-Yau threefold to obtain the BPS partition function of a 5D gauge theory on the dual 5-brane web \cite{Iqbal:2002we, Iqbal:2003ix, Iqbal:2003zz, Eguchi:2003sj, Hollowood:2003cv, Taki:2007dh}. The topological string method known as (refined) topological vertex \cite{Aganagic:2003db, Awata:2005fa,Iqbal:2007ii, Awata:2008ed,Iqbal:2012mt} thus provides another powerful tool for computing the partition function of 5D $\mathcal{N}=1$ gauge theories.

In recent years, there has been much progress on understanding of 5D $\mathcal{N}=1$ gauge theories from the perspective of Type IIB 5-brane webs 
\cite{Hayashi:2013qwa,Bergman:2013aca,Bergman:2014kza,Hayashi:2015fsa,Bergman:2015dpa,Zafrir:2015rga,Hayashi:2015zka,
Jefferson:2017ahm,Jefferson:2018irk}, revealing even larger class of 5D $\mathcal{N}=1$ theories can be realized by 5-brane webs. For instance, 5D $SU(2)$ theories with $5\le N_f\le 7$ hypermultiplets in the fundamental representation (flavors) can be obtained by Higgsing of a 5-brane web for $T_3$, $T_4$ and $T_6$ theories, respectively \cite{Benini:2009gi}. It is straightforwardly generalized to 5D $SU(N)$ theory with $2N+1\le N_f \le 2N+3$ flavors. The dual diagrams for these 5-brane webs are generically non-toric, as they can be understood as a Higgsed diagram of certain quiver theories \cite{Bergman:2014kza,Kim:2015jba,Hayashi:2015fsa}.  
5D $SO(N)$/$Sp(N)$ (quiver) gauge theories with the number of hypermultiplet in the fundamental representation exceeding the bound in \cite{Intriligator:1997pq} can also be constructed from 5-brane webs with the orientifold planes by Higgsing \cite{Bergman:2015dpa,Hayashi:2015vhy}.
Moreover, $Sp(N)$ gauge theories with the hypermultiplet in the antisymmetric representation \cite{Bergman:2015dpa} and $SO(N)$ gauge theories with hypermultiplets in the spinor representation \cite{Zafrir:2015ftn} are also constructed. Even $G_2$ gauge theories with $N_f\le 6$ flavors are represented by 5-brane webs \cite{Hayashi:2018bkd}. Dual diagrams for such 5-brane webs are generically non-toric as they can be obtained by a Higgsing of some certain quiver theory and also by introduction of the orientifolds, implying that the corresponding Calabi-Yau threefolds are non-toric.

The topological vertex formalism also has been implemented to non-toric Calabi-Yau threefolds \cite{Hayashi:2013qwa, Hayashi:2014wfa, Hayashi:2015xla} by tuning K\"ahler parameters \cite{Dimofte:2010tz, Taki:2010bj, Aganagic:2011sg, Aganagic:2012hs}. Even for a non-toric diagram with an $O5$-plane, (unrefined) topological vertex formalism was newly proposed \cite{Kim:2017jqn}, which enables one to compute  the partition function for 5D $G_2$ gauge theories based on 5-brane web with an $O5$-plane \cite{Hayashi:2018bkd}, which agrees with the field theory results up to two instanton contributions \cite{Benvenuti:2010pq,Keller:2011ek, Hanany:2012dm, Keller:2012da, Cremonesi:2014xha, Kim:2018gjo}. Though the topological vertex formalism is applicable to 5D gauge theories of various gauge groups, application of the topological vertex to theories with hypermultiplet other than fundamental hypermultiplet is still limited. 

In this paper, we utilize the topological vertex to compute the partition function for 5D $Sp(N)$ theory with an antisymmetric hypermultiplet and $N_f\le 7$ flavors. The theory has the fixed point at UV where the global symmetry is enhanced to $E_{N_f+1}\times SU(2)$ \cite{Seiberg:1996bd, Morrison:1996xf,Douglas:1996xp, Ganor:1996pc, Intriligator:1997pq}. The partition function for 5D $Sp(N)$ theory with massless antisymmetric hypermultiplet and $5\le N_f\le 7$ flavors was already computed based on web diagrams of Higgsed $T_N$ theories \cite{Hayashi:2015xla}. Here, we consider 5D $Sp(N)$ theories with {\it massive} antisymmetric hypermultiplet and $N_f$ flavors, whose diagrams are obtained from Higgsing of a certain quiver gauge theory \cite{Bergman:2015dpa, Hayashi:2015zka}. To obtain the Nekrasov partition function from the corresponding non-toric diagram, one needs to properly tune K\"ahler parameters associated with the Higgsing. We find a proper tuning for the K\"ahler parameters by comparing the partition function obtained from the topological vertex result with the known result from localization. As such Higgsed parts of the 5-brane web are locally a $T_2$ diagram, such tuning can be also determined by considering tuning of a $T_2$ diagram ($T_2$-tuning). Following \cite{Mitev:2014jza}, global symmetry enhancement can be shown by redefining the gauge theory parameters to make the fiber-base duality manifest.

The paper is organized as follows. In section \ref{sec:5-branes}, we discuss 5-brane configurations for 5D $Sp(N)$ gauge theories with one antisymmetric hypermultiplet and flavors. In section \ref{sec:TVT2}, we review the refined topological vertex formalism and discuss a special tuning of K\"ahler parameters for 5-brane diagram for $T_2$ theory, which is associated with the Higgsing of 5-brane webs giving rise to  $Sp(N)$ gauge theories with an antisymmetric hypermultiplet.  In section \ref{sec:PF}, we compute the instanton partition function for $Sp(2)$ gauge theories with an antisymmetric and $N_f\le 4$ flavors, and also discuss $Sp(3)$ gauge theory as an example of generalization to higher rank gauge group. We then conclude with some remarks in section \ref{sec:conclusion}. 

A Mathematica package for refined topological vertex for generic toric diagrams is accompanied and available at the arXiv website or \cite{github}. The package would be used for more complicated toric diagrams. 
\section{5-brane configurations for \texorpdfstring{$Sp(N)$}{Sp(N)} gauge theory with antisymmetric matter}\label{sec:5-branes}
\begin{figure}
\centering
\subfigure{
\includegraphics[width=7.2cm]{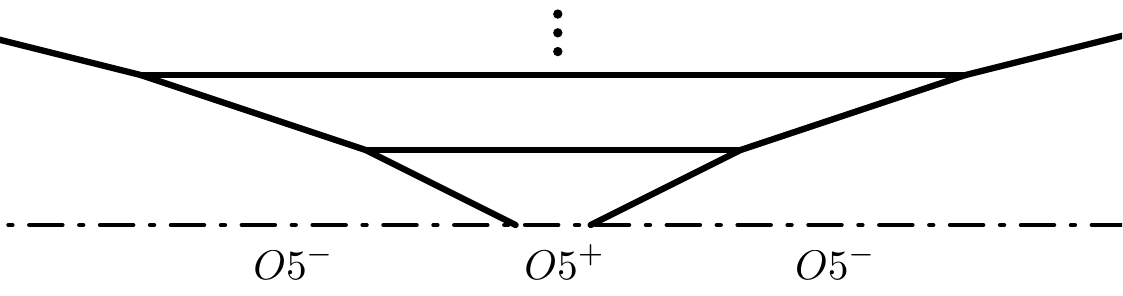}\label{fig:SpNO5}}\quad
\subfigure{
\includegraphics[width=7.2cm]{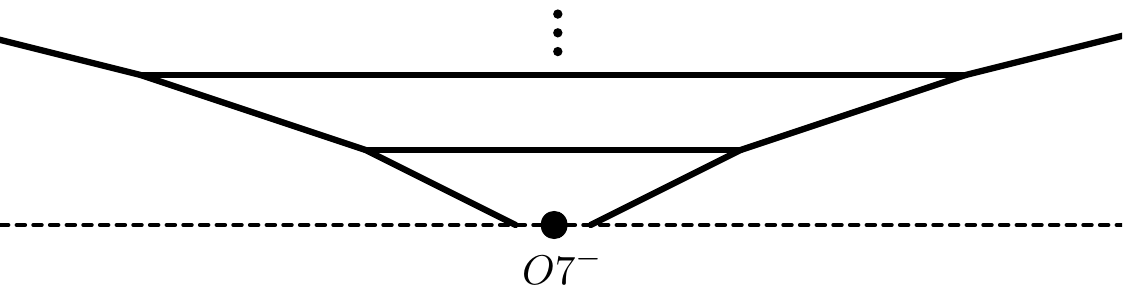}\label{fig:SpNO7}}
\caption{(a): 5-brane web for $Sp(N)$ with an $O5$-plane. (b): 5-brane web for $Sp(N)$ with an $O7^-$-plane.}
\label{fig:SpNweb}
\end{figure}
\begin{figure}
\centering
\subfigure{
\includegraphics[width=7.2cm]{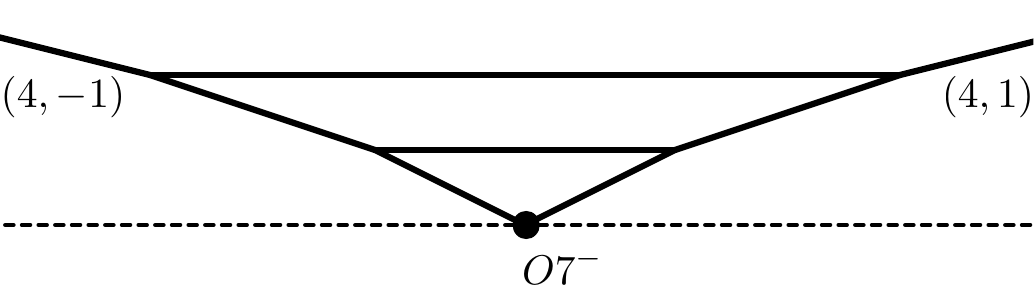}\label{fig:SpNO5def}}\quad
\subfigure{
\includegraphics[width=7.2cm]{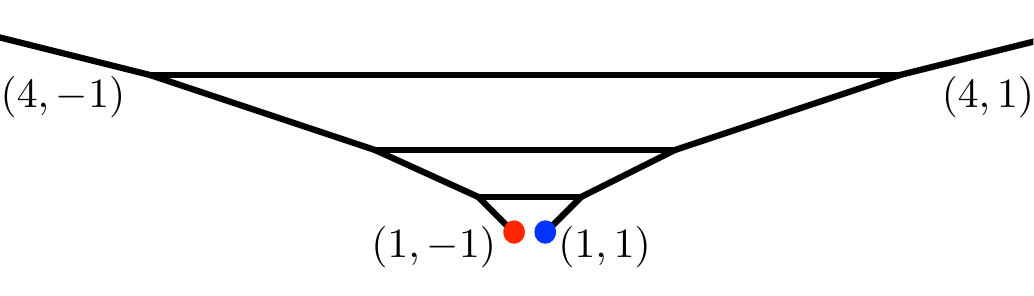}\label{fig:SpNO7resol}}
\caption{(a): A deformation of a 5-brane web with an $O7^-$-plane for pure $Sp(2)$. (b): Resolution of an $O7^-$-plane into a pair of 7-branes of the charge $(1,1)$ and $(1,-1)$. The resulting diagram is pure $SU(5)$ theory of Chern-Simons level -5.}
\label{fig:Sp2resol}
\end{figure}

 From the perspective of Type I' string theory,  5D $Sp(N)$ gauge theories with $N_f$ hypermultiplets in the fundamental representation (flavors) and one hypermultiplet in the antisymmetric (antisymmetric) are realized as $N$ D4-branes near the $N_f$ D8 branes on top of a single $O8^-$ orientifold plane. The theory has superconformal fixed point that arises in the infinite coupling limit of the gauge theory. It exhibits $SO(2N_f) \times  U(1)_I \times SU(2)_{\rm antisym.}$ global symmetry of flavors, instanton number, and an antisymmetric. At the UV fixed point, the global symmetry is enhanced to  \cite{Seiberg:1996bd}
\begin{align}
E_{N_f+1}\times SU(2)_{\rm antisym.}\supset
 SO(2N_f)\times U(1)_I\times SU(2)_{\rm antisym.},
\end{align}
where $E_n$ refer to $E_8,E_7$, and $E_6$; $E_5=Spin(10)$, $E_4=SU(5)$, $E_3 = SU(3)\times   SU(2)$, $E_2 = SU(2)\times U(1)$ and $E_1 = SU(2)$.
The enhancement of global symmetry is explicitly checked from the superconformal index based on the ADHM method \cite{Kim:2012gu,Hwang:2014uwa}.
Without any flavors, $Sp(N)$ gauge theory with or without antisymmetric hypers has the discrete theta parameters (angles) associated with $\pi_4 (Sp(N))=\mathbb{Z}_2$, referred to as $\theta=0, \pi$. Hence, there are two inequivalent pure $Sp(N)$ gauge theories: one with $\theta=0$, denoted as $Sp(N)_0$, and the other with $\theta=\pi$, denoted as $Sp(N)_\pi$. The origin of the discrete theta parameters from Type I' theory is discussed in \cite{Bergman:2013ala}. 
 We note that $Sp(N)_0$ theory with an antisymmetric enjoys enhanced global symmetry $SU(2)_I\times SU(2)_{\rm antisym.}$ at the UV fixed point, while $Sp(N)_\pi$ theory with an antisymmetric has $U(1)_I\times SU(2)_{\rm antisym.}$. Without antisymmetric matter, both theories have only $U(1)_I$ global symmetry, except for the $Sp(1)$ theory where the global symmetry is enhanced to $SU(2)_I\supset U(1)_I$.

A 5D $Sp(N)$ gauge theory can also be understood from Type IIB string theory. In fact, a wide range of 5D $\mathcal{N}=1$ theories can be described by Type IIB string theory, which provides not only qualitative understanding but also quantitative aspects for 5D gauge theories. To describe 5D $Sp(N)$ gauge theory in Type IIB string theory. One can introduce an $O5$-plane or an $O7^-$-plane. As a representative example, 5-brane webs for pure $Sp(N)$ gauge theory is depicted in Figure \ref{fig:SpNweb}. In 5-brane webs with an $O5$-plane, when one changes the coupling of the pure $Sp(N)$ theory, the brane configurations are deformed in two different ways. These two different phases distinguish the discrete theta angles for the pure $Sp(N)$ theory \cite{Hayashi:2017btw}. One can also compute the (unrefined) partition function of $Sp(N)$ theory with $N_f\leq 2N+6$ flavors based on a 5-brane web using topological vertex method \cite{Kim:2017jqn}. 

5-brane configurations with an $O7^-$-plane are in particular interesting. An $O7^-$-plane can be resolved into a pair of two 7-branes of the same monodromy \cite{Sen:1996vd}. For instance, suppose one resolves an $O7^-$-plane in Figure \ref{fig:SpNO5def}, and then the resulting 5-brane configuration becomes a 5-brane configuration for an $SU(N+1)_\kappa$ theory with the Chern-Simons level $\kappa= 2N+6-2|\kappa|$ as depicted in Figure \ref{fig:SpNO7resol}, and, hence provides a diagrammatical account for the duality between 5D $Sp(N)$ gauge theory with $N_f\leq 2N+6$ flavors and $SU(N+1)$ gauge theory with the same number of $N_f$ flavors \cite{Gaiotto:2015una,Hayashi:2015fsa}, where flavors in 5-brane webs are represented by D7-branes.

\begin{figure}
\centering
\includegraphics[width=12cm]{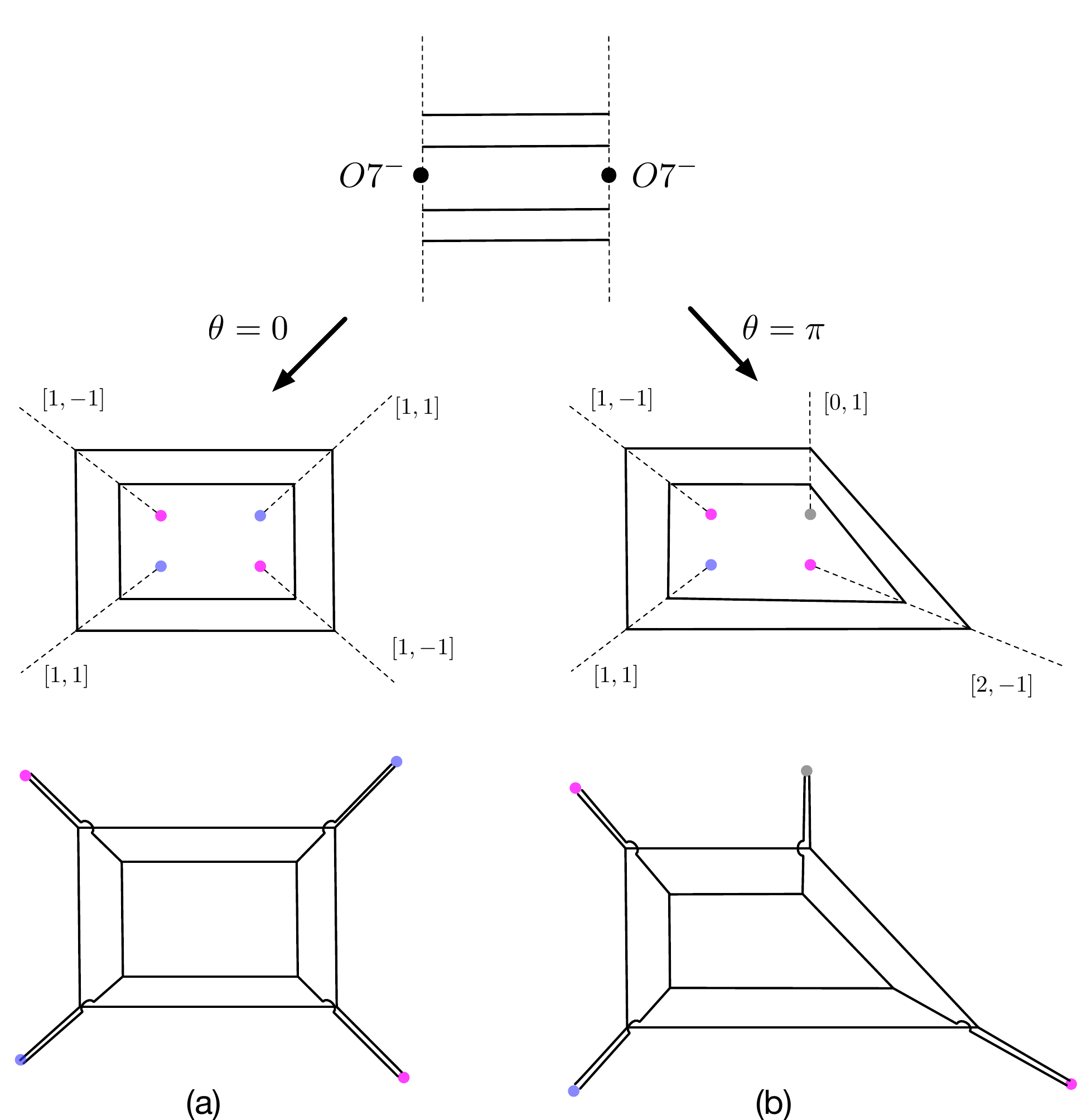}
\caption{Massless case: (a) 5-brane configuration for $Sp(2)_0+1\mathbf{AS}$, where resolving two $O7^-$-planes into the same types of a 7-brane pair (in this case, those of the 7-brane charges $[1,1]$ and $[1,-1]$) yields the discrete theta angle $\theta=0$. 
(b) 5-brane configuration for $Sp(2)_\pi+1\mathbf{AS}$, where resolving two $O7^-$-planes into the different types of 7-brane pairs (in this case, $([1,1], [1,-1])$  and $([2,-1], [0,1])$) yields the discrete theta angle $\theta=0$. 
}
\label{fig:massless}
\end{figure}
\begin{figure}
\centering
\includegraphics[width=12.2cm]{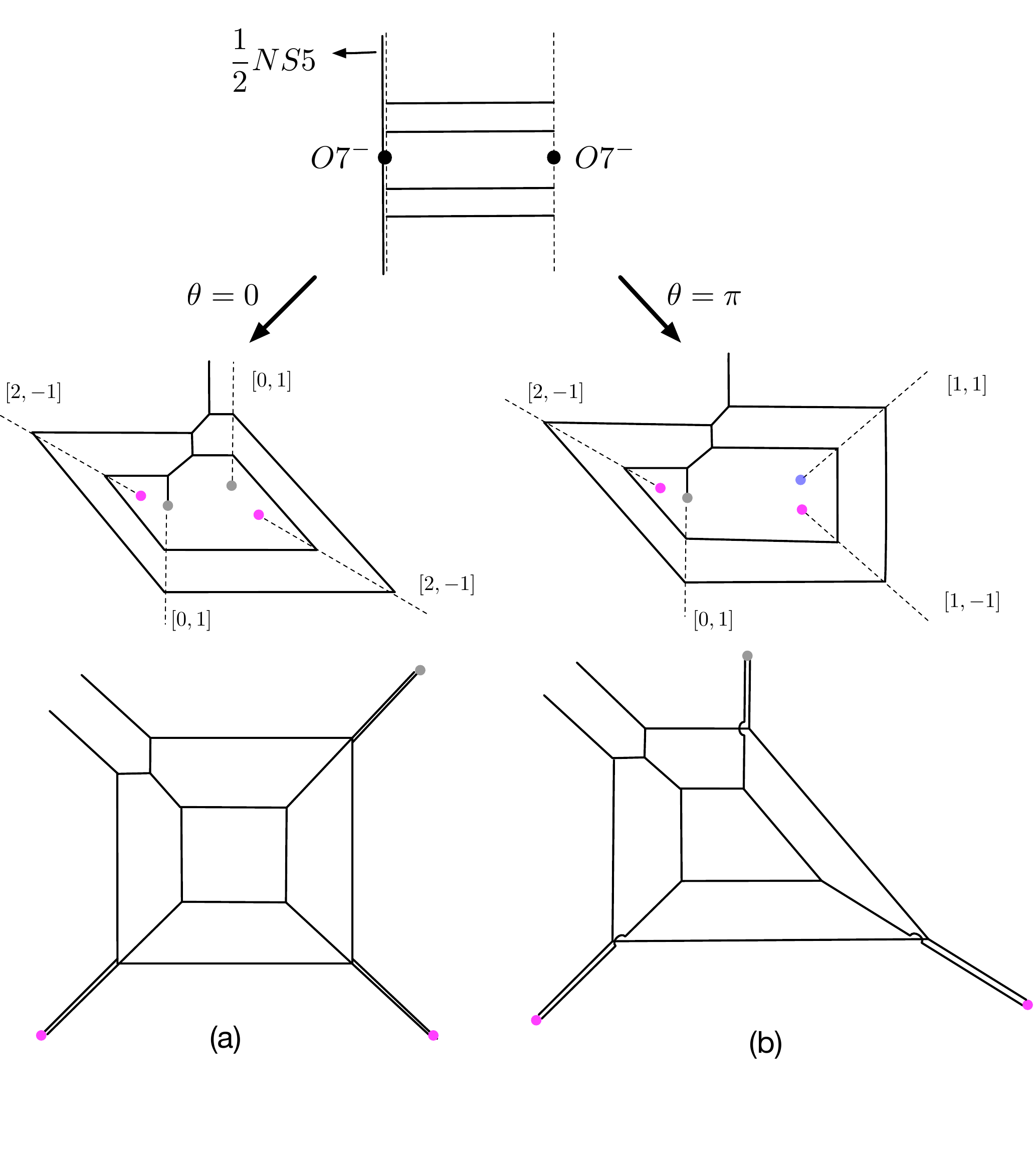}
\caption{Massive cases: (a) A 5-brane configuration for $Sp(2)_0+1\mathbf{AS}$.
(b) A 5-brane configuration for $Sp(2)_\pi+1\mathbf{AS}$. The 5-brane diagrams in the bottom are obtained after pulling out 7-branes from 5-brane loops and performing $SL(2,\mathbb{Z})$ transformations.}
\label{fig:massive}
\end{figure}

For $Sp(N)$ gauge theory with an antisymmetric hypermultiplet, it is still a challenge to describe the theory based on 5-brane webs with an $O5$-plane. It is however possible to describe antisymmetric matter of $Sp(N)$ theory using $O7^-$-planes. To realize an $Sp(N)$ theory with an antisymmetric, one introduces two $O7^-$-planes horizontally separated on a 5-brane web and $N$ D5-branes are placed parallel to two $O7^-$-planes as depicted in Figure \ref{fig:massless}. An alternative description is to introduce a half NS5-brane stuck on one of $O7^-$-planes \cite{Bergman:2015dpa} as in Figure \ref{fig:massive}. 
While the 5-brane description in Figure \ref{fig:massless} corresponds to an $Sp(N)$ theory with a massless antisymmetric hypermultiplet \cite{Benini:2009gi}, the 5-brane in Figure \ref{fig:massive} describe  an $Sp(N)$ theory with a massive antisymmetric hypermultiplet, where the mass of an antisymmetric is parameterized by the vertical distance between two $O7^-$-planes.

The discrete theta parameters for $Sp(N)$ gauge theory in this 5-brane webs with $O7^-$-planes are realized as two different resolutions of an $O7^-$-plane into a pair of 7-branes \cite{Bergman:2015dpa}. For instance, $O7^-$ can be resolved either into a pair of 7-branes of the charges $[1,-1]$ and $[1,1]$, or into a pair of 7-branes of the charges $[2,-1]$ and $[0,1]$. If one resolves two $O7^-$-planes into the same types of 7-brane pairs, then it gives the discrete theta angle $\theta=0$, while the resolution into two different types of 7-brane pairs leads to the discrete theta angle $\theta=\pi$.  One can summarize 5-brane configurations for $Sp(N)$ theory with an antisymmetric hypermultiplet with the discrete theta angle $\theta=0$  ($Sp(N)_0+1\mathbf{AS}$) and that with the discrete theta angle $\theta=\pi$ ($Sp(N)_\pi+1\mathbf{AS}$) as follows: For massless antisymmetric hypermultiplet, it is depicted in Figure \ref{fig:massless}. For massive antisymmetric hypermultiplet, it is depicted in Figure \ref{fig:massive}.

Flavors can be introduced by adding D7-branes. We list some representative 5-brane webs for 5D $Sp(N)$ gauge theory with one antisymmetric hypermultiplet and $N_f$ flavors ($Sp(N)+1\mAS+ N_f \mF$) in Figure \ref{representativewebs}. For web diagrams for $Sp(2)+ 1\mAS+ N_f(\le 8)\mF$, see Appendix in \cite{Hayashi:2018lyv}.
\begin{figure}[]
	\centering
	\includegraphics[width=5in]{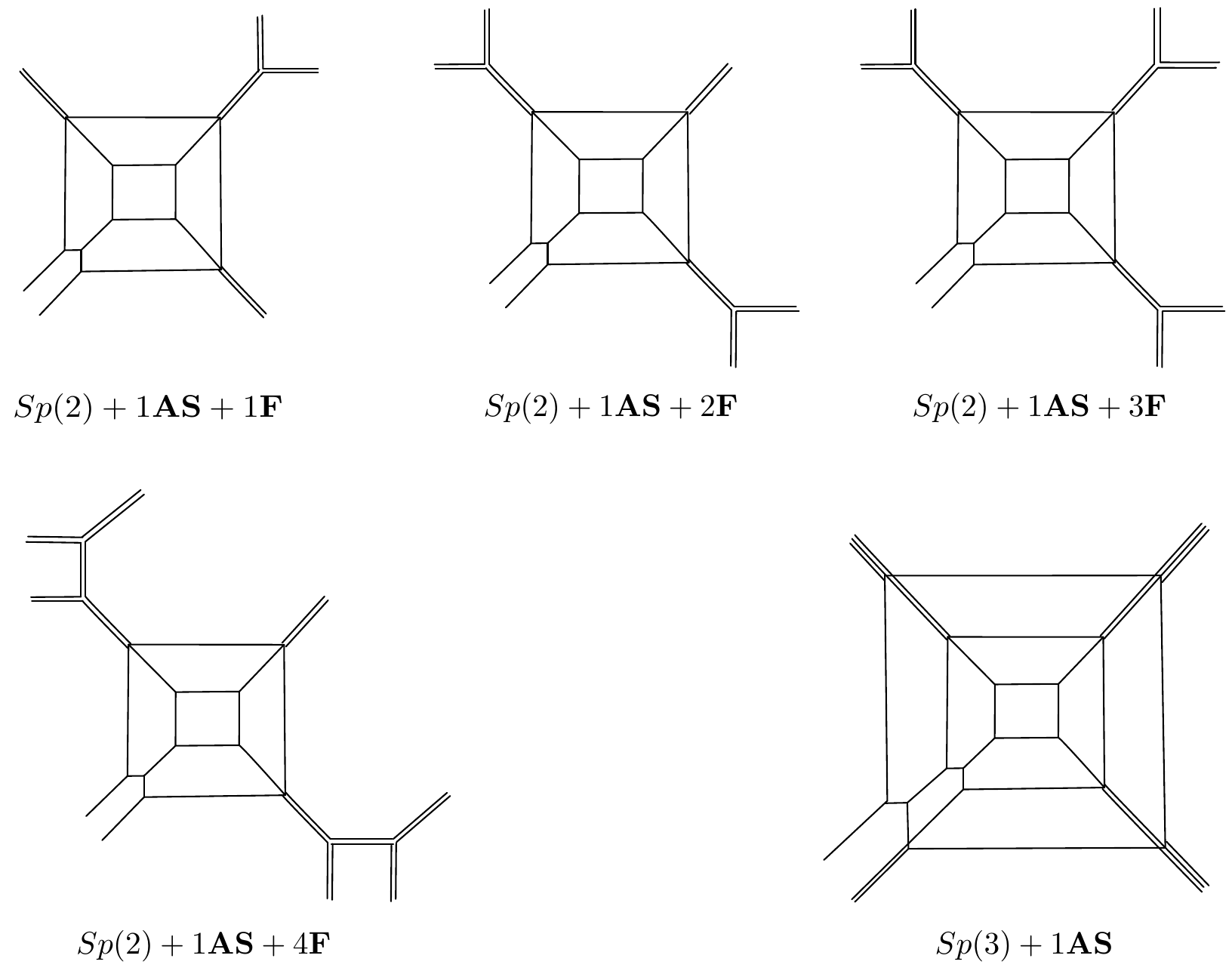}
	\caption{Some representative examples of 5-brane webs $Sp(N)$ gauge theories with one antisymmetry hypermultiplet and flavors, which are also considered in section \ref{sec:PF}.}
	\label{representativewebs}
\end{figure}

\begin{figure}
\centering
\includegraphics[width=10cm]{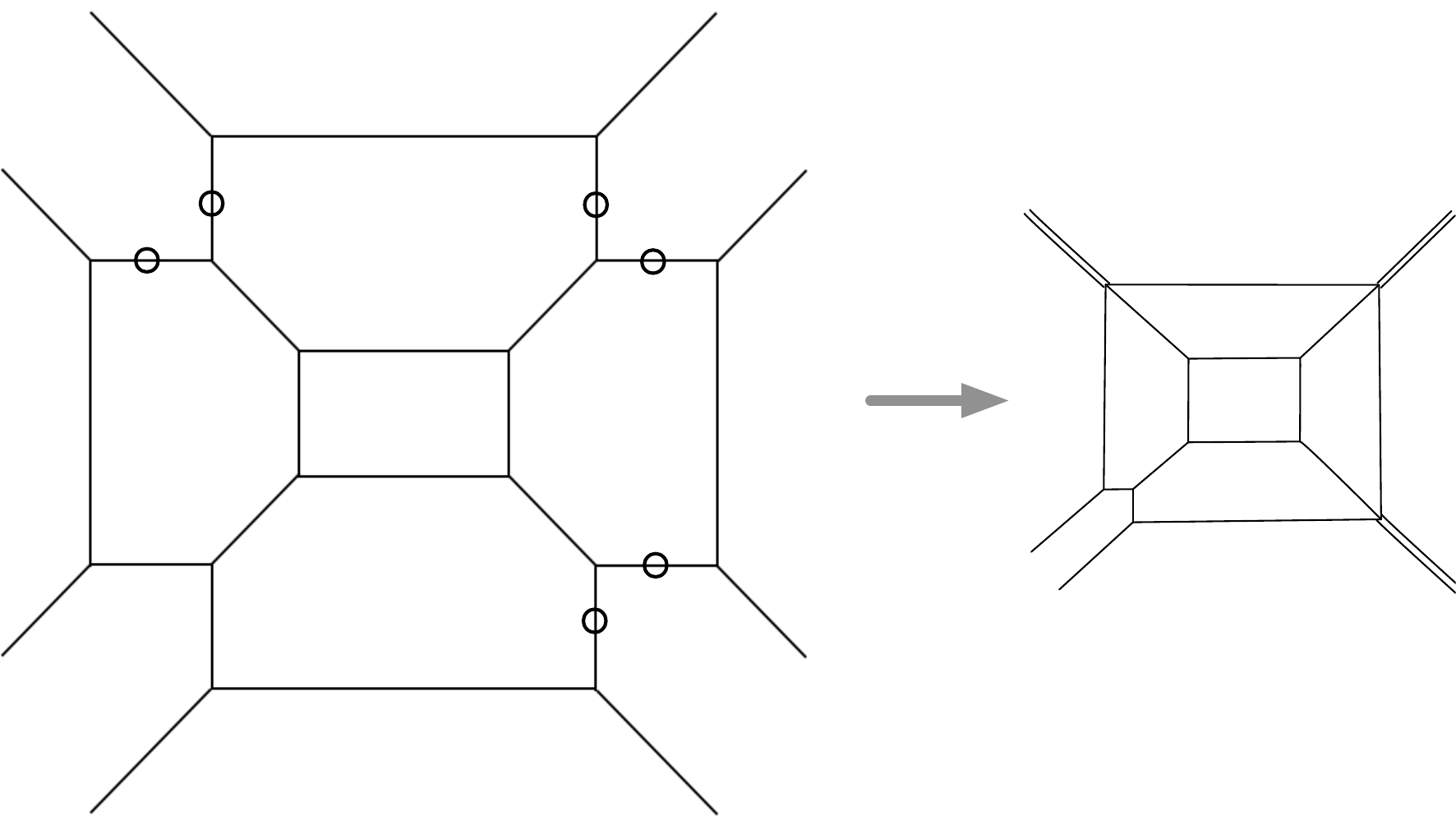}\label{fig:su2su4su2}
\caption{A 5-brane web diagram for 5D $SU(2)\times SU(4)\times SU(2)$ quiver gauge theory and its Higgsed diagram giving rise to a 5-brane web for 5D $Sp(N)_0+1\mAS$.}
\label{fig:Higgsed}
\end{figure}

As one can see 5-brane web diagrams for 5D $Sp(N)$ theories with an antisymmetric have jumps on $(p,q)$-plane. In other words, the corresponding dual diagrams are non-toric. Such a 5-brane web can be regarded as a Higgsed web diagram from some other (quiver) gauge theories. For instance, as we will see in the later sections, a 5-brane web for 5D $Sp(2)_0$ theory with an antisymmetric hypermultiplet can be obtained from a Higgsing of a $SU(2)\times SU(4)\times SU(2)$ quiver theory as in Figure \ref{fig:Higgsed}. Another example that we will discuss is a 5-brane web for 5D $Sp(3)_0$ theory with an antisymmetric which can be obtained from a Higgsing of a $SU(2)\times SU(4)\times SU(6)\times SU(4)\times SU(2)$ quiver theory. Likewise, 5D $Sp(N)_0$ theory with an antisymmetric can be obtained from a Higgsing of an $SU(2)\times SU(4)\times\cdots\times SU(2N)\times\cdots\times SU(4)\times SU(2)$ quiver theory.

\section{Topological vertex and \texorpdfstring{$T_2$}{T2}-tuning}
\label{sec:TVT2}

In this section, we set up our convention and very briefly review the refined topological vertex formalism, which enables one to compute the Nekrasov partition function for 5D $\mathcal{N}=1$ gauge theories, via geometric engineering \cite{Hollowood:2003cv,Iqbal:2007ii}. We also discuss Higgsing procedures associated $Sp(2)$ gauge theories with antisymmetric matter. Our convention closely follows that used in \cite{Bao:2013pwa}. 

\subsection{Brief review of topological vertex}\label{vertexformula}
\begin{figure}[]
	\centering
	\includegraphics[width=4.5in]{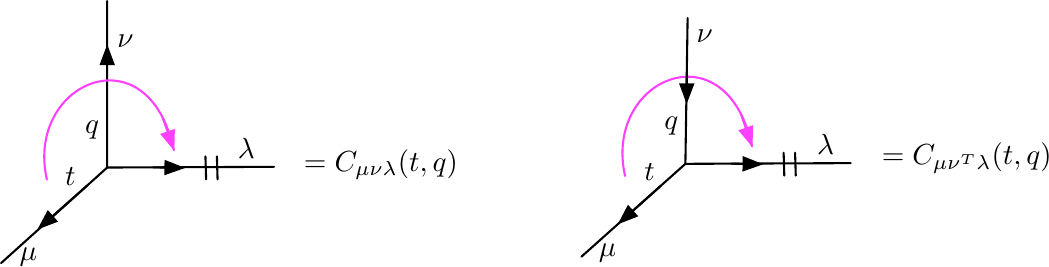}
	\caption{Vertex factor assignment. The direction of the arrow on the edges can be chosen arbitrary, and the associated Young diagrams get transposed when the arrow is flopped. 
	Here $q,t$ are the $\Omega$ deformation parameters $q=e^{-\epsilon_2},~t=e^{\epsilon_1}$ } 
	\label{fig:Cuvr}
\end{figure}

\begin{figure}[]
	\centering
	\includegraphics[width=4.5in]{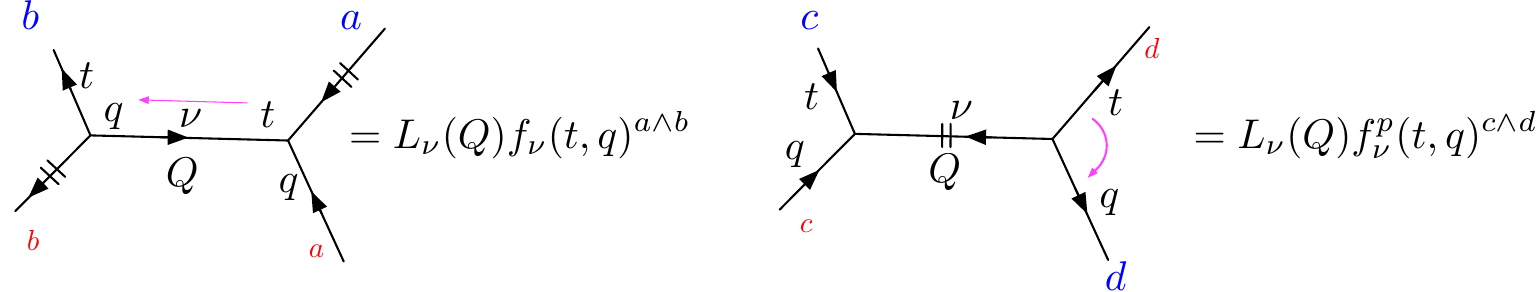}
	\caption{ The left figure is for non-preferred edges and the second for preferred edges. The colors of letters do not matter, as the blue framing number $a \wedge b$ equals  red  $a \wedge b$, and blue $c \wedge d$ equals red  $c \wedge d$. }
	\label{fig:framing}
\end{figure}

 5D $\mathcal{N}=1$ gauge theory in a general $\Omega$-background can be engineered by some local toric Calabi-Yau threefold \cite{Hollowood:2003cv}. It is also described by Type IIB 5-brane web diagrams. Through various dualities, such 5-brane webs are equivalent to toric diagrams for local Calabi-Yau threefold in A-model \cite{Leung:1997tw}. The Nekrasov partition functions can thus be given by topological string partition functions. In what follows, we may use 5-brane webs and toric diagrams in an interchangeable way. In the topological vertex utilizing toric diagrams, one chooses the preferred direction denoted by $||$, assign Young diagram $( \mu, \nu,\cdots)$ and K\"{a}hler parameter $Q_\bullet$ to edges, and the vertex factor to vertices, and then glues and performs the Young diagrams to get the topological string partition functions
\begin{align}\label{vertexformu}
Z^{\rm top} = \sum_{\lambda_i}\,\prod\, ({\rm Edge ~factor}) \cdot \prod \,({\rm Vertex~ factor})
\end{align}
The assignment of the vertex factor and the edge factor is illustrated in Figure \ref{fig:Cuvr} and Figure \ref{fig:framing}. 
With the $\Omega$ deformation parameters $q=e^{-\epsilon_2},~t=e^{\epsilon_1}$, 
the vertex factor is defined as 
\begin{align}
&C_{\la \u \v } (t,q ) := q^{ \frac{||\u||^2+ ||\v||^2} {2 } } t^{ -\frac {||\u^T||^2} {2}}  \tZ_{\v} (t,q) \sum_{\h}  \left(\frac{q}{t} \right)^{ \frac{|\h| + |\la| -|\u|}{2}  } s_{\la^T/ \h } (t^{-\p} q^{-\v} ) s_{\u /\h}(q^{-\p} t^{-\v^T})\,,
\end{align}
where
\begin{align}
& \tZ_{\v}(t,q):=\prod\limits_{(i,j)\in \v   } 
\l(     1-q^{  \v_i-j} t^{\v_j^T -i+1}       \r)^{-1}   \,,
\end{align}
and $s_{\lambda/\eta}$ are skew Schur functions. 
The edge factor is defined as 
\begin{align}
{f^{\bullet}_{\v}(t, q)}^{\text{framing num.}} L_{\v}(Q)\,,
\end{align}
with
\begin{align}
&f^p_\v(t,q):= (-1)^{|\v|}t^{\frac{||\v^T||^2}{2} } q^{-\frac{||\v||^2}{2} },~~f_\v(t,q):=\l(\frac{q}{t}\r)^{-\frac{|\v|}{2}  } f^p_\v(t,q),~~
L_\v(Q):= (-Q)^{|\v|}\,,
\end{align}
where $f^p(t, q)$ is for the edges along the preferred direction and $f(t, q)$ for other edges for non-preferred directions.

After summing over Young diagrams along non-preferred directions by Cauchy identities \eqref{app:Cauchy1}-\eqref{app:Cauchy2}, topological string partition function \eqref{vertexformu} generically takes the following form
\begin{eqnarray}\label{Z^M}
Z^{\text{top}}(Q_i, t,q )&=&Z^{M}\cdot Z^{\text{sum}}\,,
\end{eqnarray}
where $Z^{M}$ is a product of $M(Q_i, t, q)$'s 
\begin{align}
Z^{M}=\frac{\prod M(Q_i, t,q)}{\prod M(Q_j, t,q)}\,,
\end{align}
with 
\begin{align}
			M(Q, t,q):=& \prod\limits_{i, j=1}^\inf  (1-Q~q^i t^{j-1})  \, 
\end{align}
and $Z^{\rm sum}$ is the terms which contain the Young diagram sum along the preferred directions, which has the following structure
\begin{align}\label{Z^sum}
& Z^{\text{sum}}=\sum_{\u_\cdot} Q_i^{|\u_{\cdot}|}  \prod\limits_{\u_\cdot} ||\tZ_{\u_{\cdot}}(t,q)||^2  \dfrac{  \prod N_{\v_{\cdot}}^{\rm half,-
	}{(Q_i,t^{-1},q^{-1}) }N_{\u_{\cdot } \v_{\cdot}}(Q_i,t^{-1},q^{-1})   }{  \prod N_{\u_{\cdot } \v_{\cdot}}(Q_i,t^{-1},q^{-1})   } \,,
\end{align}
where 
\begin{align}
		||\tZ_{\u}(t,q)||^2:=& \tZ_{\u^T}(t,q) \tZ_{\u}(q,t)\,,\\
	N_{\u \v}(Q; t, q) :=&   \prod\limits_{i, j=1}^\inf \frac{1- Q~ q^{\v_i -j}~ t^{\u_j^T-i+1}	}{	1- Q~ q^{-j} ~t^{-i+1}	}\,,  
\end{align}
and 
\begin{align}
	&N_{\v}^{\rm half,-}(Q; t,q):= N_{\v \0}\l(Q\sqrt{ \frac{q}{t}}, t,q\r) \,,\\  
	&	N_{\v}^{\rm half,+}(Q; t,q):= N_{\0 \v}\l(Q\sqrt{ \frac{q}{t}}, t,q\r)  \,. 
\end{align}
One can also think of $Z^M$ as the overall factor multiplied to the terms which have the Young diagram sum. In other words, $Z^M$ is the term that is obtained by setting the Young diagrams along the preferred directions to $\emptyset$, or $Z^{\rm top}|_{\mu_i=\emptyset}$.

\subsection{\texorpdfstring{$T_2$-diagram and $T_2$-tuning}{T2-diagram and T2-tuning}}\label{T2tuning}
\begin{figure}[]
	\centering
	\includegraphics[width=1.8in]{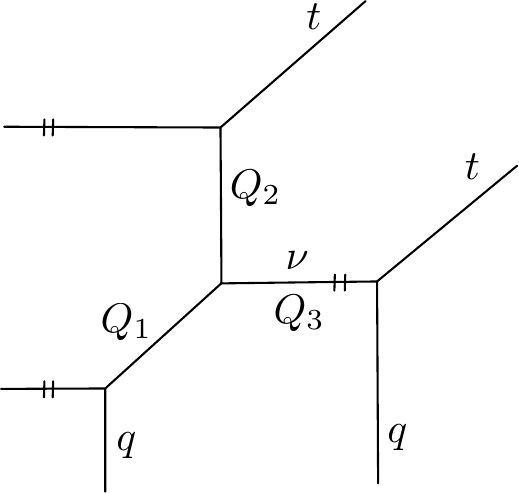}
	\caption{A $T_2$-diagram. The preferred directions are denoted by $||$ along the horizontal edges. $Q_i$ are the K\"{a}hler parameters assigned to the internal edges, $\nu$ is the Young diagram along the edge associated with $Q_i$. The empty Young diagram is given to six external edges.  
	}
	\label{fig:T2basic}
\end{figure}

\begin{figure}[]
	\centering
	\includegraphics[width=6in]{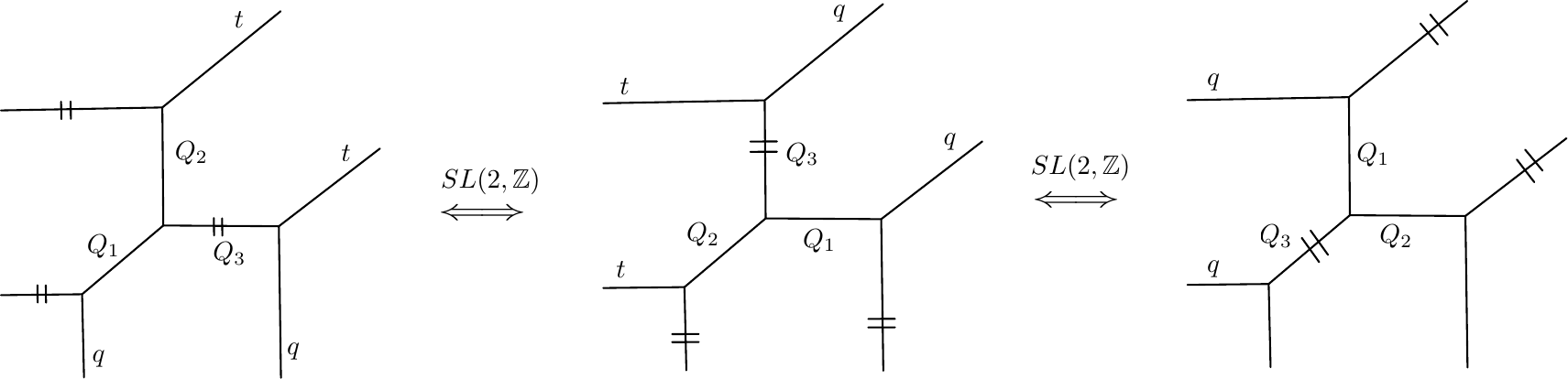}
	\caption{$T_2$-diagrams related through $SL(2, \mathbb{Z})$ transformation. }
	\label{fig:T2sl2z}
\end{figure}

As an instructive example, the 5-brane web or toric diagram for 5D $T_2$-theory is depicted in Figure \ref{fig:T2basic}. 
The topological string partition function for the $T_2$-theory is straightforward to compute and is given by
\begin{align}
	Z_{T_2}^{\rm top}(Q_1,Q_2,Q_3;t,q)= Z^{M}_{T_2}   \cdot Z^{\text{sum}}_{T_2}  \,,
\end{align}
where 
\begin{align}
	Z^{M}_{T_2} &= \frac{ M(Q_1 \sqtq, t,q ) M( Q_2 \sqtq, t,q  )   }{ M( Q_1Q_2,q,t )    }  \,.
	\end{align}
$T_2$-theory is, in fact, special in that the Young diagram sum part $Z^{\text{sum}}_{T_2} $ can be performed, yielding a compact form
	\begin{align}\label{T2sum}
	Z^{\text{sum}}_{T_2}  &=\sum_{\v} (-Q_3)^{|\v|} q^{\frac{ || \v||^2 }{2} } t^{\frac{ || \v^T||^2 }{2} }  ||\tZ_{\v}(q,t)||^2 N_{\v^T}^{\rm half,-}(Q_1; t^{-1},q^{-1}  )
	 N_{\v^T}^{\rm half,+}(Q_2; t^{-1},q^{-1}  ) \nn \\
&= \dfrac{ M(Q_3 \sqrt{\frac{t}{q}}, t,q) M(Q_1 Q_2 Q_3 \sqrt{\frac{t}{q}}, t,q)  }{M(Q_1Q_3,t,q)  M(Q_2Q_3,q,t)  }  \,.   
	 \end{align}
We thus have 
	 \begin{align} \label{eq:T2top}
	&Z^{\rm top}_{T_2}=\dfrac{M(Q_1 \sqrt{\frac{t}{q}}, t,q) M(Q_2 \sqrt{\frac{t}{q}}, t,q) M(Q_3 \sqrt{\frac{t}{q}}, t,q) M(Q_1 Q_2 Q_3 \sqrt{\frac{t}{q}}, t,q)  }{M(Q_1Q_3,t,q)  M(Q_2Q_3,q,t) M(Q_1Q_2,q, t) }\,.
\end{align}
We note that the partition function $Z^{\rm top}_{T_2}$ is $SL(2,\mathbb{Z})$ invariant; hence $T_2$-diagrams with different preferred directions in Figure \ref{fig:T2sl2z} have the same partition function. 

\begin{figure}
	\centering
	\includegraphics[width=5.5in]{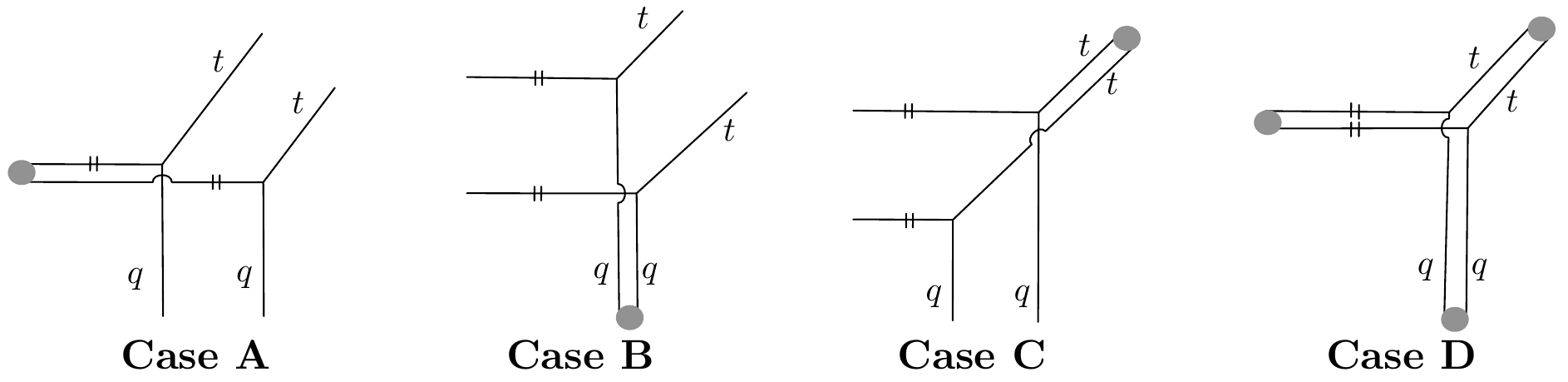}
	\caption{Four possible Higgsed $T_2$-diagram. Each jump on $T_2$-diagram denotes a particular tuning of the K\"{a}hler parameters for Higgsing.	}
	\label{fig:t2cases}
\end{figure}

As discussed in section \ref{sec:5-branes}, 5D $Sp(2)$ gauge theories with an antisymmetric hypermultiplet can be obtained by Higgsing a quiver gauge theory. The Higgsing here is locally the Higgsing of $T_2$-diagram which serves as building blocks. When the preferred direction is chosen, there are four possible Higgsings on a $T_2$-diagram. For convenience, we call them \textbf{Cases A}, \textbf{B}, \textbf{C} 
and \textbf{D}  
as shown in Figure \ref{fig:t2cases}. In particular, \textbf{Case D} is a typical configuration when flavors are added. With the assignment of K\"ahler parameters in Figure \ref{fig:T2basic},  
\textbf{Case A} is achieved by tuning the K\"ahler parameters $Q_1,Q_2$ to a special value,  \textbf{Case B} is by tuning $Q_1,Q_3$, and  \textbf{Case C} by tuning $Q_2,Q_3$. \textbf{Case D} requires tuning of all three K\"ahler parameters $Q_1,Q_2$, and $Q_3$.

\begin{figure}[]
	\centering
	\includegraphics[width=4in]{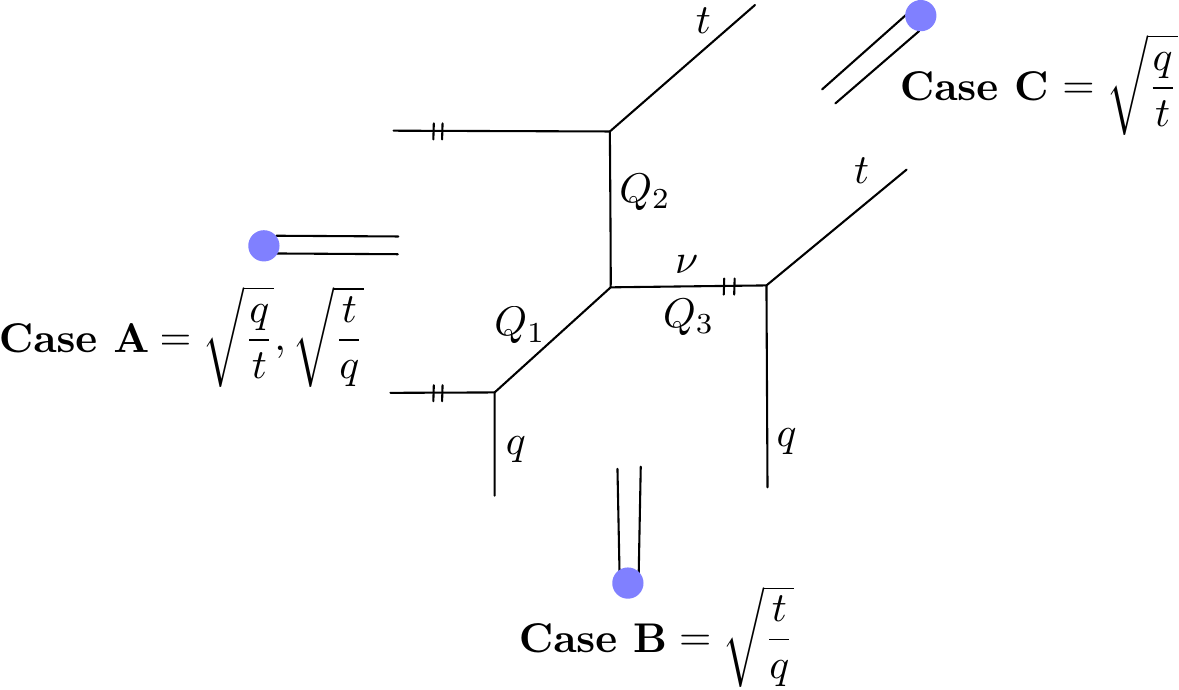}
	\caption{ $T_2$-tuning: \textbf{Case A, B, C}.
}
	\label{fig:T2tuning}
\end{figure}

This Higgsing procedure corresponds to certain geometric transitions \cite{Dimofte:2010tz,Taki:2010bj,Aganagic:2011sg,Aganagic:2012hs}, and the K\"ahler parameters responsible for the Higgsings are tuned to be either $\sqqt$ or $\sqtq$ \cite{Hayashi:2013qwa,Hayashi:2015xla}. We found the suitable choices for tuning K\"ahler parameters that reproduce the partition functions for 5D $Sp(N)+1\mAS +N_f \mF$: 
\begin{align}
	&\begin{cases}
		\textbf{Case A:} \, \quad\quad & Q_1=Q_2=\sqqt~{\rm or}~ \sqtq,\\
		\textbf{Case B:} \,  \quad\quad &Q_1=Q_3=\sqtq,\\
		\textbf{Case C:}\,  \quad\quad &Q_2=Q_3=\sqqt,  
		\\
	\textbf{Case D:} \, \qquad\quad  &Q_1=Q_2=Q_3  \,,		\end{cases}
		\label{caseABC}
\end{align}
which is consistent with the result obtained from the ADHM method.
Here, for \textbf{Case A}, either choice of K\"ahler parameters $\sqqt$ or $\sqtq$ is allowed and leads to the same result. As 5-brane configuration in \textbf{Case D} needs to be glued to either of \textbf{Case A, B, C} Higgsed diagrams, the tuning of K\"ahler parameter for \textbf{Case D} is the same as the value of K\"ahler parameter for the Higgsed diagram to which \textbf{Case D} is connected. As we will frequently refer to when we compute the partition function in the next section, we call these special tunings of K\"ahler parameters  \eqref{caseABC} collectively ``$T_2$-tuning". A pictorial version of the $T_2$-tuning is presented in Figure \ref{fig:T2tuning}. 

We remark that the Young diagram sum part of the partition function, $Z^{\rm num}_{T_2}$ should be trivial for the Higgsed $T_2$-diagrams depicted in Figure \ref{fig:t2cases}, and indeed $T_2$-tuning satisfies
\begin{align}
Z^{\rm sum}_{T_2}\big|_{\textbf{Case A, B, C, D}} =1\,.
\end{align}

\begin{figure}
	\centering
	\includegraphics[width=3in]{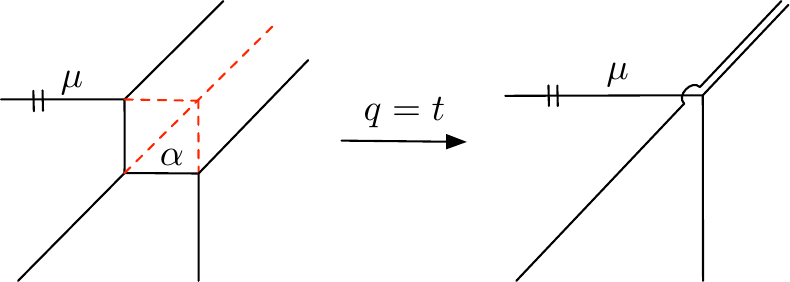}
	\caption{A unHiggsed diagram in the unrefined limit reproduces the jump. In the diagram on the left, the Young diagram sum is constrained such that 
	$\u \,\> \,\a$, while $\u = \a$, in the unrefined limit. } 
	\label{fig:hops}
\end{figure}

When applying the $T_2$-tuning to the partition function computations, we found the following identities related to geometric transitions \cite{Dimofte:2010tz} useful:
\begin{equation}\label{constraint1}
N_{\v}^{\rm half, +} \l(\sqqt;~ t^{-1},q^{-1} \r) =
\begin{cases}
1   & \v=\0\\
0   & \v\neq\0\,,
\end{cases}
\qquad	N_{\v^T}^{\rm half, -} \l(\sqtq;~ t^{-1},q^{-1} \r) =
\begin{cases}
1  & \v=\0\\
0  & \v\neq\0\, ,
\end{cases}
\end{equation} 
and many simplifications take place due to the following relations 
\begin{eqnarray}\label{constraint2}
N_{\u \a}( 1; t^{-1},q^{-1} )\neq 0,~~\text{only if}~ \u \,\> \,\a, 
\qquad N_{\u \a}( \frac{t}{q}; t^{-1},q^{-1} )\neq 0,~~\text{only if}~ \u\, \preccurlyeq  \, \a. 
\end{eqnarray}
It follows that in the unrefined limit $t=q$, as illustrated in Figure \ref{fig:hops},  \eqref{constraint2} becomes
\begin{eqnarray}\label{newcon}
N_{\u \a}( 1; t^{-1},t^{-1} )\neq 0,~~\text{only if}~ \u = \a\,,
\end{eqnarray} 
and thus further simplifies $Z^{\rm sum}$ so that the partition functions for 5D $Sp(2)$ theories with one massless antisymmetric hypermultiplet, are written as a product of two $Sp(1)$ theories \cite{Kim:2014nqa}.

\begin{figure}
	\centering
	\includegraphics[width=0.8\textwidth]{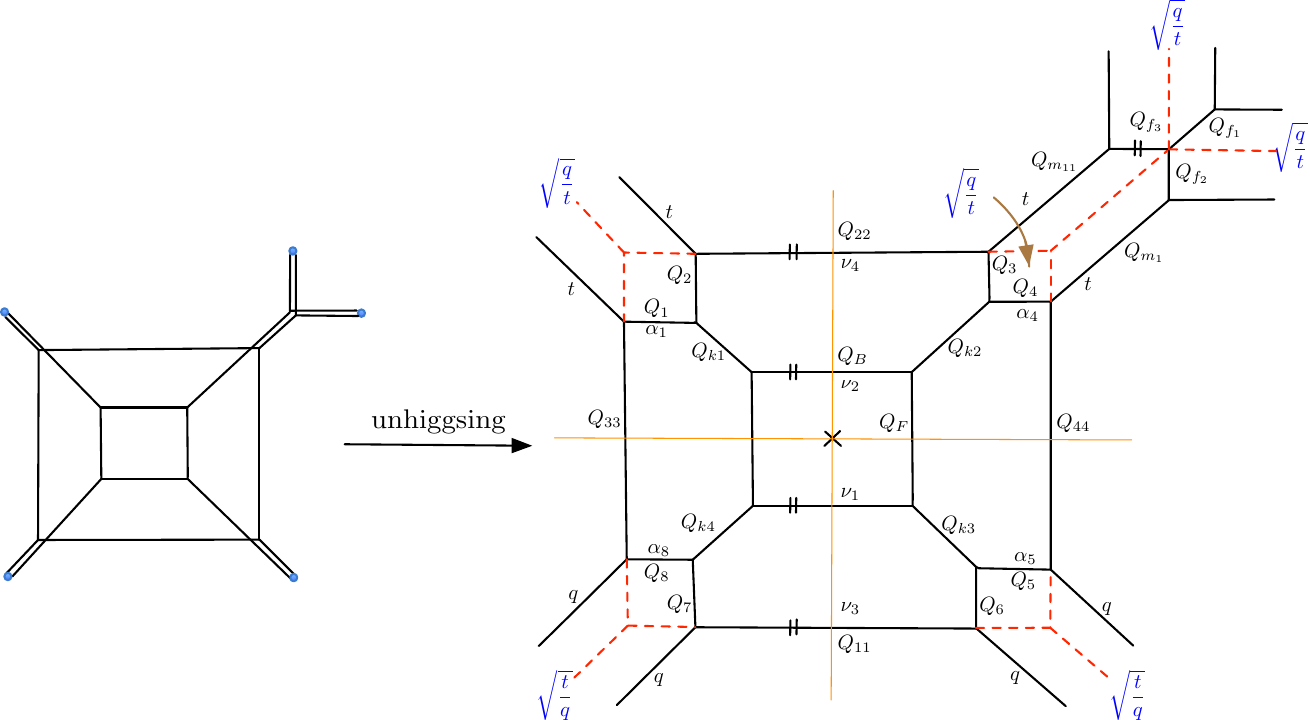}
	\caption{
		UnHiggsing procedure for 5-brane web for $Sp(2)+1\mAS+1\mF$. Following $T_2$-tuning in \eqref{caseABC}, one can assign values to Higgsed edges, where $\sqrt\frac{q}{t}$ or $\sqrt\frac{t}{q}$ in blue are tuned K\"ahler parameters. 
	}
	\label{Fig:rank2Nf1}
\end{figure}
\begin{figure}
	\centering
	\includegraphics[width=2.5in]{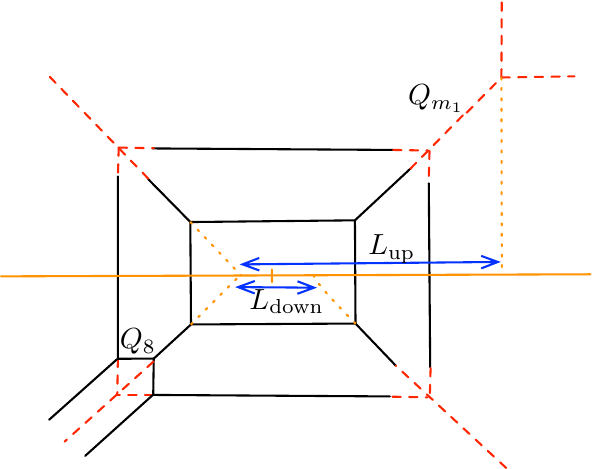}
	\caption{Auxiliary lines (in red) and instanton fugacity  		 which are projected lines when all edges are Higgsed. The instanton fugacity $u$ is then obtained via the conventional way as $L_{\rm up} \times L_{\rm down} = u^2$.}
	\label{virtualnf1}
\end{figure}
In summary, 5-brane web diagrams for 5D $Sp(2)+1\mAS+1\mF$, which can be understood as a Higgsed web diagram of a 5-brane web for the quiver gauge theory discussed in section \ref{sec:5-branes}. To compute topological string partition function, we consider the corresponding unHiggsed 5-brane web diagrams and implement refined topological vertex. We then perform the Higgsing on the unHiggsed 5-brane web diagrams by tuning K\"ahler parameters via $T_2$-tuning \eqref{caseABC}, which yields that the topological string partition function as a Young diagram sum over the preferred directions $Z^{\rm top}=Z^M\,Z^{\rm sum}$. Finally, by properly identifying K\"ahler parameters with 5D gauge theory parameters, we obtain the Nekrasov partition function as an expansion of the instanton fugacity. In Figure \ref{Fig:rank2Nf1}, we depict this procedure of unHiggsing and $T_2$-tuning for a typical 5-brane web of $Sp(2)$ gauge theory with one antisymmetric hypermultiplet and one flavor ($Sp(2)+1\mAS+1\mF$). We note that one can easily associate K\"ahler parameters with gauge theory parameters by introducing auxiliary lines which are projected lines when an unHiggsed diagram is Higgsed back. For instance, the instanton fugacity for $Sp(2)+1\mAS+1\mF$ is obtained in the conventional way as $L_{\rm up} \times L_{\rm down} = u^2$, as illustrated in Figure \ref{virtualnf1}.


\section{Instanton partition functions }\label{sec:PF}
In this section, we use the topological vertex method to obtain the refined Nekrasov partition function for $Sp(2)$ theory with one antisymmetric hypermultiplet and $N_f$ flavors ($Sp(2)+1 \mAS + N_f \mF$). As the corresponding 5-brane web diagrams are of non-toric, we properly apply the unHiggsing and $T_2$-tuning procedure discussed in section \ref{sec:TVT2}.

Recall that the topological string partition function obtained through the topological vertex factorizes into the perturbative part, $Z^{M}$, written in terms of $M(Q_\bullet,t,q)$, and the summation part, $Z^{\rm sum}$, summing over Young diagrams along preferred edges
\begin{eqnarray} 
	Z^{\text{top.}} =Z^{M} \cdot Z^{\rm sum}\,,
\end{eqnarray}
where each term contains a part of field theory perturbative contribution or  instanton (non-perturbative) contribution in general
\begin{align}
	Z^{ M}& =Z^{\textbf{pert-I}} (A_i, y_i)\cdot Z^{\textbf{nonpert-I}} (A_i,y_i,u)\,,\label{instsplit}\\
	Z^{\rm sum} &=Z^{\textbf{pert-II}}  (A_i,y_i) \cdot Z^{\textbf{nonpert-II}}(A_i,y_i,u)\ \label{instsplit2}.
\end{align}
Here $u$ is instanton fugacity, $A_i$ Coulomb branch parameters and $y_i$ fugacity for hypermultiplet fields. 
Nekrasov partition function $Z^{\rm Nek}$ is then obtained as topological string partition function $Z^{\text{top}}$ divided by the extra factor $Z^{\rm extra}$ that does not explicitly depend on the Coulomb branch parameters $A_i$. The resulting Nekrasov partition function factorizes into the perturbative contribution and the instanton contribution,
\begin{align}
	Z^{ \rm Nek} &=\frac{ Z^{\text{top}} (A_i,y_i, u)} { Z^{\rm extra }(y_i, u)   } = Z^{ \textbf{pert} }(A_i,y_i) \cdot Z^{\textbf{instanton} } (A_i,y_i,u)\ .
\end{align}
By factorizing the extra part,  
\begin{eqnarray}\label{pfformgeneral}
Z^{\rm extra }= Z^{\textbf{pert~extra}} (y_i)\cdot Z^{\textbf{inst~extra}} (y_i, u)\,, 
\end{eqnarray}
one can do the recombination and find that the full perturbative and instanton parts are obtained as
\begin{align}\label{pfformgeneral2}
Z^{ \textbf{pert } } &= \frac{Z^{\textbf{pert-I}} \cdot Z^{\textbf{pert-II}}      }{   Z^{\textbf{pert~extra}}  } \,,\\
Z^{ \textbf{instanton } } 
&= \frac{Z^{\textbf{nonpert-I}}  \cdot Z^{\textbf{nonpert-II}}       }{   Z^{\textbf{inst~extra}}  } 
\crcr
&=\frac{Z^{\textbf{nonpert-I}}  }{   Z^{\textbf{inst~extra}} }  \cdot \frac{  Z^{\rm sum}   }{   Z^{\textbf{pert-II}}     }   
 \crcr
&= 1+ \sum_{k=1}^{\infty}\, Z_k(A_i, y_i)~u^k,\label{instantonform}
\end{align}
where $Z_k(A_i, y_i)$ is the $k$-instanton partition function. Typically, it is computationally demanding to find higher-instanton partition functions. Here we also  present the results up to two instanton order. 

We note that, for a given number of flavors, one can have various 5-brane web configurations via  Hanany-Witten transitions as well as flop transitions. Applying the topological vertex method, hence, may give seemingly different partition functions. The partition functions are, however, related by extra factors. After removing such extra factors, one obtains the unique Nekrasov partition function for the gauge theory. 
One can therefore choose a representative 5-brane configuration for $Sp(N)+1\mathbf{AS}+N_f\mathbf{F}$ and compute the refined partition function as an expansion of the instanton fugacity. Here, we however, consider only $Sp(2)+1\mathbf{AS}+N_f\mathbf{F}$ ($N_f=0,1,2,3,4$) and $Sp(3)+1\mathbf{AS}$ as the partition functions for higher ranks or higher number of flavors takes a lot of time.
\subsection{\texorpdfstring{$Sp(2)_0+ 1\mathbf{AS}$}{Sp(2)_0+1AS}}
As depicted in Figure \ref{fig:nf0}, the 5-brane web for $Sp(2)_0+ 1\mathbf{AS}$ has three jumps  associated with Higgsing of the external edges. There  are also two edges that are not Higgsed, which are responsible for the mass of the hypermultiplet in the antisymmetric representation, given as the separation between these two edges.

\begin{figure}[]
	\centering
	\begin{tikzpicture}[scale=0.7,line width=.6pt]
	\draw (-1,-1)--(1,-1);
	\draw (2,2)--(2,-2);
	\draw (2,2)--(-2,2);
	\draw (2,-2)--(-1.5,-2);
	\draw (-2,2)--(-2,-1.5);
	\draw (1,1)--(3,3);	\draw (1.9,2)--(2.9,3);
	\draw (1,-1)--(3,-3);\draw (-1.9,2)--(-2.9,3);
	\draw (-1,1)--(-3,3);\draw (1.9,-2)--(2.9,-3);
	\draw (-1,-1)--(-1.5,-1.5);
	\draw (-2,-1.5)--node[above=0pt]{{\tiny $Q_8$}}node[below=0pt]{{\tiny $\a_8$}}(-1.5,-1.5) ;
	\draw (-1.5,-2)--(-1.5,-1.5);
	\draw (-1.5,-2)--(-2.5,-3);
	\draw (-2,-1.5)--(-3,-2.5)	;
	
	\draw (-1,-1) --(-1,1);
	\draw (1,-1) --(1,1);
	\draw (1,1) --(-1,1);
	\end{tikzpicture}
	\caption{A 5-brane web diagram for $Sp(2)_0+ 1\mAS$. The K\"{a}hler parameter $Q_8$ assigned on an edge with the Young diagram $\alpha_8$ is the mass fugacity for the antisymmetric hypermultiplet. }
	\label{fig:nf0}
\end{figure}

The 5-brane web can be obtained by Higgsing of a 5-brane web of a $SU(2)\times SU(4)\times SU(2)$ quiver gauge theory depicted in Figure \ref{fig:rank2}. 

\begin{figure}[]
	\centering
	\begin{tikzpicture}[scale=0.8,line width=.6pt]
	\draw(0.5,1.1)--(0.5,0.9);	\draw(0.4,1.1)--(0.4,0.9);
	\draw(0.4,2.9)--(0.4,3.1);\draw(0.5,2.9)--(0.5,3.1);
	\draw(0.5,-1.1)--(0.5,-0.9);	\draw(0.4,-1.1)--(0.4,-0.9);
	\draw(0.4,-2.9)--(0.4,-3.1);\draw(0.5,-2.9)--(0.5,-3.1);
	\draw(-0.1,-0.1)--(0.1,0.1);	\draw(-0.1,0.1)--(0.1,-0.1);
	\draw (-1.5,-1)--(-1.5,1)--node[above=0pt]{{\tiny $Q_{B}$}}node[below=0pt]{{\tiny $\v_2$}}(1.5,1)--node[left=0pt]{{\tiny $Q_{F}$}}(1.5,-1)--node[below=0pt]{{\tiny $\v_1$}}(-1.5,-1);
	\draw (-1.5,-1)--(-2.5,-2)--(-3.5,-2)--(-3.5,2)--(-2.5,2)--(-1.5,1);
	\draw (1.5,1)--(2.5,2)--(3.5,2)--(3.5,-2)--(2.5,-2)--(1.5,-1);
	\draw (2.5,2)--(2.5,3)--(-2.5,3)--(-2.5,2);
	\draw (2.5,-2)--(2.5,-3)--(-2.5,-3)--(-2.5,-2);
	
	\draw (3.5,-2)--(5.5,-4);\draw (2.5,-3)--(4.5,-5);
	
	\draw (-1.5,-1)--node[left=4pt,above=0pt]{{\tiny $Q_{k4}$}}(-2.5,-2)--node[below=-2pt]{{\tiny $Q_8$}}node[above=-2pt]{{\tiny $\a_8$}}(-3.5,-2)--node[left=-2pt]{{\tiny $Q_{33}$}}(-3.5,2)--node[above=-2pt]{{\tiny $Q_{1}$}}node[below=0pt]{{\tiny $\a_1$}}(-2.5,2)--node[left=4pt,below=0pt]{{\tiny $Q_{k1}$}}(-1.5,1);
	\draw (1.5,1)--node[right=4pt,below=0pt]{{\tiny $Q_{k2}$}}(2.5,2)--node[above=-2pt]{{\tiny $Q_4$}}node[below=-2pt]{{\tiny $\a_4$}}(3.5,2)--node[right=-2pt]{{\tiny $Q_{44}$}}(3.5,-2)--node[above=-2pt]{{\tiny $\a_5$}}node[below=-2pt]{{\tiny $Q_5$}}(2.5,-2)--node[right=4pt,above=0pt]{{\tiny $Q_{k3}$}}(1.5,-1);
	\draw (2.5,2)--node[right=-4pt]{{\tiny $Q_{3}$}}(2.5,3)--node[above=0pt]{{\tiny $Q_{22}$}}node[below=0pt]{{\tiny $\v_4$}}(-2.5,3)--node[left=-4pt]{{\tiny $Q_{2}$}}(-2.5,2);
	\draw (2.5,-2)--node[right=-4pt]{{\tiny $Q_{6}$}}(2.5,-3)--node[below=0pt]{{\tiny $Q_{11}$}}node[above=0pt]{{\tiny $\v_3$}}(-2.5,-3)--node[left=-4pt]{{\tiny $Q_{7}$}}(-2.5,-2);

	\draw (-2.5,-3)--node[above=0pt, right=0pt]{{\tiny $q$}}(-4.5,-5); 	\draw (-3.5,-2)--node[ left=0pt,above=0pt]{{\tiny $q$}}(-5.5,-4); 
	\draw (2.5,-3)--node[above=0pt,left=0pt]{{\tiny $q$}}(4.5,-5); 	\draw (3.5,-2)--node[ left=0pt,above=0pt]{{\tiny $q$}}(5.5,-4); 
	\draw (-2.5,3)--node[above=0pt, right=0pt]{{\tiny $t$}}(-4.5,5); 	\draw (-3.5,2)--node[ left=0pt,below=0pt]{{\tiny $t$}}(-5.5,4); 
	\draw (2.5,3)--node[above=0pt,left=0pt]{{\tiny $t$}}(4.5,5); 	\draw (3.5,2)--node[ right=0pt,below=0pt]{{\tiny $t$}}(5.5,4); 
	
	\node [above, blue] at (5.5,4.5) {$\sqqt$};
	\node [above] at (7,4.5) {$\textbf{ Case C}$};
	\node [above] at (7,-5.2) {$\textbf{ Case B}$};
	\node [above] at (-7.5,4.5) {$\textbf{ Case B}$};

	\node [above left, blue] at (-5,4.5) {$\sqqt$};\node [ below right, blue] at (4.7,-4.2) {$\sqtq$};

	\draw(-2.5,3)--(-4.5,5);\draw(-3.5,2)--(-5.5,4);\draw[dashed,red](-2.5,2)--(-5,4.5); \draw[dashed,red](-3.5,2)--(-3.5,3)--(-2.5,3);
	\draw[dashed,red](2.5,2)--(5,4.5);
	
	\draw[dashed,red](3.5,2)--(3.5,3)--(2.5,3);
	\draw(-2.5,-3)--(-4.5,-5);\draw(-3.5,-2)--(-5.5,-4);

	\draw[dashed,red](2.5,-2)--(5,-4.5);

	\draw[orange](-6,0)--(6,0);\draw[orange](0,-6)--(0,6); 
	
	\draw[red,dashed](2.5,-3)--(3.5,-3)--(3.5,-2);
			\draw[dashed,red](-2.5,-2)--(-5,-4.5);
	\draw[red,dashed](-2.5,-3)--(-3.5,-3)--(-3.5,-2);	
	\end{tikzpicture}
	\caption{An unHiggsed diagram for $Sp(2)+1\mAS$, which is 5-brane web for an $SU(2)\times SU(4) \times SU(2)$ quiver gauge theory. By Higgsing or tuning the K\"{a}hler parameters associated with external edges, one reproduces the 5-brane web given in Figure \ref{fig:nf0}.}
	\label{fig:rank2}
\end{figure}
On each edge of the unHiggsed web diagram in Figure \ref{fig:rank2}, we assigned the Young diagrams $\alpha_i$ and the K\"{a}hler parameters $Q_i= e^{-iL_i}$ where $L_i$ is the length of the corresponding edge. It is easy to see that not all K\"{a}hler parameters are independent. With the convention that 
\begin{align}
Q_{i,i, \cdots l}:= Q_i Q_j \cdots Q_l,	
\end{align}
we denote ten independent K\"{a}hler parameters by $Q_i$ ($i=1,\cdots,6,8, k4, F, B)$. Then other K\"{a}hler parameters are expressed as 
\begin{eqnarray}\label{relation}
&&Q_{k1}=\frac{Q_{8,k4}}{Q_1},~~Q_{k2}= \frac{ Q_{2,8,k4}  }{ Q_{1,3} },~~Q_{k3}= \frac{ Q_{2,4,8,k4}  }{ Q_{1,3,5}},~~Q_7=\frac{ Q_{2,4,6,8}  }{ Q_{1,3,5}},~~~~~~~~~\\
&&Q_{11}= \frac{ Q_{2,4,8,B, k4,k4}  }{ Q_{1,3,5} },~~Q_{22}= \frac{ Q_{2,8,8,B,k4,k4} }{ Q_{1,1,3} },~~Q_{33}= \frac{ Q_{8,F,k4,k4}  }{ Q_1 },~~Q_{44}= \frac{ Q_{2,2,4,8,8,F,k4,k4} }{ Q_{1,1,3,3,5}}\,. \nn
\end{eqnarray}

To obtain the Nekrasov partition function for $Sp(2)_0+ 1\mathbf{AS}$, one first needs to properly tune the K\"{a}hler parameters associated with the external edges, which reduces the unHiggsed diagram in Figure \ref{fig:rank2} to the 5-brane web for $Sp(2)_0+ 1\mathbf{AS}$ given in Figure \ref{fig:nf0}.  Recalling the $T_2$-tuning in section 3, we found that the correct tuning of the  K\"{a}hler parameters is given by
\begin{align}
	Q_{1} =Q_{2}=Q_{3}=Q_{4}=\sqrt{\frac{q}{t}} \,,\qquad  Q_{5} = Q_{6}=\sqrt{\frac{t}{q}} \ ,
\end{align}
which correspond to two \textbf{Case B} and one \textbf{Case C}. 
Next, the gauge theory parameters, two Coulomb branch parameters $A_1, A_2$, the instanton fugacity $u$, and the mass fugacity for an antisymmetric $Q_8$ are assigned to the following K\"{a}hler parameters
\begin{align}\label{sub0F}
Q_F={A_2}^2,\qquad Q_{k4}=\frac{A_1 }{A_2 ~Q_8},\qquad Q_B=u~{A_2}^2.
\end{align}

 $Z^{\text{sum}}_{Sp(2)_0+1 \mAS }$ as the summation  over the Young diagrams along preferred edges is given by
\begin{align}
Z^{\text{sum}}_{Sp(2)_0+1 \mAS } &=\sum_{\substack{\v_{1},\v_{2},\\\v_{3},\v_{4}}\, } \,\,\sum_{\substack{\a_{1},\a_{4}\\\a_{5},\a_{8}}\, }  
Z^{\bf sum}_\textbf{Sp(2)+1$\mAS$}\cr
&= \sum_{\substack{\v_{1},\v_{2},\\\v_{3},\v_{4}}\, } \,\,\sum_{\substack{\a_{1},\a_{4}\\\a_{5},\a_{8}}\, }  Z[\v_1,v_2,v_3,v_4;\a_1,\a_4,\a_5,\a_8]
 \,,
\end{align}
with shorthand notation
\begin{align}\label{sumpartsp2AS}
&Z^{\bf sum}_\textbf{Sp(2)+1$\mAS$}:=Z[\v_1,v_2,v_3,v_4;\a_1,\a_4,\a_5,\a_8]  \\
	&=
		u^{|\v_1|+|\v_2|+|\v_3|+|\v_4|}{A_2}^{2|\v_1|+2|\v_2|} {A_1}^{2|\v_3|+2|\v_4|}  Q_8^{-|\v_3|} \prod_{i=1}^{4}\tZ_{\v_i}(q,t)~\tZ_{\v_i^T}(t,q)  \nn\\
	&
\times	
q^{||\v_2||^2 +\frac{ |\v_3| }{2} -|\v_4| +||\v_4||^2    }
	t^{||\v_1||^2 -\frac{ |\v_3| }{2} +|\v_4| +||\v_3||^2    }\nn\\
	& 
\times
\frac{1}{  
		N_{\v_2 \v_1^T}\l({A_2}^{2} \r) 
		N_{\v_2 \v_1^T}\l(\frac{{A_2}^{2} ~t}{q}  \r) 
		N_{\v_2\v_3^T}\l(A_1A_2\r )
		N_{\v_2\v_3^T}\l(\frac{A_1A_2~ t}{q} \r)
		N_{\v_4\v_2}\l(\frac{A_1}{A_2}\r)
		N_{\v_4\v_2}\l(\frac{A_1 t}{A_2 q}\r)
	}  \nn\\
	&
\times
	\frac{1}{
		N_{\v_4\v_1^T}\l( A_1A_2\r)
		N_{\v_4\v_1^T}\l( \frac{A_1A_2 t}{q}\r) 
		N_{\v_4\v_3^T}\l( \frac{A_1}{A_2}\r)
		N_{\v_4\v_3^T}\l( \frac{A_1 t}{A_2 q} \r)
		N_{\v_1^T\v_3^T}\l( \frac{A_1}{A_2}\r)
		N_{\v_1^T\v_3^T}\l( \frac{A_1 t}{A_2 q} \r)
	}    \nn\\
	&
\times
	(-1)^{|\a_1|+|\a_4|+|\a_5|+|\a_8|  } Q_8^{|\a_8| } 
	\prod_{j=1,4,5,8}\tZ_{\a_j}(q,t)~\tZ_{\a_j^T}(t,q)
	\nn\\
	&
\times
	q^{\frac{ |\a_1|+||\a_1||^2 +|\a_4|+||\a_4||^2 -|\a_5|+||\a_5^T||^2+ ||\a_8^T||^2    }{2}   }\,\, 
	t^{\frac{ -|\a_1|+||\a_1^T||^2 -|\a_4|+||\a_4^T||^2 +|\a_5|+||\a_5||^2+ ||\a_8||^2     }{2}   }\nn\\
	&
\times
 \frac{ N_{\a_1\v_2}\l(\frac{A_1 t}{A_2 q}\r)
		N_{\a_1\v_1^T}\l(\frac{A_1 A_2t}{ q}\r)
		N_{\a_1\v_3^T} \l(\frac{{A_1}^2 t}{ q}\r)
		N_{\a_4\v_2} \l(\frac{A_1 t}{A_2 q}\r)
		N_{\a_4\v_1^T} \l(\frac{A_1 A_2t}{q}\r)
		N_{\a_4\v_3^T} \l(\frac{{A_1}^2 t}{q}\r)
		N_{\v_2\a_5^T} \l(A_1A_2\r)
	}
	{ N_{\a_1\a_8^T} \l(\frac{{A_1}^2 }{Q_8 } \sqtq \r)
		N_{\a_1\a_8^T} \l(\frac{{A_1}^2 t }{Q_8 q} \sqtq \r)
	} \nn\\
	&
\times
	\frac{
		N_{\v_2\a_8^T} \l(\frac{A_1 A_2 }{Q_8 } \sqtq \r)
		N_{\v_4\a_5^T}\l({A_1}^2 \r)
		N_{\v_4\a_8^T} \l(\frac{{A_1} ^2 }{Q_8 } \sqtq \r)
		N_{{\a_8}^T\v_3^T} \l(Q_8 \sqtq \r)
		N_{\v_1^T\a_5^T} \l(\frac{A_1 }{A_2 }\r)
		N_{\v_1^T\a_8^T} \l(\frac{A_1 }{A_2 Q_8} \sqtq\r)
	}
	{ N_{\a_4\a_5^T} \l(  A_1^2 \r)
		N_{\a_4\a_5^T} \l(  A_1^2 \frac{t}{q} \r)
	}  \nn\\
	&
\times	
	 N_{\v_4\a_1}(1)
	N_{\v_4\a_4}\l(1 \r)
	N_{\a_5\v_3} \l(\frac{t}{q}  \r)
	 \,,\nonumber
\end{align}
where we used a shorthand notation $N_{\bullet\bullet}(Q)$ for $N_{\bullet\bullet}(Q;  t^{-1}, q^{-1})$, and it follows from \eqref{constraint2} that the Young diagrams of $N_{\v_4\a_1}(1)
	N_{\v_4\a_4}\l(1 \r)
	N_{\a_5\v_3} \l(\frac{t}{q}  \r) $ in the last line satisfy $\v_4 \> \a_1,~\v_4\>\a_4,~\v_3\> \a_5$.

\paragraph{Perturbative contribution.}
With these assignments of K\"{a}hler parameters, we can express the topological string partition function 
\begin{align}
Z^{\rm top}_{Sp(2)_0+1\mAS}=Z^{M}_{Sp(2)_0+1\mAS}\cdot Z^{\rm sum}_{Sp(2)_0+1\mAS}\,,	
\end{align}
where $Z^{M}_{Sp(2)_0+1\mAS}$ takes the following form 
\begin{align}\label{pert1}
Z^{M}_{Sp(2)_0+1\mAS} = Z^{\textbf{pert-I}}_{Sp(2)_0+1\mAS} \cdot Z^{\textbf{nonpert-I}}_{Sp(2)_0+1\mAS}\,,
\end{align}
where
\begin{align}
Z^{\textbf{pert-I}}_{Sp(2)_0+1\mAS}
=&\frac{ M(\frac{A_1}{A_2 Q_8}\sqtq,t,q  )M( \frac{A_1A_2}{Q_8},t,q  ) M(Q_8\sqtq , t,q ) }
{M({A_1}^2,t,q) M(\frac{A_1}{A_2},t,q ) M(A_1A_2,t,q) M({A_2}^2,t,q) M({A_2}^2,q,t ) M( \frac{{A_1}^2}{Q_8} \sqtq,q,t)}, \nn \\
Z^{\textbf{nonpert-I}}_{Sp(2)_0+1\mAS} =&1.
\end{align}
Here we neglected unimportant factors like $M(1,t,q)$  on the right-hand side of \eqref{pert1}.
We note that, in general, $Z^{M}$ contains terms depending on the instanton fugacity $u$, but in this case, $Z^{M}$ is independent of $u$. We found, in fact, that $Z^{\textbf{nonpert-I}} =1$ even with flavors up to $N_f=3$.

The perturbative contribution from the summation part of partition function is also independent of instanton fugacity and is obtained by setting ${\v_{1}=\v_{2}=\v_{3}=\v_{4}=\0}$,
\begin{eqnarray}\label{pertII}
	Z^{\textbf{pert-II}}_{Sp(2)_0+1\mAS} =\sum_{\a_1,\a_4,\a_5,\a_8}Z[\0,\0,\0,\0; \a_{1},\a_{4},\a_{5},\a_{8}],
\end{eqnarray}
where it follows from \eqref{constraint1} and \eqref{constraint2} that $\a_{1,4}\preccurlyeq \v_4=\0, \a_{5}\preccurlyeq \v_3=\0$, which yields
\begin{align}\label{eq:pert-II}
Z^{\textbf{pert-II}}_{Sp(2)_0+1\mAS}&=
\sum_{\a_8} \underbrace{   (-1)^{|\a_8|} q^{\frac{||\a_8^T||^2 }{2} } t^{\frac{||\a_8||^2 }{2} }
	 ||\tZ_{\a_8^T}(q,t)||^2 N_{\a_8^T}^{\rm half, -} (Q_8)N_{\a_8}^{\rm half, +} (Q_8)}_{T_2 -\rm partition~ function}  
\frac{ N_{\a_8^T}^{\rm half, +} (Q_{F, k4} )  }{  N_{\a_8^T}^{\rm half, +} (Q_{8, F, k4}~ \frac{t}{q}) }\,\cr
&= \frac{ M\l(Q_8 \sqqt,q,t \r)  M(  \frac{ Q_8 A_1}{A_2} \sqqt,q,t   )  M\l( Q_8 A_1A_2 \sqqt,q,t  \r) M\l(\frac{{A_1}^2}{Q_8} \sqtq,q,t  \r) }{M\l( {A_1}^2,q,t \r)M\l(\frac{A_1}{A_2},q,t  \r) M\l(A_1 A_2,q,t \r ) M\l({Q_8}^2, t, q \r)   }, 
\end{align}
where we reorganized $M(\bullet, t,q)$ to express $Z^{\textbf{pert-II}}$ as a compact form.
We note that even though we add flavors, $Z^{\textbf{pert-II}}$  is unaltered at least for $N_f\le 4$. Taking into account the extra factor associated with an antisymmetric hypermultiplet, the stringy contribution from two parallel external edges in Figure \ref{fig:nf0} is given by
\begin{align}
	Z_{Q_8}^{\rm extra}=\frac{1}{M({Q_8}^2,t,q)} \,.
\end{align}
One obtains from \eqref{pfformgeneral2} that the full perturbative part of the partition function
\begin{align}	
Z^{\textbf{pert}}_{Sp(2)_0 +1\mAS } &=\frac{ Z^{\textbf{pert-I}}_{Sp(2)_0 +1\mAS } \cdot Z^{\textbf{pert-II}}_{Sp(2)_0 +1\mAS }  } { Z_{Q_8}^{\rm extra} }\cr
	& =\frac{  M\l(Q_8 A_1 A_2\sqtq, t,q\r) M\l(\frac{A_1 A_2}{Q_8}\sqtq, t,q\r) M\l( \frac{Q_8 A_1}{A_2}\sqtq, t,q\r)   
	}{M\l(A_1 A_2,t,q\r)M\l(A_1 A_2,q,t\r)  M\l(\frac{A_1}{A_2},t,q  \r) M\l(\frac{A_1}{A_2},q,t\r)      
	}\cr
	&\times 
	\frac{
	M\l( \frac{A_1}{Q_8 A_2}\sqtq, t,q \r)      M\l(Q_8 \sqtq,t,q\r)^2   
	}{  
	M\l({A_1}^2,t,q\r) M\l({A_1}^2,q,t\r) M\l({A_2}^2,t,q\r) M\l({A_2}^2,q,t\r)      
	},
\end{align}
which exactly agrees with the result obtained from the localization computation \eqref{pertlocalization}. With a proper normalization, the full perturbative part can be expressed as 
\begin{align}
\acute{Z^{\textbf{pert}}_{Sp(2)_0+1\mAS}}=&~\frac{Z^{\textbf{pert}}_{Sp(2)_0 +1\mAS }}{M(Q_8 \sqtq,t,q)^2} \nn\\
=&~ 1+ \frac{q+ t} {(1-q)(1-t)} {A_2}^2+  \frac{q+t-\sqrt{qt} \chi^{SU(2)}_2[Q_8]  }{(1-q)(1-t)} \frac{A_1}{A_2} \nn\\
&+\frac{(1+q t)\big(q+t -\sqrt{q t} \chi^{SU(2)}_2[Q_8]\big)  }{(1-q)^2(1-t)^2  } A_1A_2 \nn
+ \mathcal{ O}({A_1}^1; {A_2}^1) \,,
\end{align}
where $\chi_2^{SU(2)}[Q_8] = Q_8+{Q_8}^{-1}$ is the character associated with the antisymmetric mass fugacity. From here on, we use the following simpler notation $\chi_n^{SU(2)} =\chi_n^{SU(2)}[Q_8]$ for the character associated with the mass of an antisymmetric in the $n$-dimensional representation of $SU(2)$.

\vskip 0.5cm
\paragraph{Instanton contribution.}
The instanton contribution is obtained from \eqref{instantonform}. For $n$-instanton partition function $Z_n$, one restricts the power of the instanton fugacity to be $n$; in other words, $u^{|\v_{1}|+|\v_{2}|+|\v_{3}|+|\v_{4}|}=u^n$.
As $Z^{\textbf{nonpert-I}}_{Sp(2)_0 +1\mAS } (u,A_i,y_i) =1$, the one-instanton contribution is given by
\begin{eqnarray}\label{oneinstsp2}
Z^\textbf{one-instanton}_{ Sp(N)+ 1 \mAS} 
=\frac{
	\sum\limits_{ \substack{  \v_1,\v_2,\v_3,\v_4\in
			\{{\begin{tikzpicture}[scale=.15]
				\draw (0,1)--(0,0);
				\draw (0,0) --(1,0);
				\draw (1,1) --(1,0);
				\draw (1,1) --(0,1);
				\end{tikzpicture}}
			,~\0\}\\
		\text{and}~
			|\v_1|+|\v_2|+|\v_3|+|\v_4|=1}}
		~	\sum\limits_{\a_{1},\a_4,\a_5,\a_8}
	Z[\v_{1},\v_{2},\v_{3},\v_{4}
	,\a_{1},\a_{4},\a_{5},\a_{8} ] } {Z^{\pertII}_{Sp(2)_0 +1\mAS }    }\,,
\end{eqnarray}
which actually is already quite lengthy if one sums over the contributions of $|\a_1|+|\a_4|+|\a_5|+|\a_8|\leqslant 6$. 
Since (\ref{constraint2}) leads to  constraints $\a_1,\a_4 \preccurlyeq \v_4$ and $\a_5 \preccurlyeq \v_3$,
and other terms do not satisfy this constraint just equal to zero, $Z_1$ can be further simplified
\begin{align}\label{oneinstsimplify}
&Z^\textbf{one-instanton}_{ Sp(N)+ 1 \mAS} \cdot Z^\pertII=\\
&	{\sum\limits_{\a_{8}}}~
Z[\underbrace{\{1\},\0,\0,\0  }_{\v_1,\v_2,\v_3,\v_4} ;~  \underbrace{ \0,\0,\0,\a_8 } _{\a_1,\a_4,\a_5,\a_8}   ]+Z[{\0,\{1\},\0,\0}; ~{ \0,\0,\0,\a_8 }  ]+Z[{\0,\0,\{1\},\0  };~ { \0,\0,\0,\a_8 }  ]
\nn\\
&+ Z[{\0,\0,\{1\},\0  }; ~{ \0,\0,\{1\},\a_8 }  ]+ Z[{\0,\0,\0,\{1\}};~ { \0,\0,\0,\a_8 }  ]+Z[{\0,\0,\0,\{1\}  }; ~{ \0,\{1\},\0,\a_8 }  ] 
\nn\\
&+Z[{\0,\0,\0,\{1\}}; ~{ \{1\},\0,\0,\a_8 }  ]+Z[{\0,\0,\0,\{1\}  }; ~{ \{1\},\{1\},\0,\a_8 }  ]\nn\,,
\end{align}
where $\{1\}$ stands for Young diagram $\begin{tikzpicture}[scale=.2]
\draw (0,1)--(0,0);
\draw (0,0) --(1,0);
\draw (1,1) --(1,0);
\draw (1,1) --(0,1);
\end{tikzpicture}$.
By expanding \eqref{oneinstsp2} with respect to the Coulomb branch parameters, we get 
\begin{align}\label{oneinst}
&Z^\textbf{one-instanton}_{ Sp(N)+ 1 \mAS}= 
\frac{q+t }{(1-q) (1-t) }(A_1^2+ A_2^2) 
 + \frac{q+t- \sqrt{q t}\chi^{SU(2)}_2 }{(1-q)(1-t) } A_2A_1\crcr
& + 
\frac{(q^2+q t+t^2)(q+t- \sqrt{q t}\chi^{SU(2)}_2 ) }{(1-q)q(1-t)t}(A_1A_2^3+A_1^3 A_2)
 +  \frac{ (q+t)^2 ( q+t \sqrt{q t} \chi^{SU(2)}_2) }{(1-q)q(1-t)t  }  A_2^2A_1^2 \cr
&+  \frac{ (q^2 + q t + t^2)^2 ( q+t- \sqrt{q t}\chi^{SU(2)}_2)  }{ (1 - q) q^2 (1 - t) t^2 } A_2^3A_1^3+\mathcal{ O}(A_1^4; A_2^4)\,.
\end{align}

We now compare the one-instanton contribution with the known result, which we summarized in \eqref{oneinstloc}. The relevant part ($N_f=0$) is given as follows:
\begin{align}\label{oneinstlocnf0}
Z^\textbf{one-instanton}_{ Sp(N)+ 1 \mAS} &=\frac{1}{2} \bigg(
\frac{ 1}{2~\sinh \frac{\epsilon_{+} \pm \epsilon_{-}} {2} } 
\frac{ 2~\sinh \frac{m \pm \a_1 }{2}  ~ 2~\sinh \frac{m \pm \a_2 }{2}  - 2~\sinh \frac{\pm \a_1+\epsilon_{+} }{2}  ~ 2~\sinh \frac{\pm \a_2+\epsilon_{+} }{2}  } { 2~\sinh \frac{\pm \a_1+\epsilon_{+} }{2}  ~2~\sinh \frac{\pm \a_2+\epsilon_{+} }{2} ~2~\sinh \frac{m \pm \epsilon_{+} }{2}  }  \nn\\
&+\frac{ 1}{2~\sinh \frac{\epsilon_{+} \pm \epsilon_{-}} {2} } 
\frac{ 2~\cosh \frac{m \pm \a_1 }{2}  ~ 2~\cosh \frac{m \pm \a_2 }{2}  - 2~\cosh \frac{\pm \a_1+\epsilon_{+} }{2}  ~ 2~\cosh \frac{\pm \a_2+\epsilon_{+} }{2}  } { 2~\cosh \frac{\pm \a_1+\epsilon_{+} }{2}  ~2~\cosh \frac{\pm \a_2+\epsilon_{+} }{2} ~2~\sinh \frac{m \pm \epsilon_{+} }{2}  }
\bigg)\,.
\end{align}
With the identification 
$A_{1}:=e^{\a_{1}}, ~A_{2}:=e^{\a_{2}},~Q_8:=e^m$ with $m$ being the mass for the antisymmetric hypermultiplet
and the Omega deformation parameters $ q=e^{-\epsilon_2},~t=e^{\epsilon_1},~\epsilon_{\pm}=\frac{\epsilon_1 \pm \epsilon_2}{2}$. The expansion of \eqref{oneinstlocnf0} in terms of $A_{1}$, and $A_2$ agrees with our result \eqref{oneinst}.

\vskip .5cm
Similarly, the two-instanton contribution is given by
\begin{equation}\label{twoinstsp2}
Z^{{{\textbf{two-instanton}}}}_{ Sp(N)+ 1 \mAS} 
=\frac
{
	\sum\limits_{\a_{1},\a_4,\a_5,\a_8}~
	\sum\limits_{ \substack{\v_1,\v_2,\v_3,\v_4 \in\{
			\raisebox{-.1cm}	{\begin{tikzpicture}[scale=.15]
				\draw (0,2)--(0,0);
				\draw (0,0) --(1,0);
				\draw (1,2) --(1,0);	\draw (1,2) --(0,2);
				\draw (1,1) --(0,1);
				\end{tikzpicture}},~
			{\begin{tikzpicture}[scale=.15]
				\draw (0,1)--(0,0);\draw (1,1) --(1,0);
				\draw (0,0) --(2,0);	\draw (2,1) --(2,0);
				\draw (2,1) --(0,1);
				\end{tikzpicture}}, ~
			{\begin{tikzpicture}[scale=.15]
				\draw (0,1)--(0,0);\draw (1,1) --(1,0);
				\draw (0,0) --(1,0);
				\draw (1,1) --(0,1);
				\end{tikzpicture}}, ~\0
			\}\\
			|\v_{1}|+|\v_2|+|\v_3|+|\v_4|=2} }  
	Z[\v_{1},\v_2,\v_3,\v_4
	;\a_{1},\a_4,\a_5,\a_8 ] }
{Z^{\textbf{pert-II}}_{ Sp(N)+ 1 \mAS}   }
\,,
\end{equation}
which can also be reduced by constraints $\a_1,\a_4 \preccurlyeq \v_4$ and $\a_5\preccurlyeq \v_3$.
With the assignment of K\"{a}hler parameters in terms of $(A_i, u, y_i)$, the two-instanton contribution can be expanded as 
\begin{align}
&Z^{{{\textbf{two-instanton}}}}_{ Sp(N)+ 1 \mAS}=
 \frac{(q+t) (q+t-\sqrt{q t}\chi^{SU(2)}_2) }{(1 - q)^2  (1 -t)^2 } (A_2^3A_1+A_2 A_1^3 )
\cr
& +\frac{(q+t)\l(  q t\chi_3^{SU(2)} - \sqrt{qt}(1+q)(1+t) \chi_2^{SU(2)} +2(q+t)(1+qt)+3 qt+q^2+t^2 \r) }{ (1 - q)^2 (1 + q) (1 - t)^2 (1 + t)} A_2^2A_1^2
 \cr
 & 
 +\mathcal{ O}(A_1^3; A_2^3)\,,
\end{align}
where $\chi_3^{SU(2)} ={Q_8}^2+1+{Q_8}^{-2}$.

\vskip .5cm
\paragraph{Enhancement of global symmetry.}
With a proper normalization, the partition function for $Sp(2)+1\mAS+ n\mF$ with enhancement symmetry can be expressed as
\begin{align} \label{eq:Zenhanced}
Z&:=  \frac{Z^{\textbf{pert}}(A_i, y_i )}{M\l(Q_8 \sqtq,t,q\r)^2} \cdot \l(1+\sum\limits_{k=1}^{\inf} Z_{k}	
(A_i,y_i)  u^k\r)
= 1+ \sum_{m,n} f( \chi[\tu,\ty_i],t,q  ) ~\tA_1^m \tA_2^n\,,
\end{align}
where $\chi[\tu,\ty_i]$ is some characters for the enhanced global symmetry written in terms of the redefined fugacities of instanton and hypermultiplets, and $\tilde{A_i}$ are redefined parameters.

As $Sp(2)_0+1\mAS$ is rank 2 $E_1$ theory, its global symmetry is enhanced to $SU(2)$ at the UV fixed point which was explicitly shown through superconformal index computation \cite{Kim:2012gu}. At the level of partition function, the enhancement of global symmetry can also be shown by taking into account the fiber-base duality. Following \cite{Mitev:2014jza}, we redefine the Coulomb branch parameters
\begin{align}
\tilde{A_1}=A_1 u^{\frac{1}{4}},\qquad \tilde{A_2}=A_2 u^{\frac{1}{4}},
\end{align}
to make the fiber-base duality manifest
\begin{align}
u\leftrightarrow u^{-1},\qquad Q_B\leftrightarrow Q_F,\qquad Q_8\leftrightarrow Q_8.	
\end{align}
For $Sp(2)_0+1\mAS$, we indeed find that \eqref{eq:Zenhanced} is expressed in terms of the characters of the enhanced $SU(2)$ global symmetry: 
\begin{align}\label{enhancement0}
&Z_{Sp(2)_0+1\mAS}=\nn\\
&1+ \frac{(q+t) \chi_2[u] }{(1-q)(1-t) }{{\tA}_2}^2+ \frac{ q+t -\sqrt{q t}\chi_3^{SU(2)}     }{(1-q)(1-t)  } \frac{\tA_1}{\tA_2} +   \frac{(1+q t) \chi_2[u]( q+t -\sqrt{q t}\chi_3^{SU(2)}  ) }{(1-q)^2(1-t)^2  } \tA_1\tA_2  \nn\\
&+\frac{ (q+t) \big( q t ~\chi_3^{SU(2)} -\sqrt{q t}( 1+q)(1+t) \chi_2^{SU(2)} + (q+t)(1+qt)+q t  \big)   } { (1 - q)^2 (1 + q) (1 - t)^2 (1 + t)  } \frac{\tA_1^2}{\tA_2^2} \nn\\
&+\mathcal{ O}({\tA_1}^2; {\tilde{A}_2}^2) , 
\end{align}
where $\chi_2[u]=\sqrt{u}+\frac{1}{\sqrt{u}}$ is the character of the enhanced $SU(2)$ global symmetry, and $\chi^{SU(2)}_n$ are the characters for the antisymmetric hypermultiplet.
From \eqref{enhancement0}, one sees the enhancement of global symmetry is $SU(2)_u \times SU(2)_{Q_8}$, as expected.

\subsection{\texorpdfstring{$Sp(2)_\pi+1\mathbf{AS}$}{Sp(2)_pi+1AS}}
\begin{figure}[]
	\centering
	\includegraphics[width=3.5in]{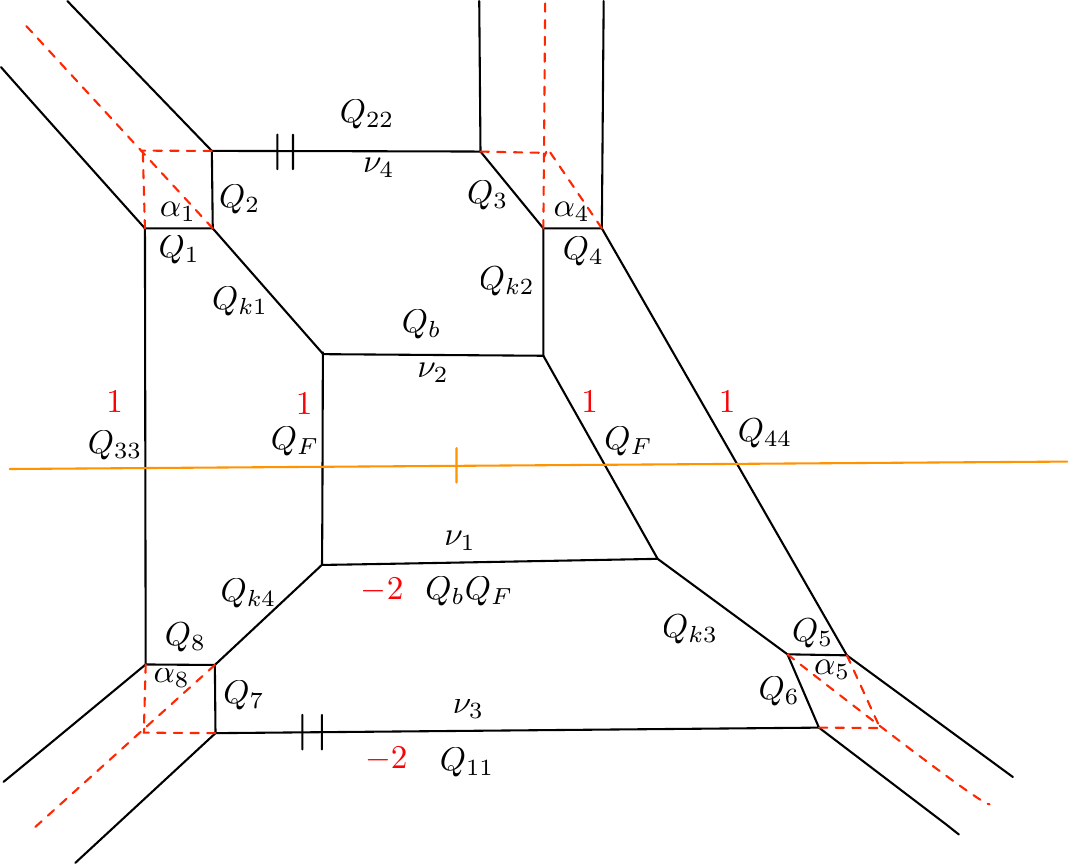}
	\caption{An unHiggsed 5-brane web diagram for $Sp(2)_\pi+1\mathbf{AS}$. The numbers in red denote framing numbers associated with edges.} 
	\label{fig:sp2pi}
\end{figure}
We now discuss $Sp(2)$ gauge theory with the discrete theta angle $\theta=\pi$ and an antisymmetric. A 5-brane web configuration is given in Figure \ref{fig:sp2pi}, from which one can read off 
the relations between K\"{a}hler parameters:
\begin{align}
	Q_{k1}&= \frac{Q_{8, k4}}{Q_1}\ ,\qquad Q_{k2}=\frac{Q_{2,8,k4}  }{Q_{1,3}}\ ,\qquad Q_{k3}=\frac{ Q_{2,4, 8, k4  } }{ Q_{1,3,5} }\ , \nn\\
	Q_7&=\frac{ Q_{2,4,6,8  } }{Q_{1,3,5}  } \ ,\qquad Q_{22}=\frac{ Q_{8, b , k4} }{ Q_{1,3} }\ ,  
	\qquad Q_{33} = \frac{ Q_{8,F, k4,k4} }{Q_1} \ ,\nn\\
	Q_{44}&=\frac{ Q_{2,2,4,8,8,F,k4,k4} }{ Q_{1,1,3,3,5}}\ ,\qquad Q_{11}=\frac{ Q_{2,2,4,4,6,8,8,b,F, k4,k4,k4} }{Q_{1,1,3,3,5,5}   }\,.
\end{align}
By the $T_2$-tuning, the correct tuning for K\"{a}hler parameter given  is as follows
\begin{equation}
	Q_1=Q_2=Q_3=Q_4=\sqrt{\frac{q}{t}}, \qquad Q_5=Q_6=\sqrt{\frac{t}{q}}\,.	
\end{equation}
Independent K\"{a}hler parameters are assigned with gauge theory parameters
\begin{equation}
	Q_F={A_2}^2,\qquad Q_{k4}=\frac{A_1}{A_2~Q_8},\qquad Q_b=u~ A_2 \,.  \label{Qk4A1A2}
\end{equation}
It is then straightforward to compute  
$Z^{\pertI}$ and  $Z^{\pertII}$, which shows that the full perturbative part $Z^{\pert}_{Sp(2)_\pi+1\mAS}$ is the same as that of $Sp(2)_0+1\mathbf{AS}$, as expected. 

\paragraph{Instanton contribution.} We now consider the instanton contribution for $Sp(2)_\pi+1\mAS$. To obtain the instanton contributions, the summation part of the topological string partition function is needed
\begin{align} 
	Z^\text{sum}_{Sp(2)_\pi+ 1 \mAS}=&\sum_{\v_1,\v_2,\v_3,\v_4}~~\sum_{\a_1,\a_4,\a_5,\a_8}
	Z^{\bf sum}_{\bf Sp(2)+ 1 AS} \cdot \bf terms_\pi\,,\\
	{\bf terms_\pi}:= &~
	(-1)^{|\v_1|+|\v_2|+|\v_3|+|\v_4|    } A_1^{|\v_3|-|\v_4|} A_2^{ |\v_1|-|\v_2| }\nn\\
	&\times 
	q^{-\frac{ ||\v_1^T||^2 +|| \v_2||^2 + || \v_3^T ||^2   +  || \v_4  ||^2      } {2}  } t^{\frac{ ||\v_1||^2 +|| \v_2^T||^2 + || \v_3 ||^2   +  || \v_4^T  ||^2      } {2}  }
	\,,\nn
\end{align}
where $Z^{\bf sum}_{\bf Sp(2)+ 1 AS} $ is defined in \eqref{sumpartsp2AS}.
\vskip .5cm
The one- and two-instanton contributions then take the form: 
\begin{align}
		&Z^\textbf{one-instanton}_{Sp(2)_\pi+ 1 \mAS}=-\frac{\sqrt{qt}}{(1-q)(1-t)} (A_1+A_2) \crcr
		&+\frac{q+t}{(1-q)(1-t)}\Big( \chi^{ SU(2)}_2(Q_8)  +  \frac{q+t}{\sqrt{q t} }       \Big) A_1A_2 (A_1 +A_2)
	+ \mathcal{O}({A_1}^2;{A_2}^2)\,.
\end{align}
	\begin{align}
	&Z^{{{\textbf{two-instanton}}}}_{Sp(2)_\pi+ 1 \mAS}=	-\frac{q t (q+t)}{(1-q)^2  (1+q)(1-t)^2(1+t) } ({A_1}^2+{A_2}^2) +  \frac{ q t}{ (1-q)^2(1-t)^2} A_1 A_2  \nn\\
		&  
		+\frac{2 (q+t)\sqrt{q t}}{ (1-q)^2(1-t)^2 } \Big( \chi^{SU(2)}_2(Q_8)+  \frac{q+t }{ q t }   \Big) {A_1}^2 {A_2}^2 +\mathcal{O}({A_1}^2;{A_2}^2)\,.
	\end{align}

\paragraph{Reduction to $SU(2)_\pi$.} 
When the antisymmetric matter $ \mAS$ is massless, the corresponding 5-brane web becomes two copies of $Sp(1)_\pi$ theories as depicted in Figure \ref{fig:massless}. This means that the corresponding partition function factorizes to 
\begin{eqnarray}\label{factorizationpi}
Z^\text{sum}_{Sp(2)_\pi+ 1 \mAS}= Z_{\textbf{inner layer}} \cdot Z_{\textbf{outer layer}} =Z_{SU(2)_{\pi}} \cdot Z_{{SU(2)_{\pi}}'}\,.
\end{eqnarray}
In fact, by applying the $T_2$-tuning, one can check that the partition $Sp(N)+1\mAS +N_f \mF$ also factorizes to $N$ copies of $Sp(N)+1\mAS +N_f \mF$ in the massless limit of antisymmetric hypermultiplet $\mAS$.  
\begin{figure}[]
	\centering
	\includegraphics[width=5in]{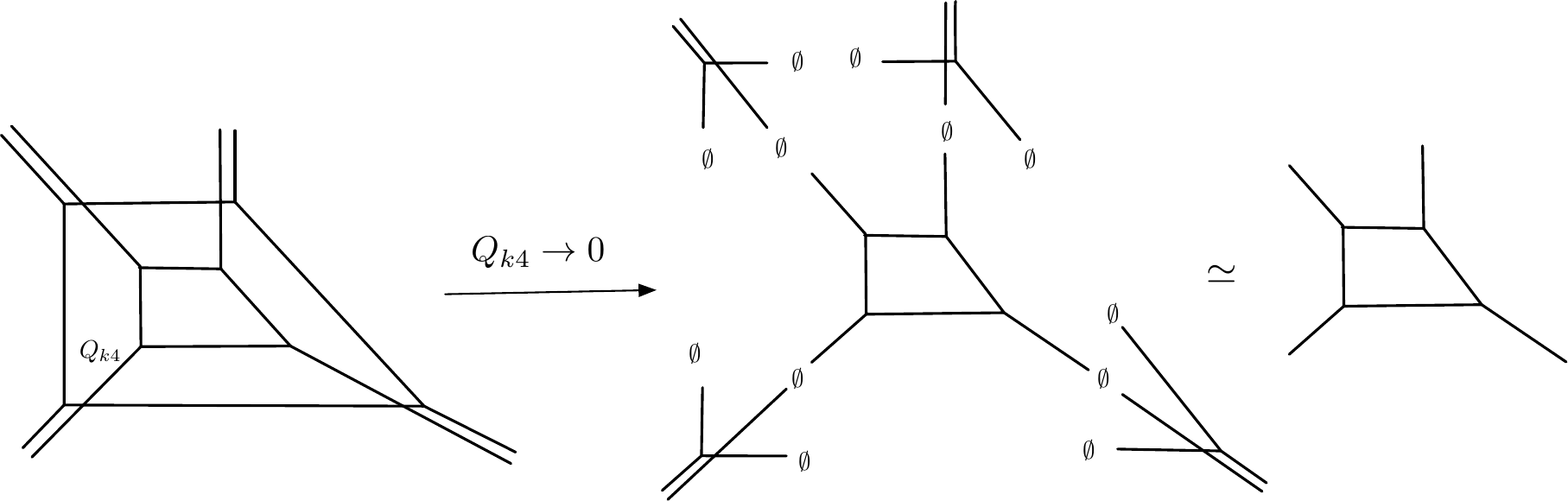}
	\caption{ 
		Sending $Q_{k4}\rightarrow 0$ implies $A_1 \rightarrow 0$. According to $T_2$-tuning, each jump on the four corners of the middle diagram provides a	 trivial contribution namely 1, or extra factors if the $\mAS$ is massive, so $Sp(2)_\pi+ 1 \mAS $ reduces to $SU(2)_\pi$. } 
	\label{fig:su2reduce}
\end{figure}

Regardless of the anti-symmetric matter being massless or massive, brane webs of $Sp(2)_{0,\pi}+1\mAS$ are reduced to the webs of $SU(2)_{0,\pi}$ as the Coulomb branch parameter $A_1 \rightarrow 0$ or equivalently $Q_{k4} \rightarrow 0$. 
This is because, in the topological vertex, each internal edge is associated with $(-Q)^{|\v|}$ and so if $Q\rightarrow 0$, only trivial Young diagram $\v=\0$ contributes.\footnote{This means:  $(\text{length of line} \rightarrow \inf) \simeq (Q \rightarrow 0) \simeq (\v \rightarrow \0) \simeq (\text{cut internal line} )$.} This reduction is illustrated in Figure \ref{fig:su2reduce}. By
 taking the Coulomb branch parameter $A_1\rightarrow 0$ or $Q_{k4} \rightarrow 0$, one gets the constraints $ \v_3=\0,\v_4=\0, \a_1=\0, \a_4=\0, \a_5=\0$ \footnote{ $\v_3=\0$ and $\v_4=\0$ force $\a_1=\0, \a_4=\0, \a_5=
	\0$ through (\ref{constraint1}). } for the unHiggsed diagram in Figure \ref{fig:sp2pi}. Hence, $Z_{{SU(2)\pi}'}\rightarrow 1$ and
$Z \rightarrow Z_{SU(2)_\pi}$. 
Through this reduction, we reproduce the partition function for $SU(2)_\pi$
\begin{eqnarray}\label{partfunpi}
	Z_{SU(2)_{ \pi}}^{\rm Nek.}&=& Z_{SU(2)_{ \pi}}^{\textbf{pert}}  \cdot Z_{SU(2)_{ \pi}}^{\textbf{instanton}} \,, \nn\\
	Z_{SU(2)_{ \pi}}^{\textbf{instanton}} &=& \sum_{\v_1,\v_2} (-1)^{|\v_1|+|\v_2| }  	Q_b^{|\v_1|+|\v_2|	} q^{	\frac{-||\v_1^T||^2 +||\v_2||	}{2	}	} t^{\frac{3||\v_1||^2+||\v_2^T||^2}{2}}	
	||\tZ_{\v_1} (t,q)||^2~||\tZ_{\v_2} (t,q)||^2 \nn \\
	&&\times  \frac{1}{ N_{\v_2,\v_1^T}(Q_F; t^{-1},q^{-1}) ~N_{\v_2,\v_1^T} (Q_F \frac{t}{q}; t^{-1} , q^{-1})  }   ,\nn\\
	Z_{SU(2)_{ \pi}}^{\textbf{pert}} &=& \dfrac{1}{M(Q_F;t,q)M(Q_F;q,t)}\,,
\end{eqnarray}
which is the same as the result in \cite{Iqbal:2007ii}. This is a trivial consistency check for the partition function of the $Sp(2)$ theory to satisfy. If an antisymmetric is massive, one of course needs to remove the extra factor arising from the mass of an antisymmetric. 

\subsection{\texorpdfstring{$Sp(2)+ 1\mathbf{AS}+ 1 \mathbf{F}$}{Sp(2)+1AS+1F}}
\begin{figure}[]
	\centering
	\includegraphics[width=5.5in]{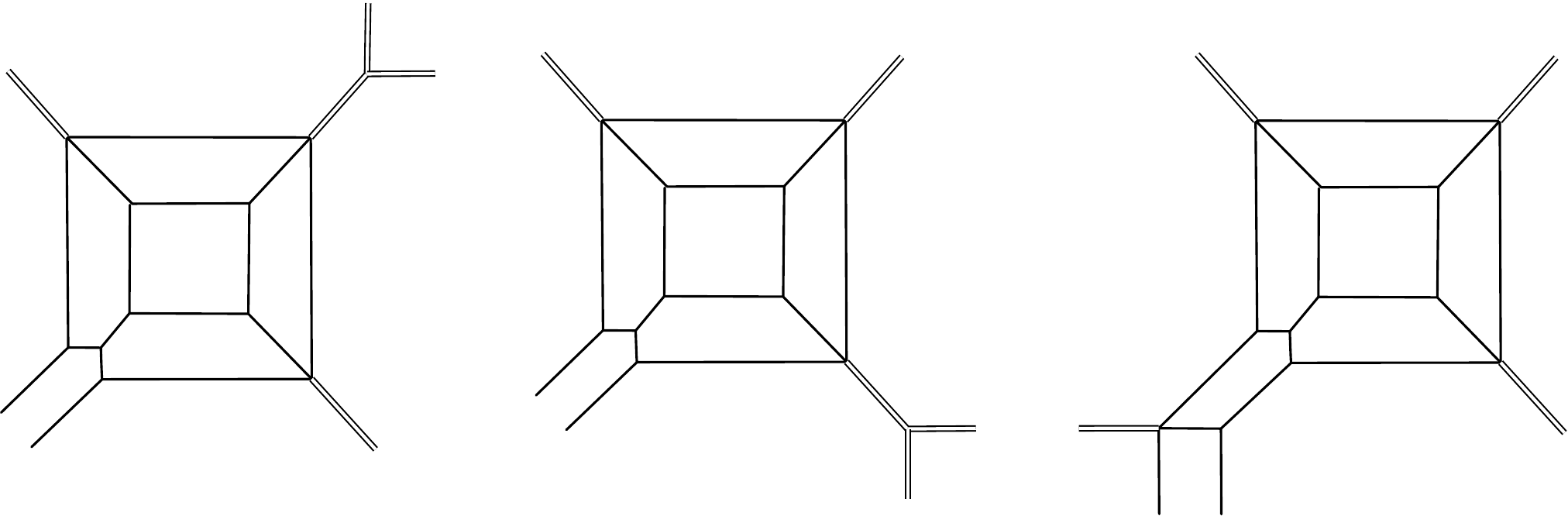}
	\caption{ Some equivalent webs for $Sp(2)+ 1\mathbf{AS}+ 1\mF$ } 
	\label{fig:1Fcases}
\end{figure}

Three equivalent webs related through Hanany-Witten moves were depicted in Figure \ref{fig:1Fcases}. 
We choose the first web in Figure \ref{fig:1Fcases} for computation as it shows fiber-base duality.
\begin{figure}[]
	\centering
	\begin{tikzpicture}[scale=0.8,line width=.6pt]
	\draw(0.5,1.1)--(0.5,0.9);	\draw(0.4,1.1)--(0.4,0.9);
	\draw(0.4,2.9)--(0.4,3.1);\draw(0.5,2.9)--(0.5,3.1);
	\draw(0.5,-1.1)--(0.5,-0.9);	\draw(0.4,-1.1)--(0.4,-0.9);
	\draw(0.4,-2.9)--(0.4,-3.1);\draw(0.5,-2.9)--(0.5,-3.1);
	\draw(-0.1,-0.1)--(0.1,0.1);	\draw(-0.1,0.1)--(0.1,-0.1);
	\draw (-1.5,-1)--(-1.5,1)--node[above=0pt]{{\tiny $Q_{B}$}}node[below=0pt]{{\tiny $\v_2$}}(1.5,1)--node[left=0pt]{{\tiny $Q_{F}$}}(1.5,-1)--node[below=0pt]{{\tiny $\v_1$}}(-1.5,-1);
	\draw (-1.5,-1)--(-2.5,-2)--(-3.5,-2)--(-3.5,2)--(-2.5,2)--(-1.5,1);
	\draw (1.5,1)--(2.5,2)--(3.5,2)--(3.5,-2)--(2.5,-2)--(1.5,-1);
	\draw (2.5,2)--(2.5,3)--(-2.5,3)--(-2.5,2);
	\draw (2.5,-2)--(2.5,-3)--(-2.5,-3)--(-2.5,-2);

	\draw (2.5,3)--node[left=5pt, above=0pt]{{\tiny $Q_{m_{11} }$}} (3.5,4)--node[above=0pt]{{\tiny $Q_{f_{3}}$}}node[below=0pt]{{\tiny $\v_5$}}(4.5,4)--node[right=0pt]{{\tiny $Q_{f_{2}}$}}(4.5,3)--node[right=0pt]{{\tiny $Q_{m_1 }$}}(3.5,2);
	\draw (4.5,3)--(6.5,3);\draw (3.5,4)--(3.5,6);\draw (4.5,4)--node[right=2pt]{{\tiny $Q_{f_1 }$}} (5.5,5)--(5.5,6);\draw (5.5,5)--(6.5, 5);
	
	\draw (3.5,-2)--(5.5,-4);\draw (2.5,-3)--(4.5,-5);
	
	\draw (-1.5,-1)--node[left=4pt,above=0pt]{{\tiny $Q_{k4}$}}(-2.5,-2)--node[below=-2pt]{{\tiny $Q_8$}}node[above=-2pt]{{\tiny $\a_8$}}(-3.5,-2)--node[left=-2pt]{{\tiny $Q_{33}$}}(-3.5,2)--node[above=-2pt]{{\tiny $Q_{1}$}}node[below=0pt]{{\tiny $\a_1$}}(-2.5,2)--node[left=4pt,below=0pt]{{\tiny $Q_{k1}$}}(-1.5,1);
	\draw (1.5,1)--node[right=4pt,below=0pt]{{\tiny $Q_{k2}$}}(2.5,2)--node[above=-2pt]{{\tiny $Q_4$}}node[below=-2pt]{{\tiny $\a_4$}}(3.5,2)--node[right=-2pt]{{\tiny $Q_{44}$}}(3.5,-2)--node[above=-2pt]{{\tiny $\a_5$}}node[below=-2pt]{{\tiny $Q_5$}}(2.5,-2)--node[right=4pt,above=0pt]{{\tiny $Q_{k3}$}}(1.5,-1);
	\draw (2.5,2)--node[right=-4pt]{{\tiny $Q_{3}$}}(2.5,3)--node[above=0pt]{{\tiny $Q_{22}$}}node[below=0pt]{{\tiny $\v_4$}}(-2.5,3)--node[left=-4pt]{{\tiny $Q_{2}$}}(-2.5,2);
	\draw (2.5,-2)--node[right=-4pt]{{\tiny $Q_{6}$}}(2.5,-3)--node[below=0pt]{{\tiny $Q_{11}$}}node[above=0pt]{{\tiny $\v_3$}}(-2.5,-3)--node[left=-4pt]{{\tiny $Q_{7}$}}(-2.5,-2);
	
	\draw (-2.5,-3)--node[above=0pt, right=0pt]{{\tiny $q$}}(-4.5,-5); 	\draw (-3.5,-2)--node[ left=0pt,above=0pt]{{\tiny $q$}}(-5.5,-4); 
	\draw (2.5,-3)--node[above=0pt,left=0pt]{{\tiny $q$}}(4.5,-5); 	\draw (3.5,-2)--node[ left=0pt,above=0pt]{{\tiny $q$}}(5.5,-4); 
	
	\node [above, blue] at (4.5,6) {$\sqqt$}; \node [right, blue] at (6.5,4) {$\sqqt$}; \node [above left, blue] at (-5,4.5) {$\sqqt$};\node [ below right, blue] at (4.7,-4.2) {$\sqtq$};

	\draw(-2.5,3)--(-4.5,5);\draw(-3.5,2)--(-5.5,4);\draw[dashed,red](-2.5,2)--(-5,4.5); \draw[dashed,red](-3.5,2)--(-3.5,3)--(-2.5,3);
	\draw[dashed,red](2.5,2)--(4.5,4);\draw[dashed,red](4.5,4)--(6.5,4); \draw[dashed,red](4.5,4)--(4.5,6);  \draw[dashed,red](3.5,2)--(3.5,3)--(2.5,3);
	\draw(-2.5,-3)--(-4.5,-5);\draw(-3.5,-2)--(-5.5,-4);
	\draw[dashed,red](2.5,-2)--(5,-4.5);
	\draw[orange](-6,0)--(6,0);\draw[orange](0,-6)--(0,6); 
		\draw[dashed,red](-2.5,-2)--(-5,-4.5);
	\draw[red,dashed](-2.5,-3)--(-3.5,-3)--(-3.5,-2);
	\end{tikzpicture}
	\caption{ An unHiggsed diagram for $Sp(2)+1\mAS+1\mF$. }
	\label{fig:reducerank2}
\end{figure}
Its unHiggsed diagram is depicted in Figure \ref{fig:reducerank2}. The relations between K\"{a}hler parameters are given by
\begin{align}
&Q_{k1}=\frac{Q_{8,k4}}{Q_1},&&Q_{k2}= \frac{ Q_{2,8,k4} }{ Q_{1,3}},&&Q_{k3}=\frac{ Q_{k4,2,4,8} }{ Q_{1,3,5} },&&Q_7=\frac{ Q_{2,3,6,8}  }{ Q_{1,3,5}},&\nn\\
&Q_{11}= \frac{ Q_{2,4,5,B,k_4,k_4} }{ Q_{1,3,5}},&&Q_{22}= \frac{ Q_{2,8,8,B,k_4,k_4}}{ Q_{1,1,3}},&&Q_{33}= \frac{ Q_{8,F,k_4,k_4} }{ Q_1 },&&Q_{44}= \frac{ Q_{2,2,4,8,8,F,k_4,k_4} }{ Q_{1,1,3,3,5}},& \nn\\
&Q_{m_{11}}= \frac{ Q_{f_2,m_{1} }}{Q_3 },& &Q_{f_3}= \frac{ Q_{3,4}}{Q_{f_2} } \,.&  &&   && 
\end{align}
Following $T_2$-tuning, we assign values to tuned K\"ahler parameters
\begin{equation}
Q_1=Q_2=Q_3=Q_4=\sqrt{\frac{q}{t}}, ~~ Q_5=Q_6=\sqrt{\frac{t}{q}}, ~~ Q_{f_1}=Q_{ f_2} =Q_{f_3}=\sqqt\, .
\end{equation}
which corresponds to two \textbf{Case B}, one  \textbf{Case C}, and one  \textbf{Case D}.
Next, we apply the conventional method on auxiliary lines to find the relations between K\"{a}hler parameters and instanton fugacity $u$, and $L_{\rm up}$ and $L_{\rm down}$ can be read off from the diagram, as discussed in Figure \ref{virtualnf1}, 
\begin{align}\label{conventionalrelation}
&L_{\rm up}=\frac{Q_B}{\sqrt{Q_F }} \cdot Q_{k2} Q_{4} Q_{m_1}= \frac{Q_B}{\sqrt{Q_F }} \cdot Q_8 Q_{k4} Q_{m_1},~~L_{\rm down}= \frac{ Q_B} {Q_F}, ~~ \nn\\
& L_{\rm up}\cdot L_{\rm down} =u^2~~\Rightarrow ~~Q_B= \frac{Q_F^{3/4} u }{\sqrt{Q_8 Q_{k4} Q_{m_1 } } }\,.
\end{align}
Hence, in order to reproduce Nekrasov partition function, independent K\"{a}hler parameters should be assigned with
\begin{equation}
Q_F={A_2}^2,~~Q_{k4}=\frac{A_1}{A_2~Q_8},~~Q_B=\frac{u~ {A_2}^2}{\sqrt{y_1}},~~Q_{m_1}=\frac{y_1}{A_1} \,,
\end{equation}
where $y_1=e^{-i m_1}$ and $m_1$ is the mass of the flavor.

In this case, 
\textbf{Case D} comes from adding flavor $\mF$ and leads to a geometric transition-related term \eqref{constraint1} in the partition function, which gives the constraint $\v_5=\0$. Thus the summation part $Z^\text{sum}$ can be reduced
\begin{align}
Z^{\text{sum}}_{Sp(2)+1 \mAS+1\mF } &=\sum\limits_{\substack{\v_1,\v_2,\v_3,\v_4,\v_5  }\, } \,\,\sum\limits_{\substack{\a_1,\a_2,\a_3,\a_4 }\, } Z[\v_1,\v_2,\v_3,\v_4,\v_5;\a_1,\a_4,\a_5,\a_8] \nn\\
&=\sum\limits_{\substack{\v_1,\v_2,\v_3,\v_4  }\, } \,\,\sum\limits_{\substack{\a_1,\a_2,\a_3,\a_4 }\, }
Z[\v_1,\v_2,\v_3,\v_4,\0;\a_1,\a_4,\a_5,\a_8].      
\end{align}
which differs from the summation part of $Sp(2)+1 \mAS$ by a term $\bf{terms}[y_1]$,
\begin{align}\label{partfunnf1}
&
Z^{\text{sum}}_{Sp(2)+1 \mAS+1\mF }=\sum_{\substack{\v_{1},\v_{2},\v_{3},\v_{4}}\, }\sum_{\substack{\a_{1},\a_{4},\a_{5},\a_{8}}\, }
Z^{\bf sum}_\textbf{Sp(2)+1$\mAS$} \cdot {\textbf{terms}[y_1]} \,, \\
&{\textbf{terms}[y_1]}:= 
y_1^{-\frac{|\v_{1}|+|v_2|+|\v_3|+|\v_4|    }{2}   }
N^{\rm half,-}_{ \v_2}\l( \frac{y_1}{A_2}  \r)
~	N^{\rm half,-}_{ \v_4}\l( \frac{y_1}{A_1}  \r)
~	N^{\rm half,-}_{ \v_1^T}\l( A_2 y_1  \r)
~	N^{\rm half,-}_{ \v_3^T}\l( A_1 y_1  \r)
\,,  \nn
\end{align}
where $Z^{\bf sum}_\textbf{Sp(2)+1$\mAS$}$ is defined in \eqref{sumpartsp2AS}.

\paragraph{Perturbative contribution.}
We find that the fundamental hypermultiplet does not contribute to $Z^{\pertII}$ 
 but adds to $Z^{\pertI}$ the term $Z_{ \mF}^{\textbf{pert}}$
 \begin{align}
 &Z^{\pertII}_{Sp(2)+1\mAS+n\mF}=Z^{\pertII}_{Sp(2)+1\mAS}  \,,\\
 &Z^{\textbf{pert-I}}_{Sp(2)+1\mAS+n\mF}  =Z^{\textbf{pert-I}}_{Sp(2)+1\mAS} \cdot \prod_{i=1}^n   Z_{ \mF}^{\textbf{pert}}(Q_{m_i})\,, \\
 	&Z_{ \mF}^{\textbf{pert}}(Q_{m_i})=M\left( \frac{A_1}{y_1} \sqtq, t,q\right) M\left( A_1 y_1 \sqtq, t,q\right) M\left( \frac{A_2}{y_1} \sqtq, t,q\right) M\left( A_2 y_1 \sqtq, t,q\right) \label{hyperpert}\,,
 	\end{align}
 where $Z_{ \mF}^{\textbf{pert}}(Q_{m_i})$ is the contribution of the flavor $\mF$ to the perturbative part. In total, the fully perturbative part for $Sp(2)+1\mAS+1\mF$ is given by
\begin{eqnarray}\label{pertnf1}
Z^{\textbf{pert}}_{Sp(2)+1\mAS+1\mF}=\frac{Z^{\textbf{pert-I}}_{Sp(2)+1\mAS+1\mF}   \cdot Z^{\textbf{pert-II}}_{Sp(2)_0 +1\mAS+1\mF }    } { Z_{Q_8}^{\rm extra} } = Z^{\textbf{pert}}_{Sp(2)+1\mAS}\cdot Z_{ \mF}^{\textbf{pert}}(Q_{m_1}) \,,
\end{eqnarray}
which exactly agrees with localization computation \eqref{pertlocalization}. With proper normalization, the perturbative part can be expanded as 
	\begin{align}
	&\frac{Z^{\textbf{pert}}_{Sp(2)+1\mAS+1\mF}}{M(Q_8 \sqtq,t,q)^2} = 1- \frac{ \sqrt{q t} \chi_2^{SU(2)}[y_1] }{(1-q)(1-t) }A_2 -\frac{ (q^{\frac{3}{2}} t^{\frac{3}{2}}-q t \chi_2^{SU(2)} + \sqrt{
			q t}) \chi_2^{SU(2)}[y_1] }{(1 - q)^2 (1 - t)^2} A_1
	\nn\\
	& +\frac{(q+t)-\sqrt{q t } \chi_2^{SU(2)}  }{(1-q)(1-t) } \frac{A_1}{A_2} + \frac{ q t (q + t) \chi_3^{SU(2)}[y_1]  +\l(  q t(1+qt) -(q+t)(q^2+t^2-q^2t^2-1)  \r) }{ (1 - q)^2 (1 + q) (1 - t)^2 (1 + t) }{A_2}^2 \nn\\
&
		+\mathcal{ O}({A_1}^1; {A_2}^1),
\nn	\end{align}
where $\chi_2^{SU(2)}[y_1]=y_1+y_1^{-1},~\chi_3^{SU(2)}[y_1]=y_1^2+1+y_1^{-2},~\chi_2^{SU(2)}=Q_8+Q_8^{-1} $. 

\paragraph{Instanton contribution.}
The instanton contribution is obtained by \eqref{instantonform}. As $Z^{\textbf{nonpert-I}} =1$ in this case, the one-instanton contribution is given by
\begin{eqnarray}\label{nf=1oneinst}
Z^{\textbf{one-instanton  }}_{Sp(2)+1\mAS+1\mF}=\frac{
	\sum\limits_{\a_{1},\a_4,\a_5,\a_8}~
	\sum\limits_{ \substack{  \v_{1},\v_2,\v_3,\v_{4}\in
			\{{\begin{tikzpicture}[scale=.15]
				\draw (0,1)--(0,0);
				\draw (0,0) --(1,0);
				\draw (1,1) --(1,0);
				\draw (1,1) --(0,1);
				\end{tikzpicture}}
			,~\0\}\\
			|\v_{1}|+|\v_2|+|\v_3|+|\v_4|=1}}
	Z[\v_{1},\v_2,\v_3,\v_4,\0
	;\a_{1},\a_4,\a_5,\a_{8} ]}{Z^{\pertII} _{Sp(2)+1\mAS+1\mF} }
	 \,.
\end{eqnarray}
By expanding it with respect to Coulomb branch parameters, we obtain
	\begin{align}\label{oneinst1F}
	&
Z^{\textbf{one-instanton}}_{Sp(2)+1\mAS+1\mF}= \frac{-\sqrt{q t } \sqrt{y_1} }{(1-q)(1-t)} (A_2+A_1)  + \frac{ q+t}{(1-q)(1-t) \sqrt{y_1} } (A_2^2 +A_1^2)      \nn \\
	& +\frac{q+t-\sqrt{q t}\chi_2^{SU(2)}   } {(1-q)(1-t) \sqrt{y_1} }A_2 A_1-\frac{ (q+t) (q+t-\sqrt{q t}\chi_2^{SU(2)}  ) \sqrt{y_1}  }{(1 - q) (1 - t) \sqrt{q t}  }  (A_2^2A_1+A_1 A_2^2)   \nn\\
	& +\frac{(q + t)^2 (q +t -\sqrt{q t} \chi_2^{SU(2)}    )  }{(1 - q) q (1 - t) t \sqrt{y_1} }  A_2^2A_1^2 +\mathcal{ O}(A_1^3; A_2^3)\,,
	\end{align}
which equals to the localization result \eqref{oneinstloc} with $N_f=1$ and $y_1:=e^{m_1},~Q_8:=e^m$.

\vskip .5cm
Similarly, the two-instanton contribution is given by
\begin{eqnarray}\label{nf=1twoinst}
Z^{\textbf{two-instanton  }}_{Sp(2)+1\mAS+1\mF}=
\frac{
	\sum\limits_{\a_{1},\a_4,\a_5,\a_8}~
	\sum\limits_{ \substack{\v_1,\v_2,\v_3,\v_4 \in\{
			\raisebox{-.1cm}	{\begin{tikzpicture}[scale=.15]
				\draw (0,2)--(0,0);
				\draw (0,0) --(1,0);
				\draw (1,2) --(1,0);	\draw (1,2) --(0,2);
				\draw (1,1) --(0,1);
				\end{tikzpicture}},~
			{\begin{tikzpicture}[scale=.15]
				\draw (0,1)--(0,0);\draw (1,1) --(1,0);
				\draw (0,0) --(2,0);	\draw (2,1) --(2,0);
			
				\draw (2,1) --(0,1);
				\end{tikzpicture}}, ~
			{\begin{tikzpicture}[scale=.15]
				\draw (0,1)--(0,0);\draw (1,1) --(1,0);
				\draw (0,0) --(1,0);
				
				\draw (1,1) --(0,1);
				\end{tikzpicture}}, ~\0
			\}\\
				|\v_{1}|+|\v_2|+|\v_3|+|\v_4|=2} }  
	Z[\v_{1},\v_2,\v_3,\v_4, \0
	;\a_{1},\a_4,\a_5,\a_8 ]}{Z^{\pertII}}
	 \,,
\end{eqnarray}
which can be expanded in terms of $A_1 , A_2	$
	\begin{align}\label{twoinstnf1}
	&Z^{\textbf{two-instanton  }}_{Sp(2)+1\mAS+1\mF}=  \frac{q t (q+t) y_1 }{ (1 - q)^2 (1 + q) (1 - t)^2 (1 + t)}(A_2^2+A_1^2) + \frac{ q t y_1 } {(1-q)^2(1-t)^2  } A_1A_2	     \\
	&-\frac{ 2 \sqrt{q t} (q+t) -q t\chi_2^{SU(2)}     }{(1 - q)^2 (1 - t)^2}(A_2^2A_1+A_2A_1^2)
	+\mathcal{ O}(A_1^2; A_2^2)\,.\nn
	\end{align}


\paragraph{Enhancement of global symmetry.}
 Nekrasov partition functions of $Sp(2)_0+1\mAS+N_f \mF$ enjoy global symmetry enhancement with properly shifting of parameters $(A_i, u, y_i)$.   
With the help of fiber-base duality, this shift can be found. The exchanging symmetry between $A_1, A_2$ preserves, so following the argument in paper \cite{Mitev:2014jza}, we shift 
\begin{eqnarray}\label{Ashiftnf1}
\tA_{1}=A_{1} u^{\frac{2}{7}}, \quad \tA_{2}=A_{2} u^{\frac{2}{7}}.
\end{eqnarray}
According to the webs in Figure \ref{fig:reducerank2}, the global symmetry of $Sp(2)+1\mAS+1\mF$ is supposed to be $ G=SU(2)_{Q_8} \times E_1= SU(2)_{Q_8}  \times SU(2) \times U(1)$. We define two new fugacities $u_1, u_2$ for $SU(2)_{u_1}\times U(1)_{u_2}$ respectively. The fiber-base duality
\begin{eqnarray}
Q_B  \leftrightarrow Q_F,~~ Q_8  \leftrightarrow Q_8,~~Q_{m_1}  \leftrightarrow Q_{m_1}\,,
\end{eqnarray}
through shifts, becomes
\begin{eqnarray}
u_1\leftrightarrow u_1^{-1}, ~~u_2 \leftrightarrow u_2,~~\tA_{1}\leftrightarrow\tA_{1},~~\tA_{2}\leftrightarrow\tA_{2}    \,,  
\end{eqnarray}
which along with \eqref{Ashiftnf1} determine the relations between $u$ and fugacities $u_{1,2}$ as follows
\begin{eqnarray}
\tA_{1}= u_1^{\frac{1}{2}} u_2^{ -\frac{1} {14 }}A_{1},~~\tA_{2}= u_1^{\frac{1}{2}} u_2^{ -\frac{1} {14 }}A_{2},~~u=u_1^{ \frac{7}{4}} u_2^{-\frac{1}{4}} ,~~y_1=\frac{1}{\sqrt{u_1 u_2}} . 
\end{eqnarray}
These relations are similar to relations for $SU(2)+1\mF$ in \cite{Mitev:2014jza}.
Once again, we note that $Sp(2)+1\mAS+N_f \mF$ is similar to $SU(2)+N_f\mF$.
We indeed find that shifted partition functions can be expressed in terms of the characters of enhanced $SU(2)_{Q_8} \times E_1$ global symmetry as expected:
	\begin{align}
	&Z_{Sp(2)+1\mAS+1\mF}=1- \frac{\sqrt{q t} \l( u_2^{\frac{4}{7} }+u_2^{-\frac{3}{7}}\chi_2^{SU(2)}[u_1] \r)  }{(1-q)(1-t) } \tA_2  +   \frac{q+t-\sqrt{q t}\chi_2^{SU(2)}   }{(1-q)(1-t)  } \frac{\tA_1}{\tA_2}   \\
	& -\frac{\sqrt{q t}   \l(u_2^{\frac{4}{7} }+u_2^{-\frac{3}{7}}\chi_2^{SU(2)}[u_1] \r) \l( -\sqrt{q t} \chi_2^{SU(2)} +1+ q t    \r)    }{ (1-q)^2(1-t)^2 }      \tA_1    \nn\\ 
	&+ \frac{ (q+t)  \big( q t ~\chi_3^{SU(2)} -\sqrt{q t}( 1+q)(1+t) \chi_2^{SU(2)} + (q+t)(1+qt)+q t  \big)     } { (1 - q)^2 (1 + q) (1 - t)^2 (1 + t)  } \frac{\tA_1^2}{\tA_2^2}  \nn\\
	 &+\mathcal{ O}(\tA_1^2;\tA_2^2), \nn
	\end{align}
where $\chi_2^{\rm SU(2)}[u_1]=u_1+u_1^{-1}$.

\begin{figure}[]
	\centering
	\includegraphics[width=4in]{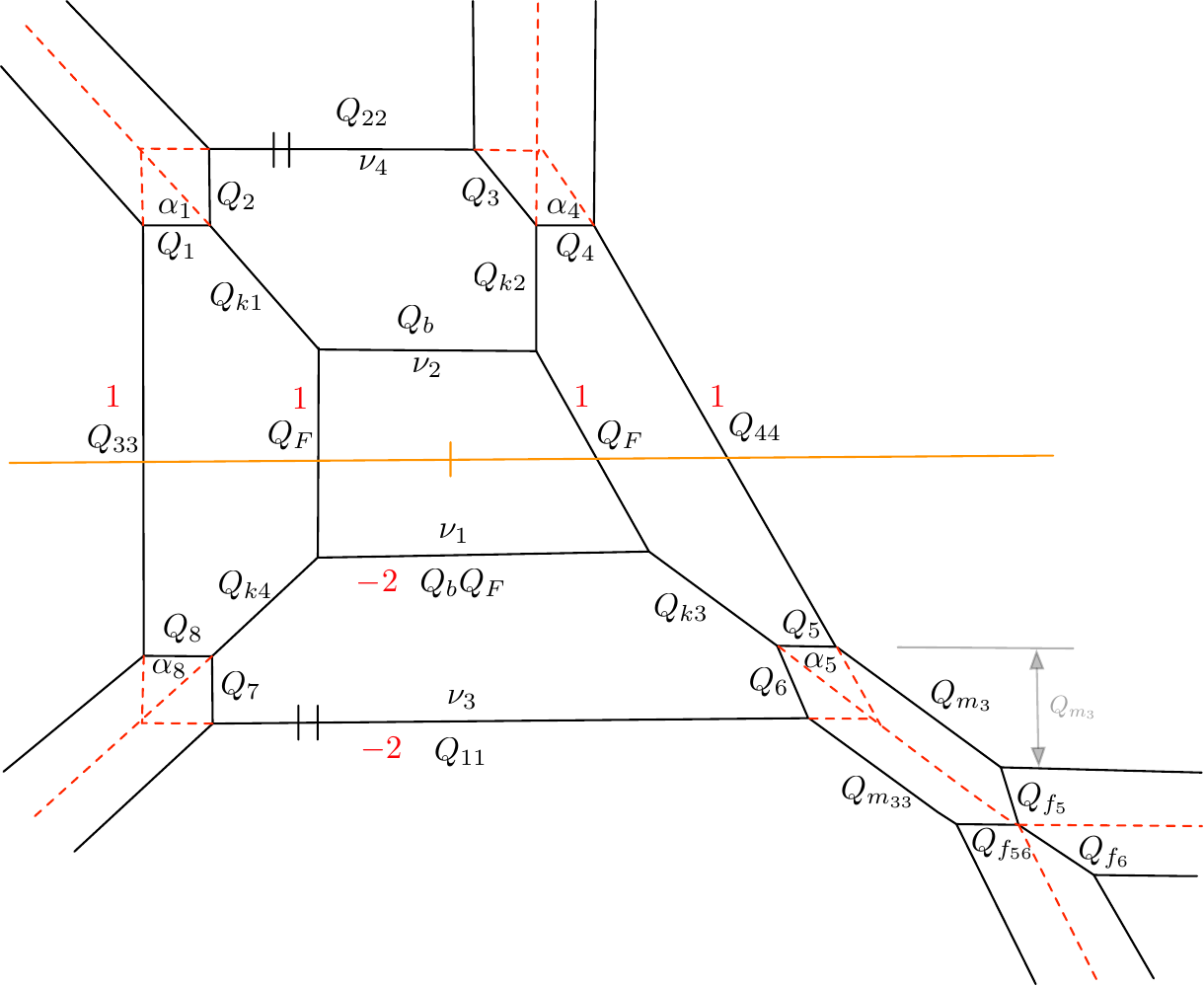}
	\caption{ The unHiggsed diagram for $Sp(2)_\pi+1\mAS+1 \mF$, with framing  numbers in red assigned. } 
	\label{fig:sp2pi1F}
\end{figure}

\paragraph{$Sp(2)_\pi+1\mathbf{AS} + 1\mathbf{F}$}
Just like the fact that $SU(2)_0+1\mathbf{F}$ and $SU(2)_{\pi}+ 1\mathbf{F}$ are equivalent up to flops, there is equivalence
\begin{eqnarray}
Sp(2)_0+1\mathbf{AS} + 1\mathbf{F}  \backsimeq Sp(2)_\pi+1\mathbf{AS} + 1\mF \,,
\end{eqnarray}
related through flops.
By taking the mass of $1 \mF$ to infinity to decouple flavor, one can obtain $Sp(2)_0+1\mathbf{AS} $ and $Sp(2)_\pi+1\mathbf{AS} $ respectively. 
\begin{figure}[]
	\centering
	\includegraphics[width=6in]{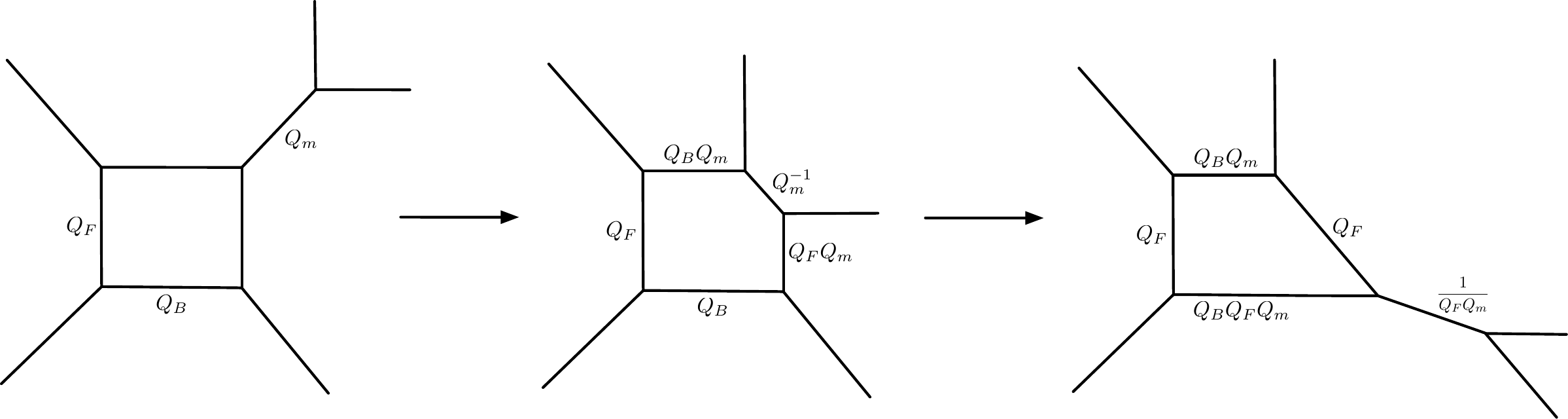}
	\caption{$SU(2)_0+1\mathbf{F}$ and $SU(2)_{\pi}+ 1 \mathbf{F}$ are equivalent via two flops. } 
	\label{fig:sp20=pi}
\end{figure}
Similar to the flops illustrated in Figure \ref{fig:sp20=pi}, K\"{a}hler parameters of $Sp(2)_0+1 \mAS+1\mathbf{F}$  and of $Sp(2)_\pi+ 1\mAS+1\mathbf{F}$ are related through transformation\footnote{The parameters on the left-hand side of arrows are for $Sp(2)_0+1\mathbf{AS} $ and the right side for $Sp(2)_\pi+11\mathbf{AS} $.} 
\begin{align}\label{relationpi1F}
	Q_F \rightarrow Q_F ,\quad Q_{k4} \rightarrow Q_{k4} ,\quad Q_B \rightarrow Q_8 Q_b Q_F Q_{k4} Q_{m_3},\quad Q_{m_1}\! \rightarrow\! ({Q_8}^2 Q_F Q_{k4} Q_{m_3}   )^{-1}\!. 
\end{align}
Following $T_2$-tuning in section 3, we find the correct tuned K\"ahler parameters for the unHiggsed diagram of $Sp(2)_\pi+1\mAS+1\mF$ depicted in Figure \ref{fig:sp2pi1F} are given as follows:
\begin{equation}
Q_1=Q_2=Q_3=Q_4=\sqrt{\frac{q}{t}}, ~~ Q_5=Q_6=Q_{f_5}=Q_{f_6}=Q_{f_{56}}=\sqrt{\frac{t}{q}}\,,	
\end{equation}
and independent K\"{a}hler parameters are assigned with gauge theory parameters
\begin{equation}
	Q_F={A_2}^2,~~Q_{k4}=\frac{A_1}{A_2~Q_8},~~Q_b=\frac{u~ A_2}{\sqrt{y_3}},~~Q_{m_3}=\frac{y_3}{A_1} \,,
\end{equation}
where $y_3=e^{-i m_3}$.

We find the perturbative part is equal to \eqref{pertnf1}, and
the one-instanton contribution obtained from 
\eqref{nf=1oneinst} can be expressed as
	\begin{align}
	&Z^\oneinst_{Sp(2)_\pi+1\mathbf{AS}+1 \mathbf{F}}=\nn\\  &-\frac{ \sqrt{q t}  }{ (1-q)(1-t) \sqrt{y_3}   } (A_1+A_2) + \frac{ (q+t) \sqrt{y_3} }{ (1-q)(1-t)   } ({A_1}^2+{A_2}^2) \nn\\
		&+ \frac{\sqrt{y_3} \l(q+t-\sqrt{q t} \chi_2^{SU(2)}  \r) }{ (1-q)(1-t)  } A_1 A_2
		- \frac{ (q+t) \l(q+t-\sqrt{q t} \chi_2^{SU(2)}  \r) }{ (1-q)(1-t) \sqrt{q t} \sqrt{y_3}} (A_1 {A_2}^2 +A_2 {A_1}^2)
		\nn\\
	&+ \frac{\sqrt{y_3} (q+t)^2 \l(q+t-\sqrt{q t} \chi_2^{SU(2)}  \r) }{ (1-q)(1-t) q t  } {A_1}^2 {A_2}^2
		+\mathcal{O} (A_1,A_2)\,, 
	\end{align}
which equals  (\ref{oneinst1F}) as expected. The two-instanton contribution was checked to agree with \eqref{twoinstnf1}. The topological string partition functions between $Sp(2)_{0,\pi}+1\mAS+1\mF$ are equal through transformation (\ref{relationpi1F}). By taking the mass $m_3$ for flavor to infinity to decouple flavor, namely $Q_{m_3} \rightarrow 0$, one can reproduce the topological string partition function for $Sp(2)_\pi+1\mathbf{AS} $.

\subsection{\texorpdfstring{$Sp(2)+ 1\mathbf{AS}+ 2 \mathbf{F} $}{Sp(2)+1AS+2F}}
\begin{figure}[]
	\centering
	\includegraphics[width=6in]{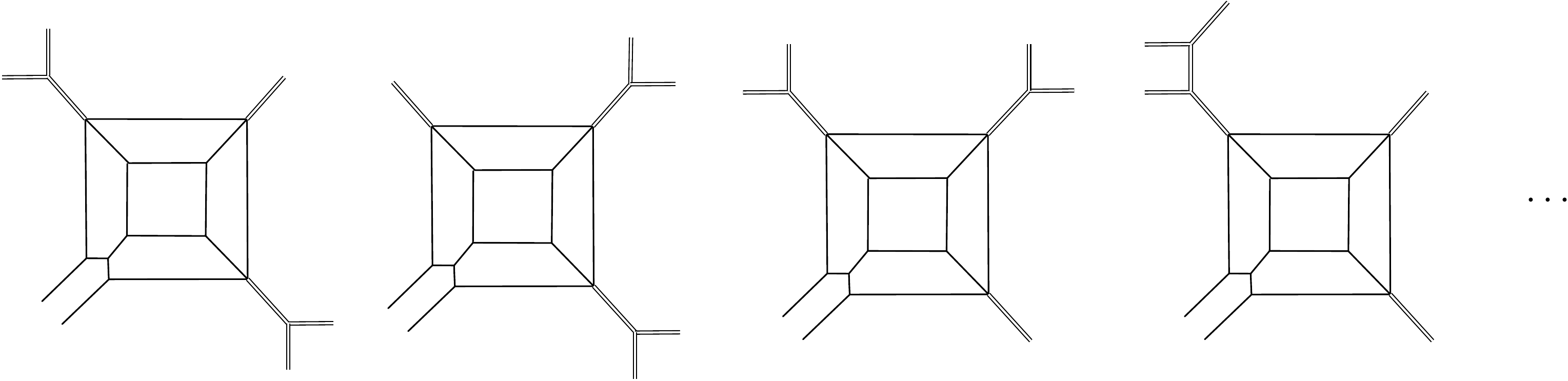}
	\caption{ Some equivalent webs related through Hanany-Witten moves for $Sp(2)+ 1\mathbf{AS}+ 2\mF$. } 
	\label{fig:2Fcases}
\end{figure}

For $Sp(2)+ 1\mathbf{AS}+ 2 \mathbf{F}  $, there are many equivalent web configurations related through Hanany-Witten moves, and some of them are depicted in Figure \ref{fig:2Fcases}. 
The Nekrasov partition functions for web configurations in Figure \ref{fig:2Fcases}  become equivalent, 
once extra factors were removed, 
\begin{align}\label{fourwebnf2}
Z^{\text{Nek}}_{Sp(2)+1\mAS+2 \mF}=&
\frac{Z^{\text{top}}_{Sp(2)+1\mAS+2 \mF^{\diagdown\!\diagdown}} }{ Z_{Q_8}^{\rm extra}}   
=\frac{Z^{\text{top}}_{Sp(2)+1\mAS+2 \mF^=}}{Z^{\rm extra}_{=} \cdot { Z_{Q_8}^{\rm extra}}}
=\frac{Z^{\text{top}}_{ Sp(2)+1\mAS+2 \mF^{||}  }}{Z^{\rm extra}_{||}\cdot { Z_{Q_8}^{\rm extra}} } =\frac{Z^{\text{top}}_{ Sp(2)+1\mAS+2 \mF^{=//}  }}{Z^{\rm extra}_{=//}\cdot { Z_{Q_8}^{\rm extra}} }
\nn\\
=&\underbrace{Z^{\textbf{pert}}_{Sp(2)+1\mAS } \prod_{i=1}^{N_f=2} Z_{\mF}^{\pert} (Q_{m_i}) }_{Z^{\pert}_{Sp(2)+1\mAS+2 \mF}   } \cdot~\underbrace{\Big(  1+\sum_{k=1}^{\inf } u^k Z_k(A_i,y_i)
	\Big)}_{Z^{\bf instanton}_{Sp(2)+1\mAS+2 \mF} }\,.
\end{align}
We checked this equivalence up to two-instanton contributions, which is consistent with the fact that 
Nekrasov partition functions are insensitive to  
Hanany-Witten moves.

As a representative example, consider 
the first web in Figure \ref{fig:2Fcases}. 
This 
contains no extra factor other than $Z^{\rm extra}_{Q_8}$ and also of reflection symmetry. 
For its unHiggsed diagram depicted in Figure $ \ref{fig:nf2diagonal}$, the associated tuned K\"ahler parameters are found to be the following
\begin{figure}[]
	\centering
	\begin{tikzpicture}[scale=0.8,line width=.6pt]
	\draw(0.5,1.1)--(0.5,0.9);	\draw(0.4,1.1)--(0.4,0.9);
	\draw(0.4,2.9)--(0.4,3.1);\draw(0.5,2.9)--(0.5,3.1);
	\draw(0.5,-1.1)--(0.5,-0.9);	\draw(0.4,-1.1)--(0.4,-0.9);
	\draw(0.4,-2.9)--(0.4,-3.1);\draw(0.5,-2.9)--(0.5,-3.1);
	\draw(-0.1,-0.1)--(0.1,0.1);	\draw(-0.1,0.1)--(0.1,-0.1);
	\draw (-1.5,-1)--(-1.5,1)--node[above=0pt]{{\tiny $Q_{B}$}}node[below=0pt]{{\tiny $\v_2$}}(1.5,1)--node[left=0pt]{{\tiny $Q_{F}$}}(1.5,-1)--node[below=0pt]{{\tiny $\v_1$}}(-1.5,-1);

	\draw (-1.5,-1)--node[left=4pt,above=0pt]{{\tiny $Q_{k4}$}}(-2.5,-2)--node[below=-2pt]{{\tiny $Q_8$}}node[above=-2pt]{{\tiny $\a_8$}}(-3.5,-2)--node[left=-2pt]{{\tiny $Q_{33}$}}(-3.5,2)--node[above=-2pt]{{\tiny $Q_{1}$}}node[below=0pt]{{\tiny $\a_1$}}(-2.5,2)--node[left=4pt,below=0pt]{{\tiny $Q_{k1}$}}(-1.5,1);
	\draw (1.5,1)--node[right=4pt,below=0pt]{{\tiny $Q_{k2}$}}(2.5,2)--node[above=-2pt]{{\tiny $Q_4$}}node[below=-2pt]{{\tiny $\a_4$}}(3.5,2)--node[right=-2pt]{{\tiny $Q_{44}$}}(3.5,-2)--node[above=-2pt]{{\tiny $\a_5$}}node[below=-2pt]{{\tiny $Q_5$}}(2.5,-2)--node[right=4pt,above=0pt]{{\tiny $Q_{k3}$}}(1.5,-1);
	\draw (2.5,2)--node[right=-4pt]{{\tiny $Q_{3}$}}(2.5,3)--node[above=0pt]{{\tiny $Q_{22}$}}node[below=0pt]{{\tiny $\v_4$}}(-2.5,3)--node[left=-4pt]{{\tiny $Q_{2}$}}(-2.5,2);
	\draw (2.5,-2)--node[right=-4pt]{{\tiny $Q_{6}$}}(2.5,-3)--node[below=0pt]{{\tiny $Q_{11}$}}node[above=0pt]{{\tiny $\v_3$}}(-2.5,-3)--node[left=-4pt]{{\tiny $Q_{7}$}}(-2.5,-2);
	\draw (2.5,3)--(4.5,5); 	\draw (3.5,2)--(5.5,4); 

	\draw (-4.5,3)--(-6.5,3);\draw (-3.5,4)--(-3.5,6);\draw (-5.5,5)--(-5.5,6);\draw (-5.5,5)--(-6.5, 5);\draw(-5.5,5)--node[left=-2pt]{{\tiny $Q_{f_{33}}$}}(-4.5,4);
	
	\draw (2.5,-3)--node[left=3pt, below=0pt]{{\tiny $Q_{m_{33} }$}}  (3.5,-4)--node[below=0pt]{{\tiny $Q_{f_{56}}$}}node[above=0pt]{{\tiny $\v_6$}}(4.5,-4)--node[right=0pt]{{\tiny $Q_{f_{5}}$}}(4.5,-3)--node[right=0pt]{{\tiny $Q_{m_3 }$}}(3.5,-2);
	\draw (4.5,-3)--(6.5,-3);\draw (3.5,-4)--(3.5,-6);\draw (5.5,-5)--(5.5,-6);\draw (5.5,-5)--(6.5, -5);\draw(5.5,-5)--node[right=0pt]{{\tiny $Q_{f_6 }$}} (4.5,-4);
	
	\draw(-2.5,3)--node[right=0pt]{{\tiny $Q_{m_{22} }$}} (-3.5,4);\draw(-3.5,2)--node[below=1pt]{{\tiny $Q_{{m_2}}$}}(-4.5,3)--node[left=-1pt]{{\tiny $Q_{f_{4}}$}}(-4.5,4)--node[above=0pt]{{\tiny $Q_{f_{34}}$}}node[below=0pt]{{\tiny $\v_7$}}(-3.5,4);
	
	\draw[dashed,red](-2.5,2)--(-4.5,4);\draw[dashed,red](-4.5,4)--(-6.5,4); \draw[dashed,red](-4.5,4)--(-4.5,6); \draw[dashed,red](-3.5,2)--(-3.5,3)--(-2.5,3);
	\draw[dashed,red](2.5,2)--(5,4.5);
	\draw[dashed,red](3.5,2)--(3.5,3)--(2.5,3);
	
	\draw (2.5,3)--node[above=0pt, left=0pt]{{\tiny $t$}}(3.5,4); 	\draw (3.5,2)--node[ right=0pt,below=0pt]{{\tiny $t$}}(4.5,3); 
	\draw (-2.5,-3)--node[above=0pt, right=0pt]{{\tiny $q$}}(-4.5,-5); 	\draw (-3.5,-2)--node[ left=0pt,above=0pt]{{\tiny $q$}}(-5.5,-4);

	\draw(-2.5,-3)--(-4.5,-5);\draw(-3.5,-2)--(-5.5,-4);
	
	\draw[dashed,red](2.5,-2)--(4.5,-4);\draw[dashed,red](4.5,-4)--(6.5,-4); \draw[dashed,red](4.5,-4)--(4.5,-6);  
	\draw[orange](-6,0)--(6,0);\draw[orange](0,-6)--(0,6); 
	
	\draw[red,dashed](2.5,-3)--(3.5,-3)--(3.5,-2);
	
	\draw[dashed,red](-2.5,-2)--(-5,-4.5);
	\draw[red,dashed](-2.5,-3)--(-3.5,-3)--(-3.5,-2);
	\end{tikzpicture}
	\caption{ The unHiggsed diagram for $Sp(2)+1\mAS+2\mF^{\setminus\setminus}$. }
	\label{fig:nf2diagonal}
\end{figure}
\begin{align} \label{tuningnf2}
Q_1=Q_2=Q_3=Q_4=Q_{f_{33}}=Q_{f_4}=Q_{f_{34}}=\sqrt{\frac{q}{t}}, ~~ Q_5=Q_6=Q_{f_5}=Q_{ f_6}=Q_{f_{56}} =\sqtq \,.
\end{align}
The relations between independent K\"{a}hler parameters and $(A_i, u, y_i)$ are
\begin{eqnarray}
Q_F={A_2}^2,~~Q_{k4}=\frac{A_1}{A_2~Q_8},~~Q_B=\frac{u~ {A_2}^2}{\sqrt{y_2}\sqrt{y_3}},~~Q_{m_2}=\frac{y_2}{A_1},~~Q_{m_3}=\frac{y_3}{A_1} \,,
\end{eqnarray}
Similar to $Sp(2)_0+1\mAS+1\mF$, geometric transition relevant terms \eqref{constraint1} caused by tuned parameters \eqref{tuningnf2} give rise to constraints on Young diagrams: $\v_6=\0$, $\v_7=\0$. Hence the summation part of the topological string partition function can be reduced to
\begin{align}\label{partfunnf2}
&Z^{\rm sum}_{Sp(2)+1\mAS+2 \mF   }=\sum_{\substack{\v_1,\v_2,\v_3,\v_4 ,\\ \a_1,\a_2,\a_3,\a_4}}
Z^{\bf sum}_\textbf{Sp(2)+1AS} \cdot 
\textbf{terms}[y_1] \cdot  \textbf{terms}[y_2]  \,, \\
&{\textbf{terms}[y_2]}:= 
y_2^{-\frac{|\v_1|+|\v_2|+|\v_3|+|\v_4|    }{2}   }
N^{\rm half,-}_{ \v_2}\l( \frac{y_2}{A_2}  \r)
~	N^{\rm half,-}_{ \v_4}\l( \frac{y_2}{A_1}  \r)
~	N^{\rm half,-}_{ \v_1^T}\l( A_2 y_2  \r)
~	N^{\rm half,-}_{ \v_3^T}\l( A_1 y_2  \r)
\,, \nn
\end{align}
where $\textbf{terms}[y_1] $ is defined in \eqref{partfunnf1}.

\paragraph{Perturbative and instanton contributions.}
Normalized perturbative part can be expressed as
\begin{align}
&\frac{Z^{\pert}_{Sp(2)+1\mAS+2 \mF   } }{M(Q_8 \sqtq,t,q)^2} = 
1
-\frac{\sqrt{q t}   \l(\chi_2^{SU(2)}[y_2]+\chi_2^{SU(2)}[y_3]  \r)    }{ (1-q)(1-t)  } A_2 +  \frac{(q+t)-\sqrt{q t } \chi_2^{SU(2)}  }{(1-q)(1-t) } \frac{A_1}{A_2}
\nn\\
&- \frac{\sqrt{q t}  \l(1+q t-\sqrt{q t } \chi_2^{SU(2)}\r) (\chi_2^{SU(2)}[y_2]+\chi_2^{SU(2)}[y_3]    ) }{(1-q)^2(1-t)^2 }  A_1
 + \mathcal{ O}(A_1^2; A_2^2),\nn
\end{align}
where  $\chi_2^{SU(2)}[y_i]=y_i+\frac{1}{y_i},~\chi_2^{SU(2)} = Q_8+Q_8^{-1},~\chi_3^{SU(2)} = Q_8^2+1+Q_8^{-2}$ .

\paragraph{Instanton contributions.}
One-instanton contribution can be expanded 
\begin{align}
&
Z^{\textbf{one-instanton}}_{Sp(2) +1\mAS+2\mF}= \frac{-\sqrt{q t } }{(1-q)(1-t)} \chi_2[y_{2/3}] (A_2+A_1)  + \frac{ q+t}{(1-q)(1-t)  } \chi_2[y_{23}] (A_2^2 +A_1^2)     \nn \\
&+ \frac{q+t-\sqrt{q t}\chi_3^{SU(2)}  } {(1-q)(1-t) }  \chi_2[y_{23}] A_1 A_2 -\frac{ (q+t) (q+t-\sqrt{q t}\chi_2^{SU(2)}  )  }{(1 - q) (1 - t) \sqrt{q t}  }    \chi_2[y_{2/3}] (A_2^2A_1+A_1 A_2^2)   \nn\\
&+ \frac{(q + t)^2 (q +t -\sqrt{q t} \chi_2^{SU(2)}   )  }{(1 - q) q (1 - t) t }   \chi_2[y_{23}] A_2^2A_1^2 +\mathcal{ O}(A_1^3; A_2^3), 
\end{align}
where we define $\chi_2[y_\cdot]=\sqrt{y_\cdot}+\frac{1}{\sqrt{y_\cdot}}$, $y_{23}:=y_2 y_3,~y_{2/3}:=\frac{y_2}{y_3}$. To match the one-instanton contribution with localization result when $N_f=2$,
we define $y_{2}=e^{m_{2}},y_{3}=e^{m_{3}},~Q_8=e^m,~A_{1}=e^{\a_{1}},~A_{2}=e^{\a_{2}}$ for \eqref{oneinstloc}.

Similarly, the two-instanton contribution can be obtained and expanded as
\begin{align}\label{twoinstonnf2}
&
Z^\twoinst_{Sp(2) +1\mAS+2\mF} =  \frac{q t( 1+q t+ (q+t))( 1+ \chi_2[y_{2/3}]   ) }{ (1 - q)^2 (1 + q) (1 - t)^2 (1 + t)}(A_2^2+A_1^2) + \frac{ q t (2+\chi_2[y_{2/3}]) } {(1-q)^2(1-t)^2  } A_2A_1   \nn\\ 
&  -\frac{ 2 \sqrt{q t} (q+t) -q t\chi_2^{SU(2)}     }{(1 - q)^2 (1 - t)^2} (\chi_2[y_2]+\chi_2[y_3]) (A_2^2A_1+A_2A_1^2) +
\mathcal{ O}(A_1^2,A_2^2), 
\end{align}
where $\chi_2[y_\cdot]=y_\cdot+\frac{1}{y_\cdot}$.

\paragraph{Enhancement of global symmetry.}
Because of enhancement, the Nekrasov partition function can be written in terms of characters of the global symmetry group. Following \cite{Mitev:2014jza}, the shifted Coulomb branch parameters for this theory are given by
\begin{eqnarray}
\tA_{1}=A_{1} u^{\frac{1}{3}},\qquad \tA_{2}=A_{2} u^{\frac{1}{3}}.
\end{eqnarray}
The global symmetry of $Sp(2)+1\mAS+2 \mF$ should be $ G=SU(2)_{Q_8} \times E_3= SU(2)_{Q_8}  \times SU(2)_{\tu} \times SU(3)_{\ty_1,\ty_2,\ty_3}$. We define new fugacities $\tu$ for $SU(2)_{\tu}$ and $\ty_1,\ty_2,\ty_3 $ for $ SU(3)_{\ty_1,\ty_2,\ty_3}$ \footnote{ For $ SU(3)$,  $\ty_1 \ty_2 \ty_3 =1$} respectively. The fiber-base duality for the diagram in Figure $ \ref{fig:nf2diagonal}$ is
\begin{eqnarray}
Q_B  \leftrightarrow Q_F,~~ Q_8  \leftrightarrow Q_8,~~Q_{m_2}  \leftrightarrow Q_{m_3}\,
\end{eqnarray}
which by new parameters $(\tA_i, \tu, \ty_i)$ can be represented as
\begin{eqnarray}
\tu\leftrightarrow \tu^{-1}, ~~	\ty_1\leftrightarrow \ty_2, ~~\ty_3 \leftrightarrow \ty_3,~~\tA_{1}\leftrightarrow\tA_{1},~~\tA_{2}\leftrightarrow\tA_{2}\,.
\end{eqnarray}
Then reparameterization can be fixed
\begin{eqnarray}\label{newparanf2}
A_{1}=\tA_{1} \sqrt{\ty_2~\ty_3},~A_{2}=\tA_{2} \sqrt{\ty_2~\ty_3},~u=\ty_2^{-\frac{3}{2}} \ty_3^{-\frac{3}{2}},~y_2=\sqrt{ \frac{\tu~\ty_2 } {\ty_3  } },~y_3=\sqrt{\frac{\ty_2  } {\tu~\ty_3   } }\,.
\end{eqnarray}
By taking these new parameters \eqref{newparanf2} into the normalized Nekrasov partition function, we indeed see the enhanced global symmetry
\begin{align}
&Z_{Sp(2) +1\mAS+2\mF}=1- \frac{\sqrt{q t} ~\chi_2^{SU(2)}[\tu] \chi_3^{SU(3)}[\ty]   }{(1-q)(1-t) } \tA_2  +   \frac{q+t-\sqrt{q t}\chi_2^{SU(2)}  }{(1-q)(1-t)  } \frac{\tA_1}{\tA_2} \nn\\ &-\frac{\sqrt{q t} ~\chi_2^{SU(2)}[\tu] \chi_2^{SU(3)}[\ty]    \l( -\sqrt{q t} \chi_2^{SU(2)} +1+ q t  )  \r)    }{ (1-q)^2(1-t)^2 }      \tA_1  
\nn\\
&   +    \frac{ \l(q t(q +t ) \chi_3^{SU(2)}[\tu]+ qt (1+q t)  \r)\chi_6^{SU(3)}[\ty]  
}{(1 - q)^2 (1 + q) (1 - t)^2 (1 + t)  }  \tA_2^2+  \nn\\
&+   \frac{ 
	\l ( (q+t)(1+q^2+t^2)-(q^3+t^3) +q t(1+q t) \chi_3^{SU(2)}[\tu]        \r)\chi_{\bar{3}}^{SU(3)}[\ty]         }{(1 - q)^2 (1 + q) (1 - t)^2 (1 + t)  }  \tA_2^2\nn\\
&+  \frac{ (q+t) \big( q t ~\chi_3^{SU(2)} -\sqrt{q t}( 1+q)(1+t) \chi_2^{SU(2)} + (q+t)(1+qt)+q t  \big)   } { (1 - q)^2 (1 + q) (1 - t)^2 (1 + t)  } \frac{\tA_1^2}{\tA_2^2}   \nn\\
&+\mathcal{ O}(\tA_1^2; \tA_2^2),\nn
\end{align}
where 
\begin{align}
&\chi_2^{SU(2)}[\tu]=\sqrt{\tu}+\frac{1}{\sqrt{\tu}},~~\qquad~~ \chi_3^{SU(2)}[\tu] =\tu+1+\tu^{-1},~~ \chi_3^{SU(3)}[\ty]=\ty_1+\ty_2+\ty_3,~\nn\\
&\chi_{\bar{3}}^{SU(3)}[\ty]=\ty_1^{-1}+\ty_2^{-1}+\ty_3^{-1},~~\chi_6^{SU(3)}[\ty] =\sum\limits_{i=1}^3 \frac{1}{\ty_i}+\ty_i \,.\nn
\end{align}

\paragraph{Connected \textbf{Case D}.}
\textbf{Case D} can be connected to other \textbf{Case D} by adding fundamental flavors $\mathbf{F}$ or Hanany-Witten moving D7-branes on brane webs. For $Sp(2)+1 \mAS+2 \mF$, the fourth brane web shown in Figure \ref{fig:2Fcases} is one typical example, whose unHiggsed diagram is depicted in Figure \ref{fig:2nftripletuning}. Apart from the same tuned K\"ahler parameters in \eqref{tuningnf2}, there is one additional \textbf{Case D} on the left top of the diagram in Figure \ref{fig:2nftripletuning}. According to the discussion in subsection \ref{T2tuning},  the tuned K\"ahler parameters for \textbf{Case D} should be determined by  either of \textbf{Case A, B, C, D} it connects. 
We notice that for this diagram the tuned K\"ahler parameters for this \textbf{Case D} is
\begin{align}
Q_{f_9}=Q_{f_{10}}=Q_{f_{91}}  =\sqrt{\frac{q}{t}}  \,.
\end{align}

We observed that if we draw a horizontal line in orange on the web, all tuned K\"ahler parameters on the upper half-plane were given the value $\sqqt$ and all tuned parameters on the lower half-plane $\sqtq$ in this assignment of $q$ and $t$ on the diagram in Figure \ref{fig:2nftripletuning}. 
The associated gauge theory parameters for the unHiggsed diagram in Figure \ref{fig:2nftripletuning}  are related to K\"ahler parameters by
\begin{eqnarray}
Q_F={A_2}^2,~~Q_{k4}=\frac{A_1}{A_2~Q_8},~~Q_B=\frac{u~ {A_2}^2}{\sqrt{y_1}\sqrt{y_2}},~~Q_{m_1}=\frac{y_1}{A_1},~~Q_{m_2}=\frac{y_2}{A_1} \,.
\end{eqnarray}
The extra factors here are of the following
\begin{align}
Z^{\rm extra}_{=//}=\frac{1}
{ M\l(\frac{y_1}{y_2},q,t   \r) M\l( \frac{t}{q}\frac{y_1}{y_2}, q,t  \r)    } \cdot 
\frac{1}{ M\l( u \sqrt{y_1 y_2 },q,t  \r)  M \l( \frac{t}{q} u \sqrt{y_1y_2},q,t  \r)    } \,,
\end{align}
where the instanton-dependent extra terms can be extracted by expanding the partition function and then picking up terms that break $q, t$-exchanging symmetry. The partition function obtained is the same as the web in Figure \ref{fig:nf2diagonal} and,  hence, we do not show the result again.
\begin{figure}[]
	\centering
	\begin{tikzpicture}[scale=0.8,line width=.6pt]
	\draw(0.5,1.1)--(0.5,0.9);	\draw(0.4,1.1)--(0.4,0.9);
	\draw(0.4,2.9)--(0.4,3.1);\draw(0.5,2.9)--(0.5,3.1);
	\draw(0.5,-1.1)--(0.5,-0.9);	\draw(0.4,-1.1)--(0.4,-0.9);
	\draw(0.4,-2.9)--(0.4,-3.1);\draw(0.5,-2.9)--(0.5,-3.1);
	\draw(-0.1,-0.1)--(0.1,0.1);	\draw(-0.1,0.1)--(0.1,-0.1);
	\draw (-1.5,-1)--(-1.5,1)--node[above=0pt]{{\tiny $Q_{B}$}}node[below=0pt]{{\tiny $\v_2$}}(1.5,1)--node[left=0pt]{{\tiny $Q_{F}$}}(1.5,-1)--node[below=0pt]{{\tiny $\v_1$}}(-1.5,-1);
	
	\draw (-1.5,-1)--node[left=4pt,above=0pt]{{\tiny $Q_{k4}$}}(-2.5,-2)--node[below=-2pt]{{\tiny $Q_8$}}node[above=-2pt]{{\tiny $\a_8$}}(-3.5,-2)--node[left=-2pt]{{\tiny $Q_{33}$}}(-3.5,2)--node[above=-2pt]{{\tiny $Q_{1}$}}node[below=0pt]{{\tiny $\a_1$}}(-2.5,2)--node[left=4pt,below=0pt]{{\tiny $Q_{k1}$}}(-1.5,1);
	\draw (1.5,1)--node[right=4pt,below=0pt]{{\tiny $Q_{k2}$}}(2.5,2)--node[above=-2pt]{{\tiny $Q_4$}}node[below=-2pt]{{\tiny $\a_4$}}(3.5,2)--node[right=-2pt]{{\tiny $Q_{44}$}}(3.5,-2)--node[above=-2pt]{{\tiny $\a_5$}}node[below=-2pt]{{\tiny $Q_5$}}(2.5,-2)--node[right=4pt,above=0pt]{{\tiny $Q_{k3}$}}(1.5,-1);
	\draw (2.5,2)--node[right=-4pt]{{\tiny $Q_{3}$}}(2.5,3)--node[above=0pt]{{\tiny $Q_{22}$}}node[below=0pt]{{\tiny $\v_4$}}(-2.5,3)--node[left=-4pt]{{\tiny $Q_{2}$}}(-2.5,2);
	\draw (2.5,-2)--node[right=-4pt]{{\tiny $Q_{6}$}}(2.5,-3)--node[below=0pt]{{\tiny $Q_{11}$}}node[above=0pt]{{\tiny $\v_3$}}(-2.5,-3)--node[left=-4pt]{{\tiny $Q_{7}$}}(-2.5,-2);
	
	\draw (-4.5,3)--(-6.5,3);\draw (-3.5,4)--(-3.5,6);\draw (-5.5,5)--(-5.5,6);\draw (-5.5,5)--(-6.5, 5);\draw(-5.5,5)--node[left=-2pt]{{\tiny $Q_{f_{33}}$}}(-4.5,4);
	
	\draw (2.5,-3)-- (4.5,-5);
	\draw (3.5,-2)-- (5.5,-4);
	
	\draw(-5.5,6)--node[left=10pt,below]{ \tiny $Q_{m_{11}}$}(-5.5,7)--node[left]{ \tiny  $Q_{f_{10}}$}(-4.5,8)--node[above=6pt,right=-6pt]{\tiny $Q_{f_{91}}$ }node[below]{ \tiny $\v_8$}(-3.5,8)--(-3.5,6)  ;\draw(-5.5,7)--(-6.5,7) ;\draw(-4.5,8)--node[left]{\tiny $Q_{f_9}$}(-4.5,9)--(-6.5,9) ;\draw(-4.5,9)--node[above,left]{ \tiny $t$}(-3,10.5)      ;\draw(-3.5,8)--node[below=2pt,right=-2pt]{ \tiny $t$}(-2,9.5) ;\draw[dashed,red](-4.5,4)--(-4.5,8)--(-6.5,8);\draw[dashed,red](-4.5,8)--(-2.5,10);\draw(-3.5,4)--node[right]{\tiny$ \frac{Q_{m_1}}{ Q_{m_2} }$ }(-3.5,8);
	
	\draw(-2.5,3)--node[right=0pt]{{\tiny $Q_{m_{22} }$}} (-3.5,4);\draw(-3.5,2)--node[below=1pt]{{\tiny $Q_{{m_2}}$}}(-4.5,3)--node[left=-1pt]{{\tiny $Q_{f_{4}}$}}(-4.5,4)--node[above=0pt]{{\tiny $Q_{f_{34}}$}}node[below=0pt]{{\tiny $\v_7$}}(-3.5,4);
	
	\draw[dashed,red](-2.5,2)--(-4.5,4);\draw[dashed,red](-4.5,4)--(-6.5,4); \draw[dashed,red](-4.5,4)--(-4.5,6); \draw[dashed,red](-3.5,2)--(-3.5,3)--(-2.5,3);
	\draw[dashed,red](2.5,2)--(5,4.5);
	\draw[dashed,red](3.5,2)--(3.5,3)--(2.5,3);
	
	\draw (2.5,3)--node[above=0pt, left=0pt]{{\tiny $t$}}(4.5,5); 	\draw (3.5,2)--node[ right=0pt,below=0pt]{{\tiny $t$}}(5.5,4); 
	\draw (-2.5,-3)--node[above=0pt, right=0pt]{{\tiny $q$}}(-4.5,-5); 	\draw (-3.5,-2)--node[ left=0pt,above=0pt]{{\tiny $q$}}(-5.5,-4);

	\draw(-2.5,-3)--(-4.5,-5);\draw(-3.5,-2)--(-5.5,-4);
	\draw[dashed,red](2.5,-2)--(5,-4.5); 
	\draw[orange](-6,0)--(6,0);\draw[orange](0,-6)--(0,6); 
	\draw[red,dashed](2.5,-3)--(3.5,-3)--(3.5,-2);

	\draw[dashed,red](-2.5,-2)--(-5,-4.5);
	\draw[red,dashed](-2.5,-3)--(-3.5,-3)--(-3.5,-2);
	\end{tikzpicture}
	\caption{The unHiggsed diagram for $Sp(2)+1\mAS+2\mF^{=//}$.}
	\label{fig:2nftripletuning}
\end{figure}

\subsection{\texorpdfstring{$Sp(2)+ 1\mathbf{AS}+ 3 \mathbf{F}  $}{}}

 The unHiggsed diagram for this theory is depicted in Figure \ref{fig:nf3}.
\begin{figure}[]
	\centering
	\begin{tikzpicture}[scale=0.8,line width=.6pt]
	\draw(0.5,1.1)--(0.5,0.9);	\draw(0.4,1.1)--(0.4,0.9);
	\draw(0.4,2.9)--(0.4,3.1);\draw(0.5,2.9)--(0.5,3.1);
	\draw(0.5,-1.1)--(0.5,-0.9);	\draw(0.4,-1.1)--(0.4,-0.9);
	\draw(0.4,-2.9)--(0.4,-3.1);\draw(0.5,-2.9)--(0.5,-3.1);
	\draw(-0.1,-0.1)--(0.1,0.1);	\draw(-0.1,0.1)--(0.1,-0.1);
	\draw (-1.5,-1)--(-1.5,1)--node[above=0pt]{{\tiny $Q_{B}$}}node[below=0pt]{{\tiny $\v_2$}}(1.5,1)--node[left=0pt]{{\tiny $Q_{F}$}}(1.5,-1)--node[below=0pt]{{\tiny $\v_1$}}(-1.5,-1);
	
	\draw[red,dashed](2.5,-3)--(3.5,-3)--(3.5,-2);
	
	\draw (-2.5,-3)--node[above=0pt, right=0pt]{{\tiny $q$}}(-4.5,-5); 	\draw (-3.5,-2)--node[ left=0pt,above=0pt]{{\tiny $q$}}(-5.5,-4); 
	
	\draw (2.5,3)--node[left=5pt, above=0pt]{{\tiny $Q_{m_{11} }$}} (3.5,4)--node[above=0pt]{{\tiny $Q_{f_{3}}$}}node[below=0pt]{{\tiny $\v_5$}}(4.5,4)--node[right=0pt]{{\tiny $Q_{f_{2}}$}}(4.5,3)--node[right=0pt]{{\tiny $Q_{m_1 }$}}(3.5,2);
	\draw (4.5,3)--(6.5,3);\draw (3.5,4)--(3.5,6);\draw (4.5,4)--node[right=2pt]{{\tiny $Q_{f_1 }$}} (5.5,5)--(5.5,6);\draw (5.5,5)--(6.5, 5);

	\draw (-1.5,-1)--node[left=4pt,above=0pt]{{\tiny $Q_{k4}$}}(-2.5,-2)--node[below=-2pt]{{\tiny $Q_8$}}node[above=-2pt]{{\tiny $\a_8$}}(-3.5,-2)--node[left=-2pt]{{\tiny $Q_{33}$}}(-3.5,2)--node[above=-2pt]{{\tiny $Q_{1}$}}node[below=0pt]{{\tiny $\a_1$}}(-2.5,2)--node[left=4pt,below=0pt]{{\tiny $Q_{k1}$}}(-1.5,1);
	\draw (1.5,1)--node[right=4pt,below=0pt]{{\tiny $Q_{k2}$}}(2.5,2)--node[above=-2pt]{{\tiny $Q_4$}}node[below=-2pt]{{\tiny $\a_4$}}(3.5,2)--node[right=-2pt]{{\tiny $Q_{44}$}}(3.5,-2)--node[above=-2pt]{{\tiny $\a_5$}}node[below=-2pt]{{\tiny $Q_5$}}(2.5,-2)--node[right=4pt,above=0pt]{{\tiny $Q_{k3}$}}(1.5,-1);
	\draw (2.5,2)--node[right=-4pt]{{\tiny $Q_{3}$}}(2.5,3)--node[above=0pt]{{\tiny $Q_{22}$}}node[below=0pt]{{\tiny $\v_4$}}(-2.5,3)--node[left=-4pt]{{\tiny $Q_{2}$}}(-2.5,2);
	\draw (2.5,-2)--node[right=-4pt]{{\tiny $Q_{6}$}}(2.5,-3)--node[below=0pt]{{\tiny $Q_{11}$}}node[above=0pt]{{\tiny $\v_3$}}(-2.5,-3)--node[left=-4pt]{{\tiny $Q_{7}$}}(-2.5,-2);

	\draw(-5.5,5)--node[left=-2pt]{{\tiny $Q_{f_{33}}$}}(-4.5,4);

	\draw(-2.5,3)--node[right=0pt]{{\tiny $Q_{m_{22} }$}} (-3.5,4);\draw(-3.5,2)--node[below=1pt]{{\tiny $Q_{{m_2}}$}}(-4.5,3)--node[left=-1pt]{{\tiny $Q_{f_{4}}$}}(-4.5,4)--node[above=0pt]{{\tiny $Q_{f_{34}}$}}node[below=0pt]{{\tiny $\v_7$}}(-3.5,4);

	\draw (2.5,-3)--node[left=3pt, below=0pt]{{\tiny $Q_{m_{33} }$}}  (3.5,-4)--node[below=0pt]{{\tiny $Q_{f_{56}}$}}node[above=0pt]{{\tiny $\v_6$}}(4.5,-4)--node[right=0pt]{{\tiny $Q_{f_{5}}$}}(4.5,-3)--node[right=0pt]{{\tiny $Q_{m_3 }$}}(3.5,-2);
	\draw (4.5,-3)--(6.5,-3);\draw (3.5,-4)--(3.5,-6);\draw (5.5,-5)--(5.5,-6);\draw (5.5,-5)--(6.5, -5);\draw(5.5,-5)--node[right=0pt]{{\tiny $Q_{f_6 }$}} (4.5,-4);
	
	\draw (-1.5,-1)--(-2.5,-2)--(-3.5,-2)--(-3.5,2)--(-2.5,2)--(-1.5,1);
	\draw (1.5,1)--(2.5,2)--(3.5,2)--(3.5,-2)--(2.5,-2)--(1.5,-1);
	\draw (2.5,2)--(2.5,3)--(-2.5,3)--(-2.5,2);
	\draw (2.5,-2)--(2.5,-3)--(-2.5,-3)--(-2.5,-2);
	\draw (2.5,3)--(3.5,4)--(4.5,4)--(4.5,3)--(3.5,2);
	
	\draw (4.5,3)--(6.5,3);\draw (3.5,4)--(3.5,6);\draw (4.5,4)--(5.5,5)--(5.5,6);\draw (5.5,5)--(6.5, 5);
	
	\draw (2.5,-3)--(3.5,-4)--(4.5,-4)--(4.5,-3)--(3.5,-2);
	\draw (4.5,-3)--(6.5,-3);\draw (3.5,-4)--(3.5,-6);\draw (5.5,-5)--(5.5,-6);\draw (5.5,-5)--(6.5, -5);\draw(5.5,-5)--(4.5,-4);
	
	\draw (-4.5,3)--(-6.5,3);\draw (-3.5,4)--(-3.5,6);\draw (-5.5,5)--(-5.5,6);\draw (-5.5,5)--(-6.5, 5);\draw(-5.5,5)--(-4.5,4);
	
	\draw(-2.5,3)--(-3.5,4);\draw(-3.5,2)--(-4.5,3)--(-4.5,4)--(-3.5,4);
	\draw[dashed,red](-2.5,2)--(-4.5,4);\draw[dashed,red](-4.5,4)--(-6.5,4); \draw[dashed,red](-4.5,4)--(-4.5,6); \draw[dashed,red](-3.5,2)--(-3.5,3)--(-2.5,3);
	\draw[dashed,red](2.5,2)--(4.5,4);\draw[dashed,red](4.5,4)--(6.5,4); \draw[dashed,red](4.5,4)--(4.5,6);  \draw[dashed,red](3.5,2)--(3.5,3)--(2.5,3);
	\draw(-2.5,-3)--(-4.5,-5);\draw(-3.5,-2)--(-5.5,-4);

	\draw[dashed,red](2.5,-2)--(4.5,-4);\draw[dashed,red](4.5,-4)--(6.5,-4); \draw[dashed,red](4.5,-4)--(4.5,-6);  
	\draw[orange](-6,0)--(6,0);\draw[orange](0,-6)--(0,6); 
	
		\draw[dashed,red](-2.5,-2)--(-5,-4.5);
	\draw[red,dashed](-2.5,-3)--(-3.5,-3)--(-3.5,-2);
	\end{tikzpicture}
	\caption{ The unHiggsed diagram for $Sp(2)+1\mAS+3\mF$. }
	\label{fig:nf3}
\end{figure}
Following the discussion in subsection \ref{T2tuning},
the tuned K\"ahler parameters are determined to be
\begin{align}\label{tuningnnf3}
&Q_{1}=Q_2=Q_3=Q_4=Q_{f_1}=Q_{f_2}=Q_{f_{33}}=Q_{f_4}=\sqrt{\frac{q}{t}}, \nn\\ &Q_{5}=Q_6=Q_{f_5}=Q_{f_6}=\sqrt{\frac{t}{q}}\,. 	
\end{align}
The maps between K\"{a}hler parameters and gauge theory parameters $(A_i, u, y_i)$ are
\begin{align}
	\label{3FKahlarA}
Q_F&={A_2}^2, &Q_{k4}&=\frac{A_1}{A_2~Q_8},&Q_B&=\frac{u~ {A_2}^2}{\sqrt{y_2}\sqrt{y_3}},\crcr
Q_{m_1}&=\frac{y_1}{A_1},&Q_{m_2}&=\frac{y_2}{A_1},&Q_{m_3}&=\frac{y_3}{A_1},
\end{align}
where $y_i=e^{-i \,m_i}$ are mass parameters.
Similar to $Sp(2)_0+1\mAS+N_f \mF\, (N_f =1, 2)$, constraints from Nekrasov factors \eqref{constraint1} appear in the partition function 
and give rise to constraints on various Young diagrams: $\v_5=\0$, $\v_6=\0$, $\v_7=\0$. Hence the summation part of topological string partition function $Z^{\text{sum}}$ can be reduced to the following
\begin{align}
&Z^{\rm sum}_{Sp(2)+1\mAS+3\mF}=\sum_{\substack{\v_1,\v_2,\v_3,\v_4 ,\\ \a_1,\a_2,\a_3,\a_4}}
Z^{\bf sum}_\textbf{Sp(2)+1AS} \cdot
{\textbf{terms}[y_1]}
\cdot
{\textbf{terms}[y_2]}
\cdot
{\textbf{terms}[y_3]}
\,,\nn \\
&{{\textbf{terms}[y_3]}}:= \nn\\
&\frac{ 
	N^{\rm half,-}_{ \a_4}\l( A_1 y_3  \r)
	~N^{\rm half,-}_{ \a_4}\l( A_1 y_3 \frac{t }{ q}  \r)
	~N^{\rm half,-}_{ \v_1^T}\l( \frac{ y_3}{A_2 }  \r)
	~N^{\rm half,-}_{ \v_2^T}\l( A_2 y_3  \r)
	~N^{\rm half,-}_{ \v_3^T}\l( \frac{ y_3}{A_1 }  \r)
	~N^{\rm half,-}_{ \v_4^T}\l( A_1 y_3  \r)
}
{ y_3^{\frac{|\v_1|+|\v_2|+|\v_3|+|\v_4|    }{2}   }
	N^{\rm half,-}_{ \a_4^T}\l( A_1 y_3  \r)
	~N^{\rm half,-}_{ \a_4^T}\l( A_1 y_3 \frac{t}{q}  \r)
} \,, \nn
\end{align}
where ${\textbf{terms}[y_1]}$ and ${\textbf{terms}[y_2]}$ are defined in \eqref{partfunnf1} and \eqref{partfunnf2} respectively.
The extra factors for this theory are given by
\begin{align}\label{extraright}
Z^{\bf extra}_{Sp(2) +1\mAS+3\mF}=&Z^{\bf pert extra}_{Sp(2) +1\mAS+3\mF} \cdot Z^{\bf inst extra}_{Sp(2) +1\mAS+3\mF}  \nn\\
=&\frac{1}{ M(Q_{44} Q_{m_1} Q_{m_3} Q_{f_1} Q_{f_2}, t,q  ) M(Q_{44} Q_{m_1} Q_{m_3} , t,q)  } \cdot
 \nn\\&\frac{1}{ M(Q_{22} Q_{m_{11}} Q_{m_{22}} , q,t  ) M(Q_{22} Q_{f_3} Q_{f_{34}} Q_{m_{11}} Q_{m_{22}} , q,t)  }     \nn\\
=&\frac{1}{M( y_1 y_3,t,q ) M(\frac{y_1 y_3 q}{t},t,q  )} \cdot  \frac{1}{M( u \sqrt{\frac{y_1 y_2}{y_3}},t,q ) M(\frac{t }{q} u \sqrt{\frac{y_1 y_2}{y_3}}  , q,t  )} \,,
\end{align}
\paragraph{Perturbative contribution.}
We obtain perturbative part 
\begin{align}\label{nf4pertstrange}
Z^{\textbf{pert}}_{Sp(2) +1\mAS+3\mF}=\frac{Z^{\textbf{pert-I}}_{Sp(2) +1\mAS+3\mF}  \cdot Z^{\textbf{pert-II}}_{Sp(2) +1\mAS+3\mF} }
{{ 
		Z_{Q_8}^{\rm extra} } \cdot Z^{\bf pert~extra}_{Sp(2) +1\mAS+3\mF}
	}
=
Z^{\textbf{pert}}_{Sp(2) +1\mAS} \prod_{i=1}^{N_f=3} Z_{\mF}^{\bf pert} (Q_{m_i})\,,
\end{align}
which can be expanded as
	\begin{align}
	&\frac{Z^{\textbf{pert}}_{Sp(2)+1\mAS+3\mF}}{M(Q_8 \sqtq,t,q)^2}=
	 1
	-\frac{\sqrt{q t}   ~\chi_6^{SU(4)}[y]    }{ (1-q)(1-t)  } A_2 +   \frac{q t ~{\chi_6^{SU(3)}[y]}^2  }{ 2(1-q)^2(1-t)^2   } 
	A_2^2
	\nn\\
	&
	  + \frac{
		(q+t)(q+t+q t+1)	+q t(2\chi_{15}^{SU(3)}[y]-{\chi_6^{SU(3)}[y]}^2  ) }{(1-q^2)( 1-t^2 )  }         A_2^2   \nn\\
	& +  \frac{(q+t)-\sqrt{q t } \chi_2^{SU(2)}  }{(1-q)(1-t) } \frac{A_1}{A_2} - \frac{\sqrt{q t}  \l(  1+q t-\sqrt{q t } \chi_2^{SU(2)}\r) ~\chi_6^{SU(4)}[y]  }{(1-q)^2(1-t)^2 }  A_1 \nn\\
	& +  \frac{ (q+t) \big( q t ~\chi_3^{SU(2)} -\sqrt{q t}( 1+q)(1+t) \chi_2^{SU(2)} + (q+t)(1+qt)+q t  \big) }{  (1 - q)^2 (1 + q) (1 - t)^2 (1 + t)} \frac{A_1^2}{A_2^2}  
\nn\\
&
	+\mathcal{ O}(A_1^2; A_2^2)\,, \nn
	\end{align}
where  $ \chi_6^{SU(4)}[y] = \sum\limits_{i=1}^{3} y_i+y_i^{-1},~\chi_{15}^{SU(4)}[y] =\sum\limits_{i<j}1+y_iy_i+\frac{1}{y_i y_j}+\frac{y_i}{y_j}+\frac{y_j}{y_i}$\footnote{This definition of character is a bit different here. One can use LieART package in \cite{Feger:2012bs} to get characters in omegabasis,  and then do a transformation $y_2\rightarrow y_2,~y_3\rightarrow \sqrt{ \frac{ y_2 y_3}{y_1} },~ y_1 \rightarrow \sqrt{y_1 y_2 y_3  }$ to get this definition. We remind that all characters in this paper are obtained by using LieART.}. 
\paragraph{Instanton contributions}
One-instanton contribution can be expanded as
	\begin{align}\label{oneinst3F}
	&
Z^{\textbf{one-instanton}}_{Sp(2)+1\mAS +3 \mF}= \frac{-\sqrt{q t } }{(1-q)(1-t)} \chi_4^{SU(4)}[y] (A_2+A_1)  + \frac{ q+t}{(1-q)(1-t)  }  \chi_{\bar{4}}^{ SU(4)}[y] (A_2^2 +A_1^2)   
	 \nn  \\
	&+\frac{q+t-\sqrt{q t}\chi_2^{SU(2)}   } {(1-q)(1-t) }  \chi_{\bar{4}}^{ SU(4)}[y] A_2 A_1-\frac{ (q+t) (q+t-\sqrt{q t}\chi_2^{SU(2)}     )  }{(1 - q) (1 - t) \sqrt{q t}  } \chi_4^{ SU(4)}[y]
	\nn\\
	& \times (A_2^2A_1+A_1 A_2^2)  
	+ \frac{(q + t)^2 (q +t -\sqrt{q t} \chi_2^{SU(2)}  )  }{(1 - q) q (1 - t) t }   \chi_{\bar{4}}^{ SU(4)}[y] A_2^2A_1^2  +\mathcal{ O}(A_1^3,A_2^3), 
	\end{align}
where 
	\begin{align}
&	\chi_4^{SU(4)}[y]=\sqrt{y_1y_2y_3 } + \frac{\sqrt{y_1} }{\sqrt{y_2}\sqrt{y_3} }   +\frac{\sqrt{y_2} }{\sqrt{y_1}\sqrt{y_3} } +\frac{\sqrt{y_3} }{\sqrt{y_1}\sqrt{y_2} } \,,
\nn\\
& \chi_{\bar{4}}^{SU(4)}[y] =\frac{1} { \sqrt{y_1y_2y_3 }} + \frac{\sqrt{y_2}\sqrt{y_3} }{\sqrt{y_1} }+   \frac{\sqrt{y_1}\sqrt{y_3} }{\sqrt{y_2} }   +\frac{\sqrt{y_1}\sqrt{y_2} }{\sqrt{y_3} }\,.\nn
	\end{align}
We checked that \eqref{oneinst3F} agrees with the  localization result in appendix \eqref{oneinstloc} and by mapping parameters
	$y_{1}=e^{m_{1}}, ~ y_{3}=e^{m_{3}},~y_2=e^{-m_2},~Q_8=e^m,~A_{1}=e^{\a_{1}},~ A_{2}=e^{\a_{2}}$.

\vskip .5cm
Similarly, we obtains two-instanton contribution
	\begin{align}
	&
Z^{\textbf{two-instanton}}_{Sp(2)+1\mAS +3 \mF}=  \frac{q t\l( (1+q t) \chi_6^{ SU(4)}[y]+ (q+t) \chi_{10}^{ SU(4)}[y]   \r) }{ (1 - q)^2 (1 + q) (1 - t)^2 (1 + t)}(A_2^2+A_1^2) + \frac{ q t \l(\chi_6^{ SU(4)}[y]+\chi_{10}^{ SU(4)}[y]  \r)} {(1-q)^2(1-t)^2  } A_2A_1   \nn\\ 
	& +\frac{  \sqrt{q t} (2 q +\sqrt{q t}+2 t    ) \chi_{15}^{ SU(4)}[y] -\big( (q+t)(1+q t)  -qt -(q^2+t^2) +q t \chi_{15}^{SU(4)}[y]  \big) (\chi_2^{SU(2)}+1)              }{(1 - q)^2 (1 - t)^2}  
	\nn\\
&\times	(A_2^2A_1+A_2A_1^2) +\mathcal{ O}(A_1^2;A_2^2),  \nn
	\end{align}
where 
\begin{align}
\chi_{10}^{SU(4)}[y] =
\sum_{i=1}^3  y_i+\frac{1}{y_i}+\frac{y_i}{y_j y_k}, \qquad
\chi_{15}^{SU(4)}[y]=3+\sum_{i\neq j} y_iy_j+\frac{1}{y_iy_j}+\frac{y_i}{y_j}, ~~i=1,2,3\,.\nn
\end{align}

\paragraph{Enhancement of global symmetry}
We shift Coulomb branch parameters to the gauge parameters
\begin{eqnarray}
\tA_{1}=A_{1}~ u^{\frac{2}{8-N_f}}=A_{1}~ u^{\frac{2}{5}},\quad \tA_{2}=A_{2}~ u^{\frac{2}{8-N_f}}=A_{2}~ u^{\frac{2}{5}}\,. 
\end{eqnarray}
The global symmetry of $Sp(2)+1\mAS+3 \mF$ should be $ G= SU(2)_{Q_8} \times E_4= SU(2)_{Q_8} \times SU(5)_{\ty_i, {i=1,\cdots,5}}$. Here we define new fugacities $\ty_{1},\cdots, \ty_5 $ for $ E_4$.\footnote{ For $ E_4=SU(4)$,  $\ty_1 \ty_2 \ty_3  \ty_4\ty_5=1$.} By observing the diagram in Figure \ref{fig:nf3}, we notice that the fiber-base duality leads to
\begin{eqnarray}\label{3FQfiberbase}
Q_B  \leftrightarrow Q_F,~~ Q_8  \leftrightarrow Q_8,~~Q_{m_2}  \leftrightarrow Q_{m_3},~Q_{m_1} \leftrightarrow Q_{m_1}  \,,
\end{eqnarray}
or it takes the following relations in terms of gauge theory parameters:
\begin{eqnarray}\label{3FAfiberbase}
\ty_1\leftrightarrow \ty_2, ~~	\ty_3\leftrightarrow \ty_4, ~~\tA_{1}\leftrightarrow\tA_{1},  ~~\tA_{2}\leftrightarrow\tA_{2}\,.
\end{eqnarray}
Combining \eqref{3FQfiberbase} and \eqref{3FAfiberbase}, we express the K\"ahler parameters in terms of the gauge theory parameters:
\begin{eqnarray}
{ Q_F=\tA_2^2 \ty_1,~Q_B=\tA_2^2\ty_2,~Q_{k4}=\frac{\tA_1}{\tA_2 A_8},~Q_{m_1}=\frac{\ty_3\ty_4 }{\tA_1},~Q_{m_2}=\frac{\ty_3\ty_5 }{\tA_1},~Q_{m_3}=\frac{\ty_4\ty_5 }{\tA_1}}\,.\qquad
\end{eqnarray}
In addition to \eqref{3FKahlarA}, we find the reparameterization of the gauge theory parameters as follows
\begin{eqnarray}\label{newparametersnf3}
A_{1}= \tA_{1}\sqrt{\ty_1},~A_{2}= \tA_{2}\sqrt{\ty_2},~ 
~u=\ty_1^{-\frac{5}{4}},
~y_1=\sqrt{ \frac{\ty_{3}\ty_4 } {\ty_{2}\ty_5	  } },~y_2=\sqrt{ \frac{\ty_{3}\ty_5 } {\ty_{2}\ty_4  } },~y_3=\sqrt{ \frac{\ty_{4}\ty_5 } {\ty_{2}\ty_3  } }\,.\qquad
\end{eqnarray}
By expressing the normalized Nekrasov partition function in terms of the parameters $(\tA_i, \tu, \ty_i)$, we see the enhancement of global symmetry as follows
	\begin{align}
	& Z_{Sp(2)+1\mAS+3\mF}=1- \frac{\sqrt{q t} ~\chi_{ \overline{10}}^{E_4}[\tu]}{(1-q)(1-t)  } \tA_2  +   \frac{q+t-\sqrt{q t}~\chi_2^{SU(2)}   }{(1-q)(1-t)  } \frac{\tA_1}{\tA_2}  +\frac{\l( qt ~\chi_2^{SU(2)} - \sqrt{q t}(1+ q t  )   \r) \chi_{ \overline{10}}^{ E_4}[\tu]   }{ (1-q)^2(1-t)^2 }      \tA_1 
	 \nn\\
	& +  \frac{  \l(   (1+q^2 t^2)(q+t) -(q^3+t^3) \r)\chi_5^{ E_4}[\ty] +q t(1+q t)  \chi_{45}^{E_4}[\ty]  +q t(q+t) \chi_{50}^{ E_4}[\ty]  }{(1 - q)^2(1+q) (1 - t)^2 (1+t)  }  \tA_2^2
	 \nn  \\
	&  +\frac{ (q+t) \big( q t ~\chi_3^{SU(2)} -\sqrt{q t}( 1+q)(1+t) \chi_2^{SU(2)} + (q+t)(1+qt)+q t  \big)   } { (1 - q)^2 (1 + q) (1 - t)^2 (1 + t)  } \frac{\tA_1^2}{\tA_2^2}   \nn\\
	&+\frac{ \sqrt{q t}  \l(    
		-q t(q+t)  \chi_3^{SU(2)}+ \sqrt{q t}\l( (1+q t)^2+(q+t)(1-(q-t)^2)    \r)  \chi_2^{ SU(2)} 
	+\cdots	
		   \r) ~\chi_{ \overline{10}}^{ E_4}[\tu]    }{ (1-q)^3 (1-t)^3 (1+q)(1+t) } \frac{\tA_1^2}{\tA_2} \nn\\
	   &+\mathcal{ O}(\tA_1^2;\tA_2^2)\,, 
	\end{align}
where characters are defined as \footnote{Here, we choose orthogonal basis in LieART.} 
	\begin{align}
&\chi_5^{ E_4}[\ty]=\sum\limits_i \ty_i,  \quad \chi_{\overline{10}}^{ E_4}[\ty]=\sum\limits_{i\neq j} \frac{1}{ \ty_i \ty_j} ,\quad \chi_{45}^{ E_4}[\ty]=3 \sum\limits_i \ty_i + \sum\limits_{i\neq j\neq k} \frac{\ty_i \ty_j }{\ty_k} ,\nn\\
&\chi_{50}^{ E_4}[\ty]=2 \sum\limits_i \ty_i + \sum\limits_{i\neq j\neq k} \frac{\ty_i \ty_j }{\ty_k}  +\sum\limits_{i\neq j} \frac{ 1}{ \ty_i \ty_j}  \,,~~	 i,j,k=1,\cdots, 5\,.\nn
\end{align}

\subsection{\texorpdfstring{$Sp(2)+ 1\mathbf{AS}+ 4 \mathbf{F}  $}{Sp(2)+1AS+4F}}
\begin{figure}[]
	\centering
	\includegraphics[width=5.5in]{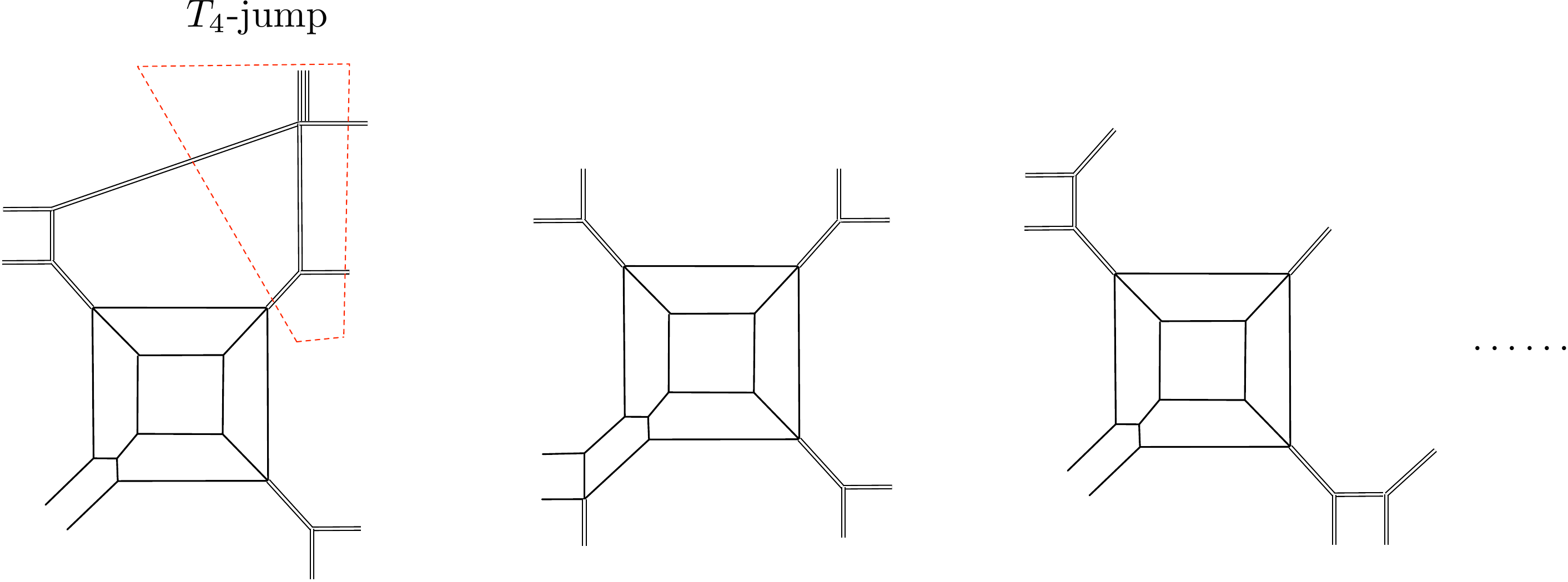}
	\caption{ Some equivalent webs related through Hanany-Witten moves for $Sp(2)+ 1\mathbf{AS}+ 4\mF$. } 
	\label{fig:4Fcases}
\end{figure}

\begin{figure}[]
	\centering
	\begin{tikzpicture}[scale=0.8,line width=.6pt]
	\draw(0.5,1.1)--(0.5,0.9);	\draw(0.4,1.1)--(0.4,0.9);
	\draw(0.4,2.9)--(0.4,3.1);\draw(0.5,2.9)--(0.5,3.1);
	\draw(0.5,-1.1)--(0.5,-0.9);	\draw(0.4,-1.1)--(0.4,-0.9);
	\draw(0.4,-2.9)--(0.4,-3.1);\draw(0.5,-2.9)--(0.5,-3.1);
	\draw(-0.1,-0.1)--(0.1,0.1);	\draw(-0.1,0.1)--(0.1,-0.1);
	\draw (-1.5,-1)--(-1.5,1)--node[above=0pt]{{\tiny $Q_{B}$}}node[below=0pt]{{\tiny $\v_2$}}(1.5,1)--node[left=0pt]{{\tiny $Q_{F}$}}(1.5,-1)--node[below=0pt]{{\tiny $\v_1$}}(-1.5,-1);
	
	\draw[dashed,red](-2.5,-2)--(-5,-4.5);
	\draw[red,dashed](-2.5,-3)--(-3.5,-3)--(-3.5,-2);
	
	\draw (2.5,3)--node[left=5pt, above=0pt]{{\tiny $Q_{m_{11} }$}} (3.5,4)--node[above=0pt]{{\tiny $Q_{f_{3}}$}}node[below=0pt]{{\tiny $\v_5$}}(4.5,4)--node[right=0pt]{{\tiny $Q_{f_{2}}$}}(4.5,3)--node[right=0pt]{{\tiny $Q_{m_1 }$}}(3.5,2);
	\draw (4.5,3)--(6.5,3);\draw (3.5,4)--(3.5,6);\draw (4.5,4)--node[right=2pt]{{\tiny $Q_{f_1 }$}} (5.5,5)--(5.5,6);\draw (5.5,5)--(6.5, 5);
	
	\draw (-1.5,-1)--node[left=4pt,above=0pt]{{\tiny $Q_{k4}$}}(-2.5,-2)--node[below=-2pt]{{\tiny $Q_8$}}node[above=-2pt]{{\tiny $\a_8$}}(-3.5,-2)--node[left=-2pt]{{\tiny $Q_{33}$}}(-3.5,2)--node[above=-2pt]{{\tiny $Q_{1}$}}node[below=0pt]{{\tiny $\a_1$}}(-2.5,2)--node[left=4pt,below=0pt]{{\tiny $Q_{k1}$}}(-1.5,1);
	\draw (1.5,1)--node[right=4pt,below=0pt]{{\tiny $Q_{k2}$}}(2.5,2)--node[above=-2pt]{{\tiny $Q_4$}}node[below=-2pt]{{\tiny $\a_4$}}(3.5,2)--node[right=-2pt]{{\tiny $Q_{44}$}}(3.5,-2)--node[above=-2pt]{{\tiny $\a_5$}}node[below=-2pt]{{\tiny $Q_5$}}(2.5,-2)--node[right=4pt,above=0pt]{{\tiny $Q_{k3}$}}(1.5,-1);
	\draw (2.5,2)--node[right=-4pt]{{\tiny $Q_{3}$}}(2.5,3)--node[above=0pt]{{\tiny $Q_{22}$}}node[below=0pt]{{\tiny $\v_4$}}(-2.5,3)--node[left=-4pt]{{\tiny $Q_{2}$}}(-2.5,2);
	\draw (2.5,-2)--node[right=-4pt]{{\tiny $Q_{6}$}}(2.5,-3)--node[below=0pt]{{\tiny $Q_{11}$}}node[above=0pt]{{\tiny $\v_3$}}(-2.5,-3)--node[left=-4pt]{{\tiny $Q_{7}$}}(-2.5,-2);
	%
	\draw (-4.5,3)--(-6.5,3);\draw (-3.5,4)--(-3.5,6);\draw (-5.5,5)--(-5.5,6);\draw (-5.5,5)--(-6.5, 5);\draw(-5.5,5)--node[left=-2pt]{{\tiny $Q_{f_{33}}$}}(-4.5,4);
	
	\draw (2.5,-3)--node[left=3pt, below=0pt]{{\tiny $Q_{m_{33} }$}}  (3.5,-4)--node[below=0pt]{{\tiny $Q_{f_{56}}$}}node[above=-2pt]{{\tiny $\v_6$}}(4.5,-4)--node[right=0pt]{{\tiny $Q_{f_{5}}$}}(4.5,-3)--node[right=0pt]{{\tiny $Q_{m_3 }$}}(3.5,-2);
	\draw (4.5,-3)--(6.5,-3);\draw (3.5,-4)--(3.5,-6);\draw (5.5,-5)--(5.5,-6);\draw (5.5,-5)--(6.5, -5);\draw(5.5,-5)--node[below=0pt]{{\tiny $Q_{f_6 }$}} (4.5,-4);
	
	\draw(-5.5,6)--node[left=10pt,below]{ \tiny $Q_{m_{44}}$}(-5.5,7)--node[left]{ \tiny  $Q_{f_{10}}$}(-4.5,8)--node[above=6pt,right=-6pt]{\tiny $Q_{f_{91}}$ }node[below]{ \tiny $\v_8$}(-3.5,8)--(-3.5,6)  ;\draw(-5.5,7)--(-6.5,7) ;\draw(-4.5,8)--node[left]{\tiny $Q_{f_9}$}(-4.5,9)--(-6.5,9) ;\draw(-4.5,9)--node[above,left]{ \tiny $t$}(-3,10.5)      ;\draw(-3.5,8)--node[below=2pt,right=-2pt]{ \tiny $t$}(-2,9.5) ;\draw[dashed,red](-4.5,4)--(-4.5,8)--(-6.5,8);\draw[dashed,red](-4.5,8)--(-2.5,10);\draw(-3.5,4)--node[right]{\tiny$ \frac{Q_{m_4}}{ Q_{m_2} }$ }(-3.5,8);
	
	\node [above] at (4.3,19.5) {$\textbf{ Case B}$};
	\node[left]at (3.5,10){\tiny $ Q_{66}$ };
	
	\draw(-3.5,8)--node[below,right]{\tiny $ Q_{77}$ }(2.5,14)--node[right=-2pt]{\tiny $ Q_{F_9}$ }(2.5,15);
	\draw(-4.5,9)--node[left=3pt,above=1pt]{\tiny $ Q_{88}$ }(1.5,15)--node[left]{\tiny $t$}(1.5,19);
	\draw(3.5,4)--(3.5, 14);
	\draw[red,dashed](-4.5,8)--(4.5,17)--(8.5,17);	\draw[red,dashed](4.5,4)--(4.5,19);\draw[red,dashed](4.5,4)--(6.5,4);
	\draw(1.5,15)--(1.5,17);	\draw(5.5,5)--node[right]{\tiny $ Q_{55}$ }(5.5,15);
	\draw(2.5,14)--node[below]{\tiny $ Q_{F_{13}}$ }node[above=-2 pt]{\tiny $ \v_{11}$ }(3.5,14)--node[right,below=4pt]{\tiny $ Q_{F_{12}}$ }(4.5,15)--node[below]{\tiny $ Q_{F_{11}}$ }node[above=-2pt]{\tiny $ \v_9$ }(5.5,15)--node[below,right]{\tiny $ Q_{F_{10}}$ }(6.5,16)--(8.5,16);
	\draw(1.5,15)--node[above]{\tiny $ Q_{F_8}$ }node[below]{\tiny $ \v_{13}$ }(2.5,15)--node[above=2pt]{\tiny $ Q_{F_7}$ }(3.5,16)--node[above=5pt,right=-7pt]{\tiny $ Q_{F_5}$ }node[below]{\tiny $ \v_{12}$ }(4.5,16)--node[above=6pt, right=-15pt]{\tiny $ Q_{F_4}$ }(5.5,17)--node[above]{\tiny $ Q_{F_3}$ }node[below]{\tiny $ \v_{10}$ }(6.5,17)--node[above=2pt]{\tiny $ Q_{F_1}$ }(7.5,18)--(8.5,18);
	\draw(3.5,16)--(3.5,19);	\draw(4.5,15)--node[right=-2pt]{\tiny $ Q_{F_6}$ }(4.5,16);\draw(5.5,17)--(5.5,19);\draw(6.5,16)--node[right=-2pt]{\tiny $ Q_{F_2}$ }(6.5,17);\draw(7.5,18)--(7.5,19);
	
	\draw(-2.5,3)--node[right=0pt]{{\tiny $Q_{m_{22} }$}} (-3.5,4);\draw(-3.5,2)--node[below=1pt]{{\tiny $Q_{{m_2}}$}}(-4.5,3)--node[left=-1pt]{{\tiny $Q_{f_{4}}$}}(-4.5,4)--node[above=0pt]{{\tiny $Q_{f_{34}}$}}node[below=0pt]{{\tiny $\v_7$}}(-3.5,4);
	
	\draw[dashed,red](-2.5,2)--(-4.5,4);\draw[dashed,red](-4.5,4)--(-6.5,4); \draw[dashed,red](-4.5,4)--(-4.5,6); \draw[dashed,red](-3.5,2)--(-3.5,3)--(-2.5,3);
	\draw[dashed,red](2.5,2)--(5,4.5);
	\draw[dashed,red](3.5,2)--(3.5,3)--(2.5,3);
	
	\draw (2.5,3)--node[above=0pt, left=0pt]{{\tiny $t$}}(3.5,4); 	\draw (3.5,2)--node[ right=0pt,below=0pt]{{\tiny $t$}}(4.5,3); 
	\draw (-2.5,-3)--node[above=0pt, right=0pt]{{\tiny $q$}}(-4.5,-5); 	\draw (-3.5,-2)--node[ left=0pt,above=0pt]{{\tiny $q$}}(-5.5,-4);

	\draw(-2.5,-3)--(-4.5,-5);\draw(-3.5,-2)--(-5.5,-4);
	
	\draw[dashed,red](2.5,-2)--(4.5,-4);\draw[dashed,red](4.5,-4)--(6.5,-4); \draw[dashed,red](4.5,-4)--(4.5,-6);  
	\draw[orange](-6,0)--(6,0);\draw[orange](0,-6)--(0,6); 
	\draw[red,dashed](2.5,-3)--(3.5,-3)--(3.5,-2);
	
	\end{tikzpicture}
	\caption{An unHiggsed diagram for $Sp(2)+1\mAS+4 \mF$. The type of tuned parameters on the right top are found to be type \textbf{Case B}. The red dashed lines on the diagram are auxiliary lines. }	\label{fig:nf4topcross}
\end{figure}

\paragraph{unHiggsed $T_4$-diagram.}

There are many equivalent webs for this theory as depicted in Figure \ref{fig:4Fcases}. We choose the first web for computation, as it contains a Higgsed $T_4$-diagram, which has not been discussed in the previous examples. We find $T_2$-tuning still works for this Higgsed $T_4$-diagram, which implies that $T_2$-tuning could be used for generic Higgsed $T_N$-diagrams.
The unHiggsed diagram for the first web in Figure \ref{fig:4Fcases} is shown in Figure \ref{fig:nf4topcross}. 
Following the discussion in subsection \ref{T2tuning},
we find the tuned K\"ahler parameters are as follows
\begin{align}\label{hignf4}
&Q_1=Q_2=Q_3=Q_4=Q_{f_1}=Q_{f_2}=Q_{f_{33}}=Q_{f_4}=Q_{f_9}=Q_{f_{10}}=\sqrt{\frac{q}{t}}\,, 
\nn\\
& Q_5=Q_6=Q_{f_5}=Q_{f_6}=\sqrt{\frac{t}{q}}, \quad Q_{F_1}=Q_{F_2}= \cdots =Q_{F_{13 }} =\sqqt  \,.
\end{align}
With the help of auxiliary lines, the relations between K\"{a}hler parameters and gauge theory parameters $(A_i, u, y_i)$ can be determined as follows
\begin{eqnarray}
Q_F={A_2}^2,~~Q_{k4}=\frac{A_1}{A_2~Q_8},~~Q_B=\frac{u~ {A_2}^2}{\sqrt{y_1 y_2y_3y_4}},~~Q_{m_i}=\frac{y_i}{A_1} \,,
\end{eqnarray}
where $y_i=e^{-im_i}, i=1, 2, 3, 4$ are mass parameters. 
Just like $Sp(2)_0+1\mAS+N_f\mF (N_f =1, 2, 3)$, geometric transitions related terms \eqref{constraint1} given by tuning K\"ahler parameters provide constraints on various Young diagrams $\v_i=\0, i=5,\cdots,13$. Hence the summation part of the partition function was reduced as
\begin{align}
&Z^{\rm sum}_{Sp(2)+1\mAS+4 \mF}=\sum_{\substack{\v_1,\v_2,\v_3,\v_4 ,\\ \a_1,\a_2,\a_3,\a_4}}
Z^{\bf sum}_{\bf Sp(2)+1 AS} \cdot
{\textbf{terms}[y_4]'} \cdot  \prod_{i=1}^3{\textbf{terms}[y_i]} \,,    \label{zsum4F}\\
&{{\textbf{terms}[y_4]}'}= \nn\\
&
\frac{  
	y_4^{-\frac{|\v_1|+|\v_2|+|\v_3|+|\v_4|    }{2}   }
	N^{\rm half,-}_{ \v_2}\l( \frac{y_4}{A_2}  \r)
	~N^{\rm half,-}_{ \v_4}\l( \frac{y_4}{A_1}  \r)
	~N^{\rm half,-}_{ \v_1^T}\l( A_2 y_4  \r)
	~N^{\rm half,-}_{ \v_3^T}\l( A_1 y_4  \r)
}
{
	N^{\rm half,-}_{ \v_2}\l(   \frac{u}{A_2}  \frac{ t^2}{ q^2} \sqrt{   \frac{y_1 y_2 y_4} { y_3}  }  \r)
	~	N^{\rm half,-}_{ \v_4}\l(   \frac{u}{A_1}  \frac{ t^2}{ q^2} \sqrt{   \frac{y_1 y_2 y_4} { y_3}  }  \r)
	~	N^{\rm half,-}_{ \v_1^T}\l(  u~ \frac{A_2  t^2}{ q^2} \sqrt{   \frac{y_1 y_2 y_4} { y_3}  }  \r)
	~	N^{\rm half,-}_{ \v_3^T}\l(  u~ \frac{A_1  t^2}{ q^2} \sqrt{   \frac{y_1 y_2 y_4} { y_3}  }  \r)
} \,. \nn
\end{align}
\paragraph{Perturbative contribution.}
We obtain the perturbative part
\begin{eqnarray}\label{nf4pert}
Z^{\textbf{pert}}_{Sp(2)+1\mAS+4\mF}=\frac{
	Z^{\textbf{pert-I}}_{Sp(2) +1\mAS+4\mF} \cdot	Z^{\textbf{pert-II}}_{Sp(2) +1\mAS+4\mF} 
 }{{ 
		Z_{Q_8}^{\rm extra} \cdot Z^{\rm pert~extra}_{=}    } }=
Z^{\textbf{pert}}_{Sp(2)+1\mAS} \prod_{i=1}^{N_f=4} Z_\mF^{ \pert} (Q_{m_i})\,, ~~~~
\end{eqnarray}
where $Z^{\rm extra}_{=} $ is given by
\begin{align}
Z^{\rm extra}_{=}
&= ~Z^{\textbf{pert extra} }_{Sp(2)+1\mAS+4\mF} \cdot Z^{\textbf{inst extra}}_{Sp(2)+1\mAS+4\mF}   \nn\\
&=\frac{1}{   M(y_1 y_3,t,q) M\l( \frac{q}{t}  y_1y_3,t,q   \r) M\l( \frac{y_4}{y_2} ,q,t  \r)  
	M\l(\frac{t}{q} \frac{y_4}{y_2} ,q,t  \r)  }    \nn\\
&\times  \frac{ 1}
{   M\l( u \sqrt{\frac{y_2y_4 }{y_1 y_3 }  },t,q    \r)
	M\l( u \frac{q}{t}\sqrt{\frac{y_2y_4 }{y_1 y_3 }  },t,q    \r)
	M\l( u \sqrt{ y_1y_2y_3y_4 },t,q    \r)
	M\l( u\frac{q}{t} \sqrt{ y_1y_2y_3y_4 },t,q    \r)
}\,.\nn
\end{align}
The normalized perturbative part can be expanded as \footnote{In orthogonal basis of LieART.}
\begin{align}
&\frac{Z^{\textbf{pert}}_{Sp(2)+1\mAS+4\mF}}{M(Q_8 \sqtq,t,q)^2} = 1	-\frac{\sqrt{q t}   ~\chi_{8_v}^{SO(8)} [y]  }{ (1-q)(1-t)  } A_2 - \frac{\sqrt{q t}  \l(1+q t-\sqrt{q t } \chi_2^{SU(2)}\r) ~\chi_{8_v}^{\rm SO(8)} [y] }{(1-q)^2(1-t)^2 }  A_1 
\nn\\
&+ \l(   \frac{q t ~\chi_{8_v}^{SO(8)} [y]  ^2  }{ 2(1-q)^2(1-t)^2   }   + \frac{ 
	q t(2\chi_{28}^{\rm SO(8)}[y]-\chi_{8_v}^{\rm SO(8)}[y]^2  ) +\cdots  }{2(1-q^2)( 1-t^2 )  
}         \r)A_2^2   
+  \frac{(q+t)-\sqrt{q t } \chi_2^{SU(2)}  }{(1-q)(1-t) } \frac{A_1}{A_2} \nn\\
&+  \frac{ (q+t) \big( q t ~\chi_3^{SU(2)} -\sqrt{q t}( 1+q)(1+t) \chi_2^{SU(2)} + (q+t)(1+qt)+q t  \big) }{  (1 - q)^2 (1 + q) (1 - t)^2 (1 + t)} \frac{A_1^2}{A_2^2}  \nn\\
& +\frac{\sqrt{ q t }  \l(    
	-q t(q+t)  \chi_3^{\rm SU(2)}+ \sqrt{q t}\l( (1+q t)^2+(q+t)(1-(q-t)^2)    \r)  \chi_2^{SU(2)}  +\cdots
	\r)   ~\chi_{8_v}^{SO(8)} [y] }{  (1 - q)^3 (1 + q)  (1 - t)^3 (1 + t)}    \nn\\ & \times \frac{A_1^2}{A_2} +\mathcal{ O}(A_1^2;A_2^2), 
\end{align}
where $ \chi_{8_v}^{ SO(8)} [y]=\sum\limits_i^4 y_i+y_i^{-1},~ \chi_{28}^{SO(8)}[y] =\sum\limits_{i\neq j}y_iy_i+\frac{1}{y_i y_j}+\frac{y_i}{y_j}, ~i=1, \cdots, 4$.

\paragraph{Instanton contributions.}
For the unHiggsed diagram in Figure \ref{fig:nf4topcross}, $Z^{M}$ contains instanton dependent terms $Z^{\textbf{nonpert-I}} \neq 1$. We remove instanton extra factors and find that the terms in the first term of 	\eqref{instantonform0} are the following, which could contribute to the instanton part:
\begin{align}
&\frac{Z^{\textbf{nonpert-I}}_{Sp(2)+1\mAS+4\mF}  }{   Z^{\textbf{inst~extra}}_{Sp(2)+1\mAS+4\mF}  } =
\frac
{ M\l(  u \sqrt{\frac{y_1y_2 }{ y_3 y_4 }}, q,t \r)  
	M\l(  u \frac{t}{q}\sqrt{\frac{y_1y_2 }{ y_3 y_4 }},q,t  \r)  
	M\l(  u \sqrt{\frac{y_1y_4 }{ y_2 y_3 }},q,t \r) 
	M\l(  u\frac{t}{q} \sqrt{\frac{y_1 y_4 }{ y_2 y_3 }} ,q,t \r)   
}
{   M\l(  \frac{u}{A_1} \frac{t^{5/2}}{q^{5/2}} \sqrt{\frac{{y_1 y_2 y_4 }}{ y_{3} }}, t,q  \r) 
	M\l(  u A_1 \frac{t^{5/2}}{q^{5/2}}
	\sqrt{\frac{{y_1 y_2 y_4 }}{ y_{3} }},t,q  \r)  
}   \nn \\
& \times 
\frac{ 
	M\l(  u \sqrt{\frac{y_2 y_4} { y_1 y_3 }} , q,t \r) 
	M\l(  u\frac{t}{q} \sqrt{\frac{y_2 y_4} { y_1 y_3 }},q,t \r)  
	M\l(  u\sqrt{ y_1 y_2 y_3 y_4}, t,q \r) 
	M\l(  u\sqrt{ y_1 y_2 y_3 y_4}, q,t \r) 
}
{
	M\l(  u A_2 \frac{t^{5/2}}{q^{5/2}}
	\sqrt{\frac{{y_1 y_2 y_4 }}{ y_{3} }} , t,q \r)  
	M\l(  \frac{u}{A_2} \frac{t^{5/2}}{q^{5/2}}
	\sqrt{\frac{{y_1 y_2 y_4 }}{ y_{3} }} , t,q \r)  
}  \,.
\end{align}
In addition, after taking into account the contributions of $Z^{\text{sum}}$ given in \eqref{zsum4F}, we obtain the one-instanton contribution which can be expressed as
\begin{align}
& Z^{\textbf{one-instanton }}_{Sp(2)+1\mAS+4\mF} = \frac{-\sqrt{q t } }{(1-q)(1-t)} 	\chi_{8_c}^{SO(8)}[y] (A_2+A_1)  + \frac{ q+t}{(1-q)(1-t)  }  	\chi_{8_s}^{\rm SO(8)}[y] (A_2^2 +A_1^2)   \nn\\  
&+ \frac{q+t-\sqrt{q t}\chi_2^{SU(2)}   } {(1-q)(1-t) }  	\chi_{8_s}^{ SO(8)}[y] A_2 A_1 -\frac{ (q+t) (q+t-\sqrt{q t}\chi_2^{SU(2)}     )  }{(1 - q) (1 - t) \sqrt{q t}  } 	\chi_{8_c}^{\rm SO(8)}[y] 
\nn\\
&\times (A_2^2A_1+A_1 A_2^2)  + \frac{(q + t)^2 (q +t -\sqrt{q t} \chi_2^{SU(2)}  )  }{(1 - q) q (1 - t) t }   	\chi_{8_s}^{ SO(8)}[y] A_2^2A_1^2  +\mathcal{ O}(A_1^3;A_2^3), 
\end{align}
where 
\begin{align}
\chi_{8_c}^{\rm SO(8)}[y]&=\sum\limits_{i\neq j\neq k\neq l}  \frac{\sqrt{y_i}}{\sqrt{y_j}  \sqrt{y_k} \sqrt{y_l} }+\frac{\sqrt{y_j}  \sqrt{y_k} \sqrt{y_l} }{\sqrt{y_i}}
,\quad i,j,k,l=1,\cdots, 4 \,, \nn\\
\chi_{8_s}^{\rm SO(8)}[y]&=\sum\limits_{i\neq j\neq k\neq l} \sqrt{y_iy_jy_k y_l }+ \frac{1} { \sqrt{y_iy_jy_k y_l }}+ \sqrt{\frac{y_i y_j}{y_k y_l} },\quad i,j,k,l=1,\cdots, 4\,.
\end{align}
In order to compare with localization result \ref{oneinstloc} with $N_f=4$, we define $Q_8=e^m, A_{1}=e^{\a_{1}}, A_{2}=e^{\a_{2}}, y_i=e^{m_i} (i=1, \cdots, 4)$ to make characters manifest.

\begin{figure}[]
	\centering
	\includegraphics[width=2in]{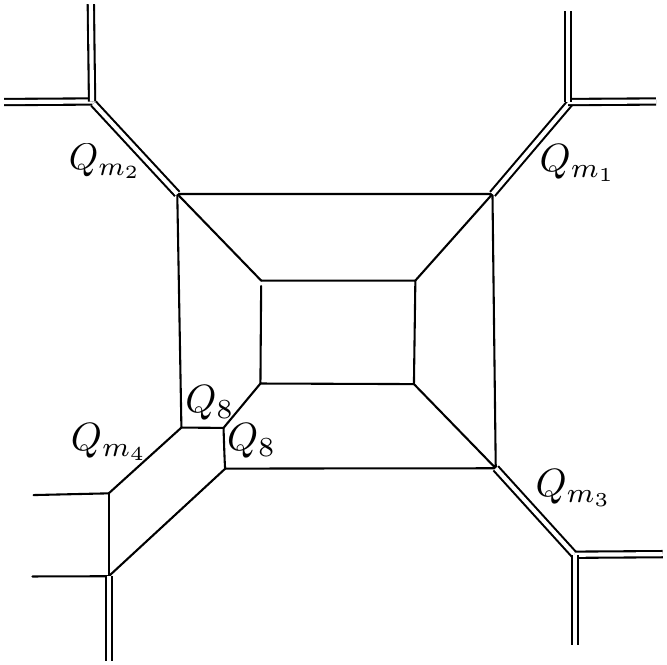}
	\caption{ Hanany-Witten moving of flavors could  create overlapped lines on the subweb that represents anti-symmetric matter $\mAS$.   }
	\label{fighwnf4}
\end{figure}

\paragraph{Cross $\mAS$ subweb.}
One can perform Hanany-Witten moves to make one D7-brane downward to infinity, which could cross the $\mAS$ subwebs as depicted in Figure \ref{fighwnf4}.  For this diagram, $T_2$-tuning is still correct and leads to correct partition function.
One can count the number of different types of Higgsed $T_2$ subwebs, and would find there are three \textbf{Case B}, one \textbf{Case C} and three \textbf{Case D} on this diagram. 
\begin{figure}[]
	\centering
	\begin{tikzpicture}[scale=0.8, line width=0.6pt ]
	\draw(0.5,1.1)--(0.5,0.9);	\draw(0.4,1.1)--(0.4,0.9);
	\draw(0.4,2.9)--(0.4,3.1);\draw(0.5,2.9)--(0.5,3.1);
	\draw(0.5,-1.1)--(0.5,-0.9);	\draw(0.4,-1.1)--(0.4,-0.9);
	\draw(0.4,-2.9)--(0.4,-3.1);\draw(0.5,-2.9)--(0.5,-3.1);
	\draw(-0.1,-0.1)--(0.1,0.1);	\draw(-0.1,0.1)--(0.1,-0.1);
	\draw (-1.5,-1)--(-1.5,1)--node[above=0pt]{{\tiny $Q_{B}$}}node[below=0pt]{{\tiny $\v_2$}}(1.5,1)--node[left=0pt]{{\tiny $Q_{F}$}}(1.5,-1)--node[below=0pt]{{\tiny $\v_1$}}(-1.5,-1);
	
	\draw (-2.5,-3)--
	node[below=4 pt,right=-4pt]{{\tiny $Q_{m_{44} }$}}(-3.5,-4)--(-4.5,-4)--(-4.5,-3); 	\draw (-3.5,-2)--
	node[left=0 pt]{{\tiny $Q_{m_{4} }$}}(-4.5,-3); 
	
	\draw (2.5,3)--node[left=5pt, above=0pt]{{\tiny $Q_{m_{11} }$}} (3.5,4)--node[above=0pt]{{\tiny $Q_{f_{3}}$}}node[below=0pt]{{\tiny $\v_5$}}(4.5,4)--node[right=0pt]{{\tiny $Q_{f_{2}}$}}(4.5,3)--node[right=0pt]{{\tiny $Q_{m_1 }$}}(3.5,2);
	\draw (4.5,3)--(6.5,3);\draw (3.5,4)--(3.5,6);\draw (4.5,4)--node[right=2pt]{{\tiny $Q_{f_1 }$}} (5.5,5)--(5.5,6);\draw (5.5,5)--(6.5, 5);
	
	\draw (-4.5,-4)--node[below=6pt,right=-6pt]{{\tiny $Q_{g_2 }$}} (-5.5,-5);
	
	\draw (-1.5,-1)--node[left=4pt,above=0pt]{{\tiny $Q_{k4}$}}(-2.5,-2)--node[below=-2pt]{{\tiny $Q_8$}}node[above=-2pt]{{\tiny $\a_8$}}(-3.5,-2)--node[left=-2pt]{{\tiny $Q_{33}$}}(-3.5,2)--node[above=-2pt]{{\tiny $Q_{1}$}}node[below=0pt]{{\tiny $\a_1$}}(-2.5,2)--node[left=4pt,below=0pt]{{\tiny $Q_{k1}$}}(-1.5,1);
	\draw (1.5,1)--node[right=4pt,below=0pt]{{\tiny $Q_{k2}$}}(2.5,2)--node[above=-2pt]{{\tiny $Q_4$}}node[below=-2pt]{{\tiny $\a_4$}}(3.5,2)--node[right=-2pt]{{\tiny $Q_{44}$}}(3.5,-2)--node[above=-2pt]{{\tiny $\a_5$}}node[below=-2pt]{{\tiny $Q_5$}}(2.5,-2)--node[right=4pt,above=0pt]{{\tiny $Q_{k3}$}}(1.5,-1);
	\draw (2.5,2)--node[right=-4pt]{{\tiny $Q_{3}$}}(2.5,3)--node[above=0pt]{{\tiny $Q_{22}$}}node[below=0pt]{{\tiny $\v_4$}}(-2.5,3)--node[left=-4pt]{{\tiny $Q_{2}$}}(-2.5,2);
	\draw (2.5,-2)--node[right=-4pt]{{\tiny $Q_{6}$}}(2.5,-3)--node[below=0pt]{{\tiny $Q_{11}$}}node[above=0pt]{{\tiny $\v_3$}}(-2.5,-3)--node[left=-4pt]{{\tiny $Q_{7}$}}(-2.5,-2);
	
	\draw[red,dashed](2.5,-3)--(3.5,-3)--(3.5,-2);
	
	\draw(-5.5,5)--node[left=-2pt]{{\tiny $Q_{f_{33}}$}}(-4.5,4);

	\draw(-2.5,3)--node[right=0pt]{{\tiny $Q_{m_{22} }$}} (-3.5,4);\draw(-3.5,2)--node[below=1pt]{{\tiny $Q_{{m_2}}$}}(-4.5,3)--node[left=-1pt]{{\tiny $Q_{f_{4}}$}}(-4.5,4)--node[above=0pt]{{\tiny $Q_{f_{34}}$}}node[below=0pt]{{\tiny $\v_7$}}(-3.5,4);

	\draw (2.5,-3)--node[left=3pt, below=0pt]{{\tiny $Q_{m_{33} }$}}  (3.5,-4)--node[below=0pt]{{\tiny $Q_{f_{56}}$}}node[above=-2pt]{{\tiny $\v_6$}}(4.5,-4)--node[right=0pt]{{\tiny $Q_{f_{5}}$}}(4.5,-3)--node[right=0pt]{{\tiny $Q_{m_3 }$}}(3.5,-2);
	\draw (4.5,-3)--(6.5,-3);\draw (3.5,-4)--(3.5,-6);\draw (5.5,-5)--(5.5,-6);\draw (5.5,-5)--(6.5, -5);\draw(5.5,-5)--node[right=0pt]{{\tiny $Q_{f_6 }$}} (4.5,-4);
	
	\draw (-1.5,-1)--(-2.5,-2)--(-3.5,-2)--(-3.5,2)--(-2.5,2)--(-1.5,1);
	\draw (1.5,1)--(2.5,2)--(3.5,2)--(3.5,-2)--(2.5,-2)--(1.5,-1);
	\draw (2.5,2)--(2.5,3)--(-2.5,3)--(-2.5,2);
	\draw (2.5,-2)--(2.5,-3)--(-2.5,-3)--(-2.5,-2);
	\draw (2.5,3)--(3.5,4)--(4.5,4)--(4.5,3)--(3.5,2);
	
	\draw (4.5,3)--(6.5,3);\draw (3.5,4)--(3.5,6);\draw (4.5,4)--(5.5,5)--(5.5,6);\draw (5.5,5)--(6.5, 5);
	
	\draw (2.5,-3)--(3.5,-4)--(4.5,-4)--(4.5,-3)--(3.5,-2);
	\draw (4.5,-3)--(6.5,-3);\draw (3.5,-4)--(3.5,-6);\draw (5.5,-5)--(5.5,-6);\draw (5.5,-5)--(6.5, -5);\draw(5.5,-5)--(4.5,-4);
	
	\draw (-4.5,3)--(-6.5,3);\draw (-3.5,4)--(-3.5,6);\draw (-5.5,5)--(-5.5,6);\draw (-5.5,5)--(-6.5, 5);\draw(-5.5,5)--(-4.5,4);
	
	\draw(-2.5,3)--(-3.5,4);\draw(-3.5,2)--(-4.5,3)--(-4.5,4)--(-3.5,4);
	\draw[dashed,red](-2.5,2)--(-4.5,4);\draw[dashed,red](-4.5,4)--(-6.5,4); \draw[dashed,red](-4.5,4)--(-4.5,6);
	\draw[dashed,red](-4.5,-4)--(-4.5,-6);\draw[dashed,red](-4.5,-4)--(-6.5,-4);\draw[red,dashed](-2.5,-2)--(-4.5,-4); \draw[dashed,red](-3.5,2)--(-3.5,3)--(-2.5,3);
	\draw[dashed,red](2.5,2)--(4.5,4);\draw[dashed,red](4.5,4)--(6.5,4); \draw[dashed,red](4.5,4)--(4.5,6);  \draw[dashed,red](3.5,2)--(3.5,3)--(2.5,3);
	
	\draw(-2.5,-3)--(-3.5,-4)--(-3.5,-6);\draw(-3.5,-2)--(-4.5,-3)--(-6.5,-3); \draw(-4.5,-4 )--(-5.5,-5)--(-5.5,-6); \draw(-5.5,-5)--(-6.5,-5);
	\draw(-3.5,-4)--node[ below=0pt]{{\tiny $Q_{g_{1} }$}} node[ above=-2pt]{{\tiny $\v_8$}} (-4.5,-4); \draw(-4.5,-4)--node[ left=-3pt]{{\tiny $Q_{g_{3} }$}} (-4.5,-3); 
	
	\draw(-4.4,-5.5)--(-4.4,-6.5);	\draw(-4.6,-5.5)--(-4.6,-6.5); 
	\draw (-4.5,-6.6) circle [radius=5pt];
	\shadedraw[bottom color=blue,top color=blue!20, draw=blue!10!black]
	(-4.5,-6.6) circle (5pt);
	\draw[dashed,red](2.5,-2)--(4.5,-4);\draw[dashed,red](4.5,-4)--(6.5,-4); \draw[dashed,red](4.5,-4)--(4.5,-6);  
	\draw[orange](-6,0)--(6,0);\draw[orange](0,-6)--(0,6); 
	\node [above] at (-4.5,-7.5) {$\textbf{ Case B}$};
	
	\end{tikzpicture}
	\caption{The unHiggsed diagram of the web depicted in Figure \ref{fighwnf4}.
		The Higgsed $T_2$ sub-diagram on the left bottom is of type \textbf{Case B}. 
	}
	\label{fig:nf4hwunhiggsing}
\end{figure}
Apart from the same tuned K\"ahler  parameters \eqref{tuningnnf3} for $Sp(2)+1\mAS + 3\mF $, there are two additional tuned K\"ahler parameters associated with the \textbf{Case B} type of Higgsed $T_2$ sub-web on the left bottom of this diagram, as shown in Figure \ref{fig:nf4hwunhiggsing}:
\begin{align}
Q_{g_1}=Q_{g_2}=\sqtq  \,.
\end{align}

\subsection{\texorpdfstring{$Sp(3)_0+ 1\mathbf{AS} $}{Sp(3)+1AS}}
For higher rank cases, as there are more Young diagrams to be summed over, the computation is more involved. Since the computation is straightforward, we skip the computations. Instead, as a representative example for higher rank cases, we discuss $Sp(3)_0+ 1\mathbf{AS}$, and the structure of the partition function.  
\begin{figure}[ht]
	\centering
	\includegraphics[width=1.8in]{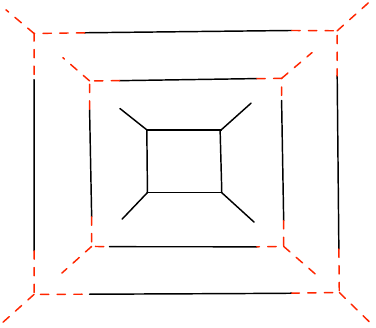}
	\caption{When $\mAS$ is massless, the web of  $Sp(3)+1\mAS$ factorizes and 	thus $Sp(3)+1\mAS$ equals $SU(2)\times SU(2) \times SU(2)$. }
	\label{fig:masslessrank3}
\end{figure}

\begin{figure}[]
	\centering
	\includegraphics[width=6in]{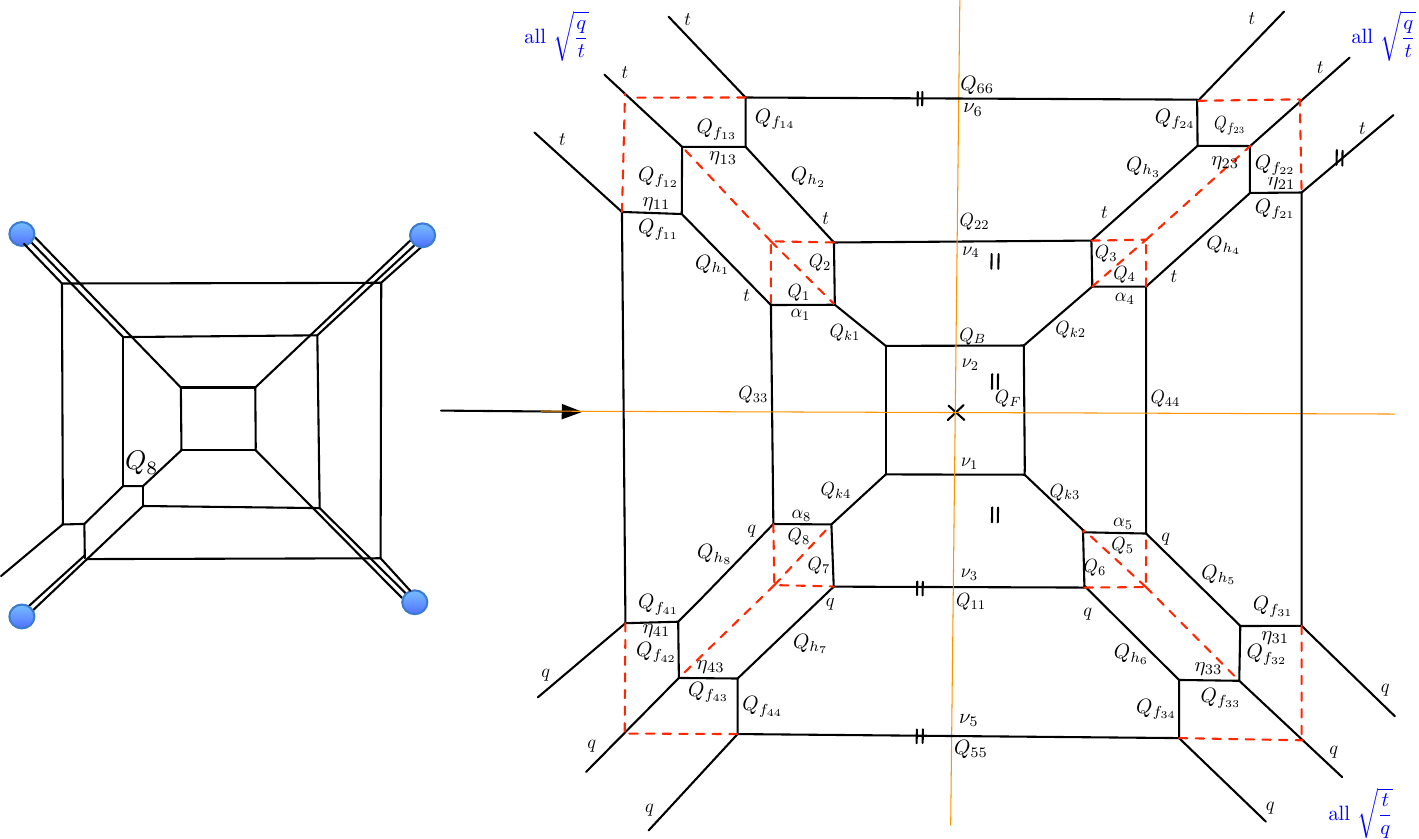}
	\caption{The left web is for the massive $\mAS$ with mass fugacity $Q_8$. The right diagram is its unHiggsed diagram. In the right diagram, we have assigned the virtual lines and the values for some tuned parameters. All tuned parameters above the horizontal orange line are given value $\sqrt{q/t}$ and all tuned parameters below this line are given $\sqrt{t/q}$.}
	\label{fig:rank3}
\end{figure}

Before discussing massive $\mAS$ cases of the partition function for $Sp(3)_0+ 1\mathbf{AS}$, let us first discuss the massless limit of $\mAS$ with $Q_8=\sqtq$. Then the partition function factorizes 
\begin{eqnarray}
&&Z_{Sp(3)+ 1\mathbf{AS} }= Z_{\textbf{first layer}} \cdot Z_{\textbf{second layer}}\cdot Z_{\textbf{third layer}} \nn \\
&&=Z_{SU(2)}\l(Q_F,Q_B\r) \cdot Z_{{SU(2)}}\l(Q_F {Q_{k4}}^2 \frac{t}{q}, Q_B {Q_{k_4}}^2 \frac{t}{q}  \r) \cdot Z_{ {SU(2)} } \l( Q_F {Q_{h_4}}^2{Q_{k_4}}^2, Q_B {Q_{h_4}}^2{Q_{k_4}}^2     \r)\,, \qquad\quad
\end{eqnarray}
where we choose $Q_F, Q_B, Q_{k4}, Q_{h_4}$ as independent parameters. At the massless limit the diagram turns out to be isolated webs as depicted in Figure \ref{fig:masslessrank3}. 
Turning on the mass parameter for $\mAS$ leads to the left web in Figure \ref{fig:rank3}. The associated tuned K\"ahler parameters to its unHiggsed diagram are given by
\begin{align}
& 
Q_1=Q_2=Q_3=Q_4=Q_{f_{11}}=Q_ {f_{12}}=Q_{f_{13}}=Q_{f_{14}}= Q_ {f_{21}}=Q_{f_{22}}=Q_{f_{23}}=Q_{f_{24}  }= \sqqt \,,\nn\\
&Q_ 5=Q_6=Q_{ f_{31}}=Q_{ f_{32}}=Q_{f_{33}}=Q_{f_{34}}=Q_{f_{43}}=Q_{f_{44}} =\sqtq  
\end{align}
Gauge theory parameters $(A_i, u)$ should be assigned to the following independent K\"{a}hler parameters
\begin{eqnarray}
Q_F={A_2}^2,\quad Q_{k4}=\frac{A_1}{A_2 Q_8},\quad Q_{h_4}=\frac{ A_3}{A_1} \sqtq,\quad Q_B={A_2}^2 u\,.
\end{eqnarray}
\paragraph{Perturbative contribution.}
We find
\begin{align}
&Z^{\pertI}_{Sp(3)+ 1\mathbf{AS} }=
 \frac{  
M\l(   \frac{ A_1}{ A_2 Q_8  } \sqtq , t,q  \r)
M\l(   \frac{ A_1 A_2 }{ Q_8 } \sqtq  ,  t,q  \r)
M\l(   \frac{ A_3}{ A_1 Q_8  } \sqtq ,  t,q  \r)
M\l(   \frac{A_1 A_3}{{Q_8}^2} ,  q,t  \r) 
}
{
M\l( {A_1}^2,  t,q  \r)  	M\l( \frac{A_1}{A_2},  t,q  \r)
M\l( A_1  A_2,   t,q  \r)  M\l( {A_2}^2 ,   t,q  \r)  
M\l( A_1  A_3,   t,q  \r)  M\l( {A_3}^2,   t,q  \r)  
} \nn\\
&\times \frac{  M\l(   {Q_8}^2,  t,q  \r) M\l(   {Q_8} \sqtq,  t,q  \r)  }
{  M\l( \frac{ Q_8 A_3 }{A_1  } \sqqt,   t,q  \r) 
	M\l( {A_2}^2,   q,t  \r)  M\l(  \frac{{A_1}^2}{Q_8} \sqtq, q, t \r)   M\l(  \frac{{A_3}^2}{{Q_8}^2} \frac{t}{q} , q, t \r) 
		 	 }\,,
\end{align}
	\begin{align}
	&Z^{\textbf{pert-II }}_{Sp(3)+ 1\mathbf{AS} }=\frac{ \textbf{numerator}}{\textbf{denormator}} \, \\
&\textbf{numerator}=
 M\l( \frac{Q_8 A_3 }{A_1} \sqqt, t,q  \r)  
	  M\l( \frac{A_1^2 }{Q_8} \sqtq , q,t  \r) 
 M\l( \frac{ A_1 A_3 }{Q_8} \sqtq , t,q  \r) 
  M\l( \frac{A_3 }{A_2 Q_8 } \sqtq, t,q  \r) 
   \nn\\
   &\times M\l( \frac{A_2 A_3 }{Q_8 } \sqtq , t,q  \r) M\l(Q_8 \sqtq , t,q  \r)^2  
     M\l( \frac{Q_8 A_1 }{A_2} \sqtq , t,q  \r) 
      M\l( Q_8 A_1 A_2  \sqtq , t,q  \r)  
       \nn\\
       &\times M\l( \frac{Q_8 A_3 }{A_1} \sqtq , t,q  \r)   
        M\l( Q_8 A_1 A_3  \sqtq , t,q  \r)  
         M\l( \frac{Q_8 A_3 }{A_2} \sqtq , t,q  \r)   
            M\l( Q_8 A_2 A_3  \sqtq , t,q  \r)   
            \nn\\
            &\times
               M\l( \frac{{A_3}^2}{{Q_8}^2} \frac{t}{q}  , q,t  \r)  \,,\nn\\
& \textbf{denormator}=
 M\l( {A_1}^2,q,t  \r) 
M\l( \frac{A_1}{A_2}, q,t  \r)
M\l( A_1 A_2,q,t  \r)
M\l( \frac{A_3}{A_1}, q,q  \r)
M\l( \frac{A_3}{A_2}, t,q  \r)
\nn\\
& \times
M\l( A_2 A_3,  t,q  \r)
M\l( {Q_8}^2,t,q  \r) 
M\l( {Q_8}^3 \sqqt,t,q  \r)
M\l( \frac{A_3}{A_1}\frac{t}{q}, t,q  \r)
M\l( A_1A_3\frac{t}{q}, t,q  \r)
\nn\\
&\times
M\l( \frac{A_3}{A_2}\frac{t}{q}, t,q  \r)
M\l( A_2A_3\frac{t}{q}, t,q  \r)
M\l( {A_3}^2\frac{t}{q}, t,q  \r)
M\l( \frac{A_1 A_3}{{Q_8}^2}\frac{t}{q}, t,q  \r) \nn\,. 
\end{align}
Then the full perturbative partition function is given by
\begin{align}
Z^{\textbf{pert}}_{Sp(3)+1\mAS} =\frac{ Z^{\pertI}_{Sp(3)+ 1\mathbf{AS} } \cdot Z^{\pertII}_{Sp(3)+ 1\mathbf{AS} } }{Z^{\rm extra}_{Q_8}}\,,
\end{align}
where the extra factor in denominator is
\begin{align}
Z_{Q_8}^{\rm extra}= \frac{1}{M \l(  {Q_8}^3 \sqqt, t,q \r)  }\,.
\end{align}
In order to match with localization computation in \eqref{pertlocalization} when $N=3$, one needs to permute $A_1\rightarrow A_2,  A_2 \rightarrow A_3,  A_3 \rightarrow A_1$.
\paragraph{Instanton contribution}
We obtain the one-instanton contribution expanded as
\begin{align}
&Z^{\textbf{one-instanton}}_{Sp(3)+1\mAS} = 
\frac{q+t }{(1-q) (1-t) } (A_1^2+A_2^2 +A_3^2) 
+ \frac{q+t- \sqrt{q t}\chi^{SU(2)}_2  }{(1-q)(1-t) } (A_2A_3 + A_1A_3+A_1A_2)     
\nn\\
&
 + \frac{(q+t)^2(q+t- \chi^{SU(2)}_2)  }{(1-q)(1-t)\sqrt{q t}}    (A_1^2A_3^2 + A_1^2A_3^2+ A_2^2A_3^2 )
   \\
  & +  \frac{(q+t)( q^2+3 q t +t^2 -2 \sqrt{q t}(q+t )    \chi_2^{SU(2)}   +  q t  \chi_3^{SU(2)}  )     }{(1-q)(1-t)q t     }    (A_1A_2A_3^2+ A_1 A_2^2 A_3 +A_1^2 A_2A_3   )
     \nn\\
     &  +   \frac{(q+t)^3( q^2+3 q t +t^2 -2 \sqrt{q t}(q+t )    \chi_2^{SU(2)}   +  q t  \chi_3^{SU(2)}  )     }{(1-q)(1-t)q^2 t^2     }     A_1^2 A_2^2 A_3^2 +\mathcal{O}(A_1^2; A_2^2; A_3^2)\,, \nn
\end{align}
which is equal to the localization result \eqref{oneinstloc}.

\section{Conclusion}\label{sec:conclusion}
In this paper, we computed the Nekrasov partition function for 5D $\mathcal{N}=1$ $Sp(2)$ gauge theories with an antisymmetric hypermultiplet and $N_f \le 4$ flavors using the refined topological vertex method based on a 5-brane web diagram with the nonzero mass of the antisymmetric matter. The corresponding 5-brane web diagram has jumps on $(p,q)$-plane (or its dual diagram is generically non-toric), and can be regarded as a Higgsed 5-brane web diagram. To implement the topological vertex method on these 5-brane webs, we considered its unHiggsed 5-brane web and properly tuned the K\"{a}hler parameters associated with the Higgsed edges. As such Higgsing can be considered locally as Higgsing on a 5-brane web diagram for $T_2$-theory, we developed systematic tunings of K\"{a}hler parameters for a $T_2$-diagram, so that such $T_2$-tuning can be applicable to various Higgsed 5-brane web diagrams.  It is also straightforward to extend the $T_2$-turning to $T_N$-tuning. As an example of $T_3$-tuning, we considered 5D $\mathcal{N}=1$ $Sp(3)$ gauge theory with an antisymmetric. We checked that our results agree with known results based on the ADHM method, up to two instanton contributions.

 By redefining the parameters of the theory to make the fiber-base duality manifest \cite{Mitev:2014jza}, the partition functions expressed in terms of new parameters explicitly show enhanced global symmetry $E_{N_f+1}\times SU(2)_{\rm antisym.}$, yielding the Gopakumar-Vafa invariants of 5D $Sp(2)$ gauge theories with an antisymmetric hypermultiplet and $N_f \le 4$ flavors.

Though we did not present the result with a higher number of flavors due to prolonged computation time with higher flavors, it is straightforward to compute the partition function for $Sp(2)$ theory with $N_f=5,6,7,8$ and for higher rank $Sp(N)$ gauge theory.

Along with new findings of 5-brane web for 5D superconformal theories with various matter fields, it would be interesting to compute the partition functions using the topological vertex method on the corresponding 5-brane webs of various matters and see enhanced global symmetries. Another interesting direction to pursue is to check various dualities of 5D rank 2 superconformal theories \cite{Jefferson:2018irk} based on 5-brane webs \cite{Hayashi:2018lyv} by computing the partition functions based on each 5-brane web.

Here, we discussed the $T_2$-tuning of the K\"{a}hler parameters for jumps. It would also be interesting to study other physical systems such as defects by tuning K\"{a}hler parameters to other values and find the corresponding tuning method.

\acknowledgments
We would like to thank Jin Chen, Hirotaka Hayashi, Hee-Cheol Kim, Kimyeong Lee, and Wenbin Yan for useful discussions. We also like to thank Futoshi Yagi for carefully reading the draft as well as valuable discussions. SC is grateful to ICTS for Asian string winter school, IHES for Localization summer school, and Babak Haghighat for long-term visiting YMSC. SC is supported by the TEAM programme of the Foundation for Polish Science co-financed by the European Union under the European Regional Development Fund (POIR.04.04.00-00-5C55/17-00).
SSK would like to gratefully acknowledge KIAS and YMSC at Tsinghua university for kind hospitality for his visit, and also like to acknowledge APCTP for hosting the Focus program "Strings, Branes and Gauge Theories." SSK is supported by the UESTC Research Grant A03017023801317.

\bigskip

\appendix 
\appendix

\section{Useful identities}\label{notation}
The Cauchy identities take the following forms:
\begin{align} 
	\sum_{\eta} s_ {\h / \la}(\bx) s_{\h/\u} (\by)&=\prod_{i,j=1}^{\infty}  (1-x_iy_j)^{-1}
	\sum_{\h} s_{\u/\h}(\bx) s_{\la/\h}(\by) \label{app:Cauchy1}, \\
	\sum_{\h} s_ {\h^T / \la}(\bx) s_{\h/\u} (\by)&=\prod_{i,j=1}^{\infty}  (1+x_iy_j)
	\sum_{\h} s_{\u^T/\h^T}(\bx) s_{\la^T/\h}(\by)\,,  \label{app:Cauchy2}
\end{align}
where the skew Schur functions satisfy 
\begin{align}
	&s_{\0/ \u}(\bx)=\delta _{\0 \u}=\delta _{\0 \u^T},~~|\u|=|\u^T|,\\
	&s_{\v/\0}(\bx)=s_{\v}(x),~~	s_{\u/ \v}(\lambda \bx) = \lambda^{|\u|-|\v|} s_{\u/\v}(\bx), \\
&	\sum\limits_{\v}(-1)^{|\v|} s_{\u/\v} (\bx) s_{\v^T/\la^T}(\bx)=(-1)^{|\u|}\delta_{\u \la}\,.
\end{align}
Using this, one can perform the Young diagram sums along non-preferred directions. 

The following functions are the functions defined with topological vertex (see also \cite{Haghighat:2013gba,Bao:2013pwa}):  
\begin{align}
	& \tZ_{\v}(t,q):=\prod\limits_{(i,j)\in \v   } 
	\l(     1-q^{  \v_i-j} t^{\v_j^T -i+1}       \r)^{-1}  \,, \nn\\
		&	||\tZ_{\u}(t,q)||^2:= \tZ_{\u^T}(t,q) \tZ_{\u}(q,t)\,,\nn\\
	&N_{\u \v}(Q; t, q) := \prod\limits_{i, j=1}^\inf \frac{1- Q~ q^{\v_i -j}~ t^{\u_j^T-i+1}	}{	1- Q~ q^{-j} ~t^{-i+1}	} 
	\nn\\
&\qquad \qquad~ \quad =
	 \prod\limits_{(i,j)\in \v}\l(1-Q~q^{\v_i - j }~t^{\u_j^T -i+1}	\r)  
	\prod\limits_{(i,j) \in \u} \left(	1-Q~q^{-\u_i+j-1}~t^{-\v_j^T+i} 	\right)\, ,	\nn\\
&	N_{\v}^{\rm half,+}(Q; t,q):= N_{\0 \v}(Q\sqrt{ \frac{q}{t}}, t,q)  \,,\nn \\
	&N_{\v}^{\rm half,-}(Q; t,q):= N_{\v \0}(Q\sqrt{ \frac{q}{t}}, t,q) \,,\nn\\  
&	M(Q, t,q):= \prod\limits_{i, j=1}^\inf  (1-Q~q^i t^{j-1}) =\text{exp} \l( -\sum\limits_{n=1}^{\inf}  \frac{ Q^n \l( \frac{q}{t} \r)^{  \frac{n}{2}}  }{ n(q^ {  \frac{n}{2}}-q^{-  \frac{n}{2}}  ) (t^{  \frac{n}{2}} -t^{  -\frac{n}{2}}   )  }  \r)
 \,. 
\end{align}
We note that for some functions, we used slightly different convention from the notations in \cite{Bao:2013pwa}
\begin{eqnarray}
	N_{\u \v}(Q; t^{-1}, q^{-1}) = \mathcal{N}_{\u \v}(Q, t,q), ~~ M(Q;t,q) =\frac{1}{\mathcal{M}(Q, t,q)}\,, 
\end{eqnarray}
where the functions on the right hand are those defined in \cite{Bao:2013pwa}.
Other useful identities are as follows
	\begin{align}
			&	||\tZ_{\u}(t,q)||^2= 
		||\tZ_{\u^T}(q,t)||^2\,,\\
			&N_{\v}^{\rm half,+}  \l(Q; t^{-1},q^{-1}  \r) =N_{\v^{T}}^{\rm half,-}\l(Q; q^{-1},t^{-1} \r)\,,\\
		& N_{\v}^{\rm half,+}  \l(Q; t^{-1},q^{-1}  \r) = N_{\0 \v}  \l(Q\sqrt{ \frac{t}{q}}, t^{-1},q^{-1}  \r), \\
		 &N_{\v}^{\rm half,-}\l(Q; t^{-1},q^{-1} \r)=
		N_{\v \0}\l(Q\sqrt{ \frac{t}{q}}, t^{-1},q^{-1} \r) \,,\\
		& N_{\v\v}\l(1; t^{-1}, q^{-1} \r)= \frac{(-1)^{|\v|}  q^{ -\frac{||\v||^2}{2} }   t^{ -\frac{||\v^T||^2}{2} }  \l(\frac{q}{t}\r)^{   \frac{|\v|}{2}   }  }{ \tZ_{\v^T}(q,t) \tZ_{\v}(t,q)  }\,, \label{iden} \\
		&  N_{\v\v}\l(\frac{t}{q}; t^{-1}, q^{-1} \r)= \frac{(-1)^{|\v|}  q^{ -\frac{||\v^T||^2}{2} }   t^{ -\frac{||\v||^2}{2} }  \l(\frac{t}{q}\r)^{   \frac{|\v|}{2}   }  }{ \tZ_{\v^T}(q,t) \tZ_{\v}(t,q)  }\,.
	\end{align}

The following relations are useful, when one expands terms in the partition functions with respect to $Q$,
\begin{align}\label{flopidentity}
	&M\l(Q \sqrt{\frac{t}{q}}, t,q \r)=
	M\l(Q \sqrt{\frac{q}{t}}, q,t \r) \,,   \\
&	M(Q^{-1}, ~t,q)= M\l(Q \frac{t}{q},~ t, q\r) = M( Q,~ q,t )\,, \\
&	N_{\u \v}\l(\sqrt{\frac{t}{q}}Q^{-1}; t^{-1},q^{-1}\r)=(-Q)^{-|\u|-|\v|}t^{\frac{-||\u^T||^2+||\v||^2}{2}}
q^{\frac{||\u||^2-||\v||^2}{2}}
N_{\v \u} \l(\sqrt{\dfrac{t}{q}}Q; t^{-1},q^{-1}  \r)    \,. 
\end{align}

\section{Result from localization}
In \cite{Kim:2012gu, Hwang:2014uwa}, the partition function for  $Sp(N)+1\mAS+N_f\mF$ was computed via localization based on the ADHM method, which takes the following form
\begin{equation}\label{pertlocalization}
Z^{\textbf{perturbative}}_{ Sp(N)+ 1 \mAS + N_f \mF}=Z_{\rm vector}^{\pert} \cdot 	Z_{\mAS}^{\pert}(y) \cdot \prod\limits_{i=1}^{N_f} Z_{\mF}^{\pert}(y_i)  .
\end{equation}
For $Sp(2)+ 1 \mAS + N_f \mF$, 
	\begin{align}\label{locpert}
Z_{\mF}^{\pert}(y_i)=& M\left( \frac{A_1}{y_i} \sqtq, t,q\right) M\left( A_1 y_i \sqtq, t,q\right) M\left( \frac{A_2}{y_i} \sqtq, t,q\right) M\left( A_2 y_i \sqtq, t,q\right) ,\nn\\
	Z_{\mAS}^{\pert}(y)= &M\left(y A_{1}A_{2}\sqtq, t,q\right) M\left( \frac{A_1A_2}{y}\sqtq, t,q\right) M\left(y \frac{A_1}{A_2}\sqtq, t,q\right) M\left(\frac{~A_1}{y A_2} \sqtq, t,q \right)  
	\nn\\
&	\times M\left(y \sqtq,t,q\right)^2
	\,,   \\
	Z_{\rm vector}^{\pert}= &\frac{1} {
		M(A_1 A_2,t,q)M(A_1A_2,q,t)  M(\frac{A_1}{A_2},t,q  ) M(\frac{A_1}{A_2},q,t)     
	} \nn\\
& \times \frac{1}{M({A_1}^2,t,q) M({A_1}^2,q,t)  M({A_2}^2,t,q) M({A_2}^2,q,t) }
\,.
	\end{align}
For  $Sp(3)+ 1 \mAS + N_f \mF$, 
	\begin{align}
	Z_{\mF}^{\pert}(y_i)=& \prod_{\a=1}^3M\left( \frac{A_{\a}}{y_i} \sqtq, t,q\right) M\left( A_{\a} y_i \sqtq, t,q\right)  ,\nn\\
	Z_{\mAS}^{\pert}(y)=&~M\left(y \sqtq,t,q\right)^3 \prod_{1\leq \a <\b\leq 3}  M\left(y A_{\a} A_{\b}\sqtq, t,q\right) M\left(\frac{ A_{\a} A_{\b}}{y}\sqtq, t,q\right)  \nn\\ 
&	\times
	M\left( \frac{y A_{\a}}{~A_{\b}}\sqtq, t,q\right) M\left( \frac{~A_{\a}}{y A_{\b}}\sqtq, t,q \right)   \,,   \nn\\
	Z_{\rm vector}^{\pert}= &\frac{1} {
		\prod\limits_{1\leq \a <\b\leq 3} M(A_{\a} A_{\b},t,q)M(A_{\a} A_{\b},q,t)  M(\frac{A_{\a}}{A_{\b}},t,q  ) M(\frac{A_{\a}}{A_{\b}},q,t)   
	 } 
 \nn\\
	&  \times  \frac{1}{   \prod\limits_{\a=1}^3 M(A_{\a}^2,t,q) M(A_{\a}^2,q,t)   }  \,.
	\end{align}

The one-instanton contribution does not involve integral and is given by
\bsmall
\begin{align}\label{oneinstloc}
Z^\textbf{one-instanton}_{ Sp(N)+ 1 \mAS + N_f \mF} &=
\frac{1}{2   } \l(
\frac{  \prod\limits_{i=1}^{N_f}2 ~\text{sinh} \frac{m_i} {2}~ }{2~\text{sinh} \frac{\epsilon_{+} \pm \epsilon_{-}} {2} } \cdot
\frac{
	\prod\limits_{i=1}^{N}2~\text{sinh} \frac{m \pm \a_i }{2}  - \prod\limits_{i=1}^{N} 2~\text{sinh} \frac{\pm \a_i+\epsilon_{+} }{2}  } 
{2~\text{sinh} \frac{m \pm \epsilon_{+} }{2}   \prod\limits_{i=1}^{N}2~\text{sinh} \frac{\pm \a_i+\epsilon_{+} }{2}  
} \right.\cr
&\qquad+\left.
 \frac{\prod\limits_{i=1}^{N_f}2~ \rm cosh \frac{m_i}{2}    ~ }{2~\text{sinh} \frac{\epsilon_{+} \pm \epsilon_{-}} {2} } \cdot
\frac{ \prod\limits_{i=1}^{N} 2~\text{cosh} \frac{m \pm \a_i }{2}    -  \prod\limits_{i=1}^{N}2~\text{cosh} \frac{\pm \a_i+\epsilon_{+} }{2}   }
{2~\text{sinh} \frac{m \pm \epsilon_{+} }{2}  \prod_{i=1}^{N} 2~\text{cosh} \frac{\pm \a_i+\epsilon_{+} }{2}    }
\r)\,,
\end{align}
\esmall
where $m_i$ are masses for fundamental flavors and $m$ is the mass for an antisymmetric hypermultiplet.

\bibliographystyle{JHEP}
\bibliography{ref}

\end{document}